\documentclass[fleqn,3p,authoryear]{elsarticle}

\usepackage{tgtermes}
\usepackage[cmintegrals,vvarbb]{newtxmath}
\DeclareMathAlphabet{\mathcal}{OMS}{cmsy}{m}{n}

\usepackage{amssymb}
\usepackage{amsmath}
\usepackage{bm}
\usepackage{graphicx}
\usepackage[dvipsnames]{xcolor}
\usepackage{tikz}
\usepackage{pgfplots}
\usepackage{subcaption}
\usepackage{adjustbox}
\usepackage[colorlinks=true]{hyperref}
\usepackage[capitalise]{cleveref}
\usepackage{braket}
\usepackage{nicefrac}
\usepackage{float}
\usepackage{multirow}
\pgfplotsset{compat=newest}

\newcommand{\dee}{\,\mathrm{d}}

\makeatletter
\AtBeginDocument{\def\@citecolor{cyan}}
\AtBeginDocument{\def\@linkcolor{cyan}}
\makeatother

\let\epsilon\varepsilon

\begin{document}

\title{High-accuracy phase-field models for brittle fracture based on \\ a new family of degradation functions}

\author[uib]{Juan Michael Sargado\corref{cor1}}
\ead{Juan.Sargado@uib.no}
\author[uib]{Eirik Keilegavlen}
\ead{Eirik.Keilegavlen@uib.no}
\author[uib,cmr]{Inga Berre}
\ead{Inga.Berre@uib.no}
\author[uib,pu]{Jan Martin Nordbotten}
\ead{Jan.Nordbotten@uib.no}

\cortext[cor1]{Corresponding author}
\address[uib]{Department of Mathematics, University of Bergen, All\'egaten 41, 5007 Bergen, Norway}
\address[cmr]{Christian Michelsen Research, Fantoftvegen 38, 5072 Bergen, Norway}
\address[pu]{Department of Civil and Environmental Engineering, Princeton University, E-208 E-Quad, Princeton, NJ 08544, USA}

\begin{abstract}
Phase-field approaches to fracture based on energy minimization principles have been rapidly gaining popularity in recent years, and are particularly well-suited for simulating crack initiation and growth in complex fracture networks. In the phase-field framework, the surface energy associated with crack formation is calculated by evaluating a functional defined in terms of a scalar order parameter and its gradients, which in turn describe the fractures in a diffuse sense following a prescribed regularization length scale. Imposing stationarity of the total energy leads to a coupled system of partial differential equations, one enforcing stress equilibrium and another governing phase-field evolution. The two equations are coupled through an energy degradation function that models the loss of stiffness in the bulk material as it undergoes damage. In the present work, we introduce a new parametric family of degradation functions aimed at increasing the accuracy of phase-field models in predicting critical loads associated with crack nucleation as well as the propagation of existing fractures. An additional goal  is the preservation of linear elastic response in the bulk material prior to fracture. Through the analysis of several numerical examples, we demonstrate the superiority of the proposed family of functions to the classical quadratic degradation function that is used most often in the literature.
\end{abstract}


\maketitle
\section{Introduction}
The accurate simulation of fracture evolution in solids is a major challenge for computational algorithms, in large part due to crack paths that are generally unknown a priori. In this regard, phase-field approaches have shown great potential with their ability to automatically determine the direction of crack propagation through minimization of an energy functional. The phase-field framework naturally handles the emergence of phenomena such as crack nucleation and branching without the need to introduce additional criteria. In particular, formulations derived from the variational theory of \cite{Francfort1998} have received a lot of attention from the applied mechanics community due to its strong ties to Griffith's theory for brittle fracture. Phase-field models belong to the category of continuum approaches for fracture propagation, utilizing a diffuse representation of cracks in place of actual discontinuities. The amount of crack regularization is controlled via a prescribed length scale $\ell$, which constitutes an additional parameter of the model.

The aim of the present work is to address two long standing issues associated with the phase-field formulation that arise in conjunction with use of the now-classical quadratic degradation function. The first has to do with premature stiffness degradation stemming from the evolution of damage around regions of stress concentration. The second and more serious issue deals with the observed dependence of simulated failure loads on the phase-field regularization parameter, a phenomenon that has largely gone unexplored in the literature until very recently. The problem is most noticeable in cases of crack growth emanating from a notch and undermines the usefulness of phase-field approaches in solving problems that involve crack initiation at a priori unknown locations \citep{Klinsmann2015}. The parameter $\ell$ was initially introduced by \cite{Bourdin2000} as a purely mathematical construct allows for the Griffith energy corresponding to a discrete crack to be recovered in the limit as $\ell$ goes to zero, in the sense of $\Gamma$-convergence \citep{Braides2006}. Of the two aforementioned issues, the first may be remedied by making use of alternative formulations as discussed in \cite{Pham2011}. On the other hand, the dependency of mechanical response on $\ell$ is not yet fully understood. It has been suggested recently \citep[e.g.][]{Bourdin2014} that the latter should be viewed as a material parameter that is closely connected to the crack nucleation stress. While we consider the  two points raised above as distinct issues, we nonetheless recognize that they are also closely inter-related, in particular because the first often exacerbates the second.

Our contribution in the following work is twofold. First, we provide a conceptual explanation of how the choice of length scale can result to either delay or acceleration of failure under quasi-static conditions. Secondly, we introduce a new family of degradation functions that allows for correctly reproducing the onset of failure for reasonably chosen arbitrary values of the regularization parameter. The latter point is important since the problem of regularization-dependent material response is not solved in the alternative formulations previously mentioned, and furthermore is not only confined to brittle fracture as demonstrated in the numerical results of \cite{Areias2016} on cracking in elastoplastic materials. Enthusiasm in the relatively new phase-field paradigm has led to a number of multi-physics applications, which include cracking in piezoelectric solids \citep{Miehe2010_jmps}, fluid-driven fracture propagation \citep{Mikelic2015,Miehe2015_jmps}, thermal shock-induced cracks \citep{Bourdin2014} and fragmentation of battery electrode particles \citep{Miehe2015_ijnme}. This underscores the need for quantitative accuracy with regard to the fracture model, particularly in the case of crack nucleation which is often the critical failure mechanism for many such applications.

The remainder of this paper is structured as follows: We begin in Section 2 with a discussion of important fundamental concepts underlying the phase-field approach as well as our motivation for pursuing the current research direction. Specifics regarding the formulation used in the present work are given in Section 3 which also includes important details with regard to numerics; in particular for the case where the phase-field evolution equation is nonlinear, we outline a linearization scheme based on a truncated Taylor series approximation that avoids the implementation of nested loops in the solution scheme. The following two sections contain the main novelties of the current work: section 4 begins with a numerical example demonstrating the apparent contradiction that is often seen between simulation results and what is expected from the $\Gamma$-convergence property of the fracture phase-field model. This is followed by a discussion which aims to explain why the latter alone is not sufficient to ensure the proper behavior of the model. In Section 5, we propose a new parametric family of degradation functions that aims to increase the accuracy of phase-field simulations by addressing key issues discussed in the previous section. Superiority of the resulting formulation over the classical model employing quadratic degradation is demonstrated via several numerical examples in Section 6. Finally, concluding remarks and outlook are given in Section 7.

\section{Theoretical aspects of phase-field modelling}
The phase-field framework was first introduced by \cite{Fix1983} and \cite{Langer1986} for modeling phase transitions in materials, and later extended to free discontinuity problems by \cite{Ambrosio1990} who worked on image segmentation. Its specific application to crack propagation in solids is much more recent, and is the result of independent work by researchers coming from the fields of physics \citep{Aranson2000,Karma2001} and applied mechanics \citep{Bourdin2000}. We adapt the latter viewpoint in this study, and furthermore note that while the original formulation introduced by Bourdin et al.\ has remained virtually unchanged in current usage, the argument on what constitutes proper solutions to the resulting equations as well as the meaning of key quantities is far from resolved. In view of this, we begin with a short review of theory pertaining to the phase-field formulation for brittle fracture along with a discussion of significant developments in the field in order to provide context for the present work.

\subsection{Brittle fracture: from Griffith to Francfort-Marigo}
\cite{Griffith1921} can be credited as being the first to formally state the thermodynamic principles governing the propagation of fractures in brittle materials that has become the foundation of modern linear elastic fracture mechanics. According to Griffith's theory, an existing crack will propagate when the rate of energy release $G$ associated with crack extension exceeds a critical value equal to the material fracture toughness, $\mathcal{G}_c$. This can be expressed via the following set of Kuhn-Tucker conditions \citep{Negri2008}:
\begin{subequations}
	\label{eq:griffith_principle}
	\begin{align}
		G - \mathcal{G}_c &\leq 0 \label{eq:griffith_criterion} \\
		\dot{a} &\geq 0 \label{eq:griffith_irreversibility} \\
		\left( G - \mathcal{G}_c \right) \dot{a} &= 0 \label{eq:slacknessCondition}
	\end{align}
\end{subequations}
where $\dot{a}$ denotes the rate of crack length increase. The first inequality precludes the case of unstable cracking where $G > \mathcal{G}_c$, while the second is an irreversibility constraint that prevents unphysical healing of fractures. Finally, condition \eqref{eq:slacknessCondition} implies that $G$ must be equal to $\mathcal{G}_c$ when the crack is growing, and conversely that a crack cannot extend when $G < \mathcal{G}_c$.  An important shortcoming of Griffith's theory is its inability to accommodate crack nucleation or predict the branching of fractures. An extension of the framework was developed by \cite{Francfort1998} in the form of a \emph{variational theory of fracture}, which is aimed at overcoming the earlier drawbacks through adoption of an energy minimization paradigm. It stipulates that the total potential energy corresponding to a linear elastic body $\Omega$ containing a set of crack points $\Gamma$ can be written as a sum of bulk and surface terms. That is,
\begin{equation}
	\label{eq:griffithFunctional}
	\Psi \left( \bm{u},\Gamma \right) = \int_{\Omega\backslash\Gamma} \frac{1}{2} \bm{\epsilon} \left( \bm{u} \right) : \mathbb{C}^e : \bm{\epsilon} \left( \bm{u} \right) \dee\Omega + \mathcal{G}_c \mathcal{H}^{n-1} \left( \Gamma \right)
\end{equation}
where $\bm{\epsilon} \left( \bm{u} \right) = \frac{1}{2} \left[ \nabla \otimes \bm{u} + \bm{u} \otimes \nabla \right]$ is the symmetric small-strain tensor, $\mathbb{C}^e$ is the standard linear isotropic elasticity tensor, and $\mathcal{H}^{n-1}$ is the $\left( n-1 \right)$-dimensional Hausdorff measure giving the surface area associated with $\Gamma$. Equation \eqref{eq:griffithFunctional} is referred to as the Griffith functional, and it is assumed that for some given boundary conditions on $\Omega$, the unknown displacements $\bm{u}$ as well as the crack set $\Gamma$ can be obtained via a \emph{global} minimization of said functional subject to the irreversibility condition
\begin{equation}
	\label{eq:irreversibility}
	\Gamma_{t + \Delta t} \supseteq \Gamma_t
\end{equation}
that is comparable to \eqref{eq:griffith_irreversibility}. In contrast, Griffith requires only stationarity of \eqref{eq:griffithFunctional}. Furthermore in the variational theory, $\Gamma$ is not restricted to consist of a single crack. Thus a body that is initially without flaw ($\Gamma = \emptyset$) may nucleate a crack if the resulting configuration has lower total energy compared to one where no crack forms. Similarly, a crack is allowed to branch if this leads to a lower potential energy than simple extension. The directions of advance are naturally obtained as those leading to minimum increase in \eqref{eq:griffithFunctional}, so that in theory the energy minimization framework is able to handle crack initiation and branching without the need to introduce additional criteria.

\subsection{Phase-field and gradient damage models}
The main difficulty in performing a direct minimization of the Griffith functional is that $\bm{u}$ is generally discontinuous across $\Gamma$, so that \eqref{eq:griffithFunctional} contains a locus of jump sets whose locations are a priori unknown and which generally do not align with the predefined domain discretization that is utilized in numerical simulations. To render the problem tractable, \cite{Bourdin2000} adopted a strategy wherein the minimization is instead performed on an approximation of the Griffith functional having regularized jump sets so that $\bm{u}$ is continuous over the entire domain. This was inspired by the earlier work of \citet{Ambrosio1990,Ambrosio1992} who solved a similar problem in image segmentation involving the functional of \cite{Mumford1989}. A scalar order parameter known as the \emph{crack phase-field} is introduced to interpolate between fractured and intact regions, with the amount of regularization controlled by a characteristic length parameter denoted by $\ell$. The validity of such as strategy rests on whether the regularized approximation tends towards the original functional as $\ell$ goes to zero, in the sense of $\Gamma$-convergence \citep{Braides2006}. Although an additional equation governing the phase-field evolution must now be solved along with the linear momentum equation, the main advantage of this approach is that numerical solutions may be obtained via classical finite element algorithms as both $\bm{u}$ and the phase-field are continuous. \cite{Bourdin2000}'s regularization of \eqref{eq:griffithFunctional} took the form
\begin{equation}
	\label{eq:bourdinFunctional}
	\Psi \left( \bm{u}, \phi \right) = \int_\Omega \frac{1}{2} \left[ \left( 1 - \phi \right)^2 + \kappa \right] \bm{\varepsilon} \left( \bm{u} \right) : \mathbb{C}^e : \bm{\varepsilon} \left( \bm{u} \right) \dee\Omega + \mathcal{G}_c \int_\Omega \left( \frac{1}{2\ell} \phi^2 + \frac{\ell}{2} \| \nabla \phi \|^2 \right) \dee\Omega
\end{equation}
where $\phi$ is the phase-field that takes on values between 0 and 1, corresponding respectively to fully intact and broken states. On the other hand, $\kappa$ is a small positive constant meant to ensure positivity of the bulk energy as $\phi \rightarrow 1$. The two important features of the above expression are (a) the replacement of $\mathcal{H}^{n-1} \left( \Gamma \right)$ by an elliptic functional that calculates the combined length of all the cracks, and (b) the coefficient $\left( 1 - \phi \right)^2 + \kappa$ known as the \emph{energy degradation function} that penalizes the material stiffness according to the value of $\phi$. \Cref{eq:bourdinFunctional} is essentially a direct adaptation of the earlier functional of \cite{Ambrosio1992}, for which a proof of $\Gamma$-convergence was subsequently given by \cite{Chambolle2004}.

\subsubsection{Alternative variational problems}
\begin{figure}
	\centering
	\includegraphics[width=0.4\textwidth]{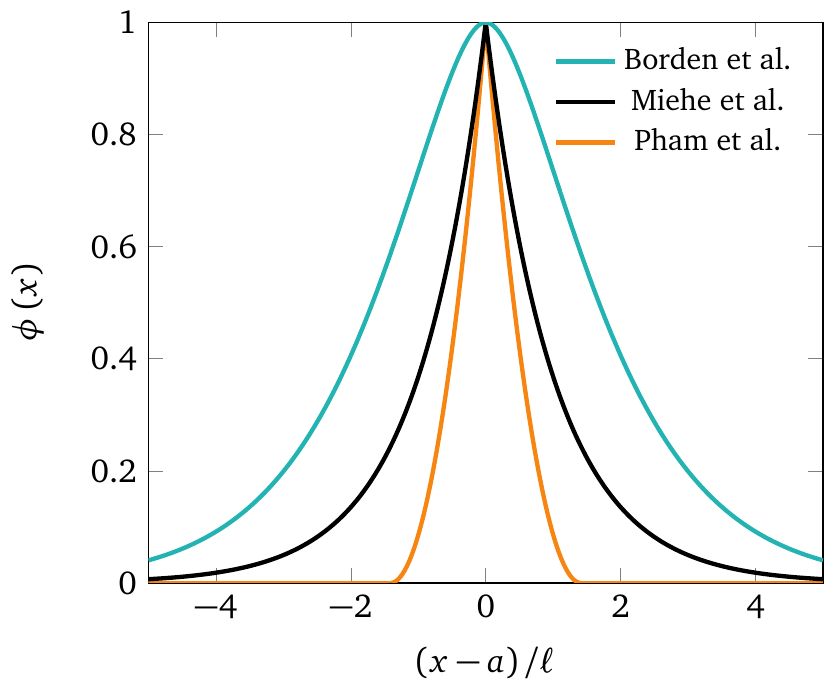}
	\captionof{figure}{Phase-field regularizations of a sharp crack at $x=a$ using various candidate functions.}
	\label{fig:phaseField_1d}
\end{figure}

It was later suggested by \cite{Miehe2010_ijnme} that development of elliptic approximations to $\mathcal{H}^{n-1} \left( \Gamma \right)$ could also be motivated from a more physical standpoint by considering the 1-dimensional case of an infinitely long bar of uniform cross section. Assuming that the bar is aligned with the $x$-axis and that a single crack fully cuts the bar at $x = a$, the phase-field profile corresponding to this sharp crack is none other than the discontinuous scalar function
\begin{equation}
	\label{eq:sharpCrack}
	\phi \left( x \right) =
	\begin{cases}
		1, & x = a \\
		0, & \text{otherwise}.
	\end{cases}
\end{equation}
A regularized approximation of the above can then be made via a function $\phi \in \Phi$, where
\begin{equation}
	\label{eq:setPhi}
	\Phi = \Set{ \phi \in H^1 \left( \mathbb{R}; \left[ 0,1 \right] \right) | 
		\begin{array}{l} 
			\phi \left( a \right) = 1 \\
			\phi^\prime \left( x \right) > 0 \text{ if } x < a  \\
			\phi^\prime \left( x \right) < 0 \text{ if } x > a  \\
			\phi \left( x \right) \rightarrow 0 \text{ as } x \rightarrow \pm\infty \end{array}}.
\end{equation}
Some candidate functions are
\begin{align}
	\phi \left( x \right) &= \exp \left( -\frac{\left| x-a \right|}{\ell} \right) \label{eq:approxExp} \\
	\phi \left( x \right) &= 
	\begin{cases}
	\left( 1 - \dfrac{\left| x-a \right|}{\sqrt{2}\ell} \right)^2, & x \in \left[ a - \sqrt{2}\ell, a + \sqrt{2}\ell \right] \\
	0 & \text{otherwise}
	\end{cases} \label{eq:approxVarIneq} \\
	\phi \left( x \right) &= \left( 1 + \frac{\left| x-a \right|}{\ell} \right) \exp \left( -\frac{\left| x-a \right|}{\ell} \right). \label{eq:approxHighOrder}
\end{align}
We note that the function \eqref{eq:approxExp} utilized by \cite{Miehe2010_ijnme} is in fact the solution obtained by minimizing the functional of \cite{Bourdin2000} in one dimension. On the other hand, \eqref{eq:approxVarIneq} is a compactly supported function that leads to a variational inequality problem \citep{Pham2011}, while \eqref{eq:approxHighOrder} is obtained by solving a 4th order governing equation and was introduced by \cite{Borden2012} with the aim of loosening the mesh size requirements with respect to $\ell$ at the same time taking advantage of numerical methods that can adequately model $C^1$-continuous solutions. \Cref{fig:phaseField_1d} shows a comparison of the three functions mentioned above. It can be observed that for a given value of $\ell$, the amount of crack diffusion is additionally dependent on the specific form of $\phi \left( x \right)$.
Higher order phase-field formulations for fracture have so far not achieved the same popularity as their lower order counterparts. This is mainly due to the higher order of continuity that they require in conjunction with numerical solutions, which is expensive to obtain with traditional frameworks such as finite elements. In addition, $\Gamma$-convergence is yet to be proven for these formulations. On the other hand, \cite{Li2015} point out that the incorporation of general anisotropic effects relating to the surface energy requires a formulation that is at least 4th order.

\subsubsection{Damage}
The connection between phase-field approaches and nonlocal damage models was explored by \cite{Pham2011}, who noted that elliptic functionals approximating \eqref{eq:griffithFunctional} can be seen as specific cases of the integral of a general state function pertaining to a gradient damage model:
\begin{equation}
	\label{eq:stateFunction}
	W_\ell \left( \bm{\epsilon} \left( \bm{u} \right), \phi, \nabla\phi \right) = \frac{1}{2} \bm{\epsilon} : \mathbb{C} \left( \phi \right) : \bm{\epsilon} + w \left( \phi \right) + \frac{1}{2} w_1 \ell^2 \nabla\phi \cdot \nabla\phi
\end{equation}
where $w \left( \phi \right)$ is a monotonically increasing function in the interval $\left[ 0,1 \right]$ with $w \left( 0 \right) = 0$ and $w \left( 1 \right) = w_1$. They suggest using the linear form  $w \left( \phi \right) = w_1 \phi$ which leads to models having a real elastic phase with no premature decrease in material stiffness, at the cost of solving a variational inequality problem for the damage evolution. In contrast, the quadratic form of $w \left( \phi \right)$ employed in \eqref{eq:bourdinFunctional} leads to material behavior having \emph{no} real elastic phase, with damage already occurring at the onset of loading \citep{Amor2009}. Bulk energy release resulting from evolution of the phase-field is facilitated through a damage-dependent elasticity tensor $\mathbb{C} \left( \phi \right)$ as seen in \eqref{eq:stateFunction}. The simplest form which leads to isotropic behavior consists of the multiplicative ansatz
\begin{equation}
	\label{eq:elasticDamage_basic}
	\mathbb{C} \left( \phi \right) = g \left( \phi \right) \mathbb{C}^e,
\end{equation}
in which $g \left( \phi \right)$ is the energy degradation function mentioned previously that is non-negative in the interval $\left[ 0,1 \right]$ with $g \left( 0 \right) = 1$ and $g \left( 1 \right) = g^\prime \left( 1 \right) = 0$. Such form however has limited applicability, since it allows for unphysical compressive cracking based on $\mathcal{G}_c$. \cite{Lancioni2009} adopted the above model to shear cracking by having $g \left( \phi \right)$ act only on the deviatoric portion of the strain. This was subsequently improved upon by \cite{Amor2009} who proposed that crack growth be driven also by the spherical part of the energy when the volumetric strain is positive. This results in more realistic anisotropic behavior where the material is allowed to crack in tension and shear, but not in compression. An alternative formulation was introduced by \cite{Miehe2010_cmame,Miehe2010_ijnme} in which degradation occurs only on tensile components of the principal strain tensor, leading to pure mode-I cracks.

\subsection{Going back to Griffith}
The functional $\Psi \left( \bm{u},\phi \right)$ in \eqref{eq:bourdinFunctional} is neither linear nor convex, which makes the task of finding global minimizers non-trivial. \cite{Bourdin2000} proposed an alternate minimization algorithm that takes advantage of the convexity of $\Psi$ with respect to either $\bm{u}$ or $\phi$ when the other is held constant. Nonetheless, solution schemes based on descent algorithms only converge to local minimizers or saddle points and are by themselves inadequate for obtaining global minimizers. It was found that naive application of the alternate minimization often resulted in solutions that exhibited unphysical dips in the total energy, particularly when crack growth is brutal. In an attempt to remedy this behavior, a heuristic backtracking scheme was developed by \cite{Bourdin2008} (with subsequent improvements by \cite{Mesgarnejad2015}) based on an additional optimality condition that enforces monotonic evolution of the total energy when the load is also monotonically increasing. From a phenomenological standpoint however, the main objection to using global energy minimization is that it allows for evolutions where the current configuration jumps over arbitrarily large energy barriers in order to reach the new configuration corresponding to the global minimizer \citep{Negri2008}. While this enables the strict preservation of energy conservation, it may also result in unphysical response where cracks propagate at lower energy release rates than $\mathcal{G}_c$ which violates Griffith's criterion. Recently, \cite{Larsen2010} introduced the notion of $\epsilon$-stability as a stepping stone towards formulations that can predict crack paths based on \emph{local} minimality, which is in turn closer to Griffith's original idea. As with Griffith's model, such solutions will also exhibit dissipation in the total energy in cases where crack propogation occurs in a brutal manner. However as pointed out by \cite{Negri2008}, brutal cracking is primarily a dynamic phenomenon which explains why the total energy cannot be completely accounted for in a quasi-static framework.

\subsection{On the treatment of $\ell$ as a material parameter}
If we settle for invoking local versus global minimality, then classical solution schemes that were previously deemed inadequate are now robust without the need to perform backtracking. We are left with the non-convexity of Griffith's functional, but this is easily dealt with via the alternate minimization algorithm as earlier mentioned. On the other hand, we end up as well with Griffith's original conundrum concerning crack initiation. It turns out that the saving grace is none other than the regularization of the functional, which as will be discussed in the later sections actually allows for crack initiation in the absence of stress singularities provided that the characteristic length $\ell$ associated with the regularization is \emph{finite}. This also brings us more in line with nonlocal damage theory where the thickness of the localization zone is usually a constant parameter associated with some physical internal length. In this context, the treatment of $\ell$ as a material parameter as suggested in \cite{Mesgarnejad2015} and \cite{Nguyen2016} makes a lot of sense since it can be shown that under the assumption of uniform damage, the peak stress $\sigma_c$ (which can be interpreted as the critical stress for crack nucleation) is dependent on $\ell$. However it has also been observed \citep[e.g.][]{Klinsmann2015} that propagation cracks explicitly modeled in the mesh exhibits a dependence on $\ell$ as well. Thus if $\sigma_c$ and $\mathcal{G}_c$ are assumed to be two independent material parameters, the interdependence of each with $\ell$ means that we can adjust the regularization based on one or the other but generally not both at the same time. Hence the idea of only relying on $\ell$ to calibrate the model is not entirely adequate. Furthermore in heterogeneous media, $\ell$ may take on a different value for each medium so that the diffuse crack becomes thicker or thinner as it passes from one material to the next. This is especially undesirable in a multiphysics setting in which the model for some overlying physical process depends directly on the phase-field. From \eqref{eq:stateFunction} and \eqref{eq:elasticDamage_basic} we can see that the degradation function is the remaining component through which we can rectify the model, and it is in fact this realization that has motivated the present work.

\section{Governing equations and numerical implementation} \label{sec:equationsAndNumerics}
For the remainder of this study, we have chosen to adopt the functional of \cite{Bourdin2000} by reason of its simplicity.  Incorporating the work done by external forces, the regularized total potential energy for a given body $\Omega$ subject to boundary conditions is given by
\begin{equation}
	\label{eq:totalEnergyPotential}
	\Pi = \Psi - W = \int_\Omega \left[ g \left( \phi \right) \psi \left( \bm{\epsilon} \right) + \mathcal{G}_c \left( \frac{1}{2\ell}\phi^2 + \frac{\ell}{2} \nabla\phi \cdot \nabla\phi \right) \right] \dee\Omega - \int_\Omega \bm{b} \cdot \bm{u} \,\dee\Omega - \int_{\partial\Omega^t} \bm{t} \cdot \bm{u} \,\dee S
\end{equation}
where $\psi \left( \bm{\epsilon} \right) = \frac{1}{2} \bm{\epsilon} \!:\! \mathbb{C}^e \!:\! \bm{\epsilon}$ is the Helmholtz free energy density and $\partial\Omega^t$ denotes the part of the boundary for which Neumann (i.e.\ traction) conditions are prescribed. The quantity $\bm{b}$ represents the body force, while $\bm{t}$ is the vector of prescribed tractions acting on the Neumann boundary. By imposing the stationarity of $\Pi$ we obtain the variational equation
\begin{multline}
	\label{eq:stationarity}
	\delta\Pi = \int_\Omega g \left( \phi \right) \frac{\partial\psi}{\partial\bm{\epsilon}} : \delta\bm{\epsilon} \dee\Omega 
	- \int_\Omega \bm{b} \cdot \delta\bm{u} \,\dee \Omega - \int_{\partial\Omega^t} \bm{t} \cdot \delta\bm{u} \,\dee S \\
	+ \int_\Omega g^\prime \left( \phi \right) \psi \left( \bm{\epsilon} \right) \delta\phi \dee\Omega + \mathcal{G}_c  \int_\Omega \left( \frac{1}{\ell}\phi \,\delta\phi + \ell \nabla \phi \cdot \delta\nabla\phi \right) \dee \Omega = 0.
\end{multline}
The above equality must hold for arbitrary values of $\delta \bm{u}$ and $\delta\phi$, implying that
\begin{subequations}
	\label{eq:weakForm}
	\begin{align}
	&\int_\Omega g \left( \phi \right) \frac{\partial\psi}{\partial\bm{\epsilon}} : \delta\bm{\epsilon} \,\dee\Omega = \int_\Omega \bm{b} \cdot \delta\bm{u} \,\dee \Omega + \int_{\partial\Omega^t} \bm{t} \cdot \delta\bm{u} \,\dee S 
	\label{eq:weakForm_momentum} \\
	&\int_\Omega g^\prime \left( \phi \right) \psi \left( \bm{\epsilon} \right) \delta\phi \dee\Omega + \mathcal{G}_c  \int_\Omega \left( \frac{1}{\ell} \phi \delta\phi + \ell \nabla\phi \cdot \nabla \delta\phi \right) \dee \Omega = 0.
	\label{eq:weakForm_phaseField}
	\end{align}
\end{subequations}
These constitute the weak form of the governing equations. Noting that $\bm{\sigma} = \partial\psi/\partial\bm{\epsilon}$, the equivalent strong formulation may be obtained by applying Gauss' divergence theorem yielding the following coupled system:
\begin{subequations}
	\label{eq:govEq_StrongForm}
	\begin{align}
	\nabla \cdot \left[ g \left( \phi \right) \bm{\sigma} \right] + \bm{b} &= \bm{0} \text{ on } \Omega \label{eq:linearMomentum} \\
	g \left( \phi \right) \bm{\sigma} \cdot \bm{n} &= \bm{t} \text{ on } \partial\Omega^t \label{eq:linearMomentum_Neumann} \\
	\bm{u} &= \bar{\bm{u}} \text{ on } \partial\Omega^u \label{eq:linearMomentum_Dirichlet} \\
	\mathcal{G}_c \ell_0 \nabla^2\phi - \frac{\mathcal{G}_c}{\ell_0} \phi &= g^\prime \left( \phi \right) \psi \left( \bm{\varepsilon}\right) \text{ on } \Omega \label{eq:evolution} \\
	\nabla\phi \cdot \bm{n} &= 0 \text{ on } \partial\Omega. \label{eq:evolution_bc}
	\end{align}
\end{subequations}
\Cref{eq:linearMomentum,eq:linearMomentum_Neumann,eq:linearMomentum_Dirichlet} comprise the linear momentum equation and its corresponding boundary conditions, while \eqref{eq:evolution} is the phase-field evolution equation with the associated boundary condition given by \eqref{eq:evolution_bc}. As mentioned earlier, $\phi$ should go to zero away from the crack. Thus it is tacitly assumed that the domain is of sufficient size to provide adequate separation between the regularized crack and the boundary, allowing the phase-field to decay to values that are small enough to approximate this condition.

Extension of the irreversibility condition \eqref{eq:irreversibility} to the regularized case is not immediately obvious, since the intermediate states $0 < \phi < 1$ do not have a straightforward physical interpretation. The natural course from the perspective of damage mechanics is to enforce the condition
\begin{equation}
	\label{eq:phaseField_growth}
	\phi \left( \bm{x} \right)_{t + \Delta t} \geq \phi \left( \bm{x} \right)_t \quad \forall \bm{x} \in \Omega.
\end{equation}
This can be imposed via an additional penalty term in the phase-field evolution equation \citep{Miehe2010_ijnme}, or alternatively through a history variable $\mathcal{H}$ that replaces the quantity $\psi \left( \bm{\epsilon} \right)$ in \eqref{eq:evolution}, defined as \citep{Miehe2010_cmame}
\begin{equation}
	\label{eq:histField}
	\mathcal{H} \left( \bm{x},t \right) = \max\limits_{s \in \left[ 0,t \right]} \psi \left( \bm{\epsilon} \left( \bm{x},s \right) \right)
\end{equation}
A closer look at the phase-field localization process however reveals that \eqref{eq:phaseField_growth} may not be the best extension of \eqref{eq:irreversibility}. For example in the 1-dimensional case \cite{Kuhn2015} demonstrate that localization of the phase-field at the center of a diffuse crack involves both the growth of $\phi$ near the crack tip, as well as a \emph{decrease} of the same in adjacent regions, which enables the phase-field to correctly settle to the exponential profile given in \eqref{eq:approxExp}. Based on this observed behavior, a strict imposition of \eqref{eq:phaseField_growth} may lead to an overestimate of the crack length. Consequently, we introduce a modified version of \eqref{eq:histField} in which imposes irreversibility only when $\phi$ exceeds a certain threshold, i.e.
\begin{equation}
	\label{eq:histField_threshold}
	\mathcal{H} \left( \bm{x},t \right) =
	\begin{cases}
	\max\limits_{s \in \left[ 0,t \right]} \psi \left( \bm{\epsilon} \left( \bm{x},s \right) \right) & \text{if } \phi > \phi_c \\[1em]
	\psi \left( \bm{\epsilon} \left( \bm{x},t \right) \right) & \text{otherwise.} \end{cases}
\end{equation}
The parameter $\phi_c$ represents the maximum value for damage that is allowed to heal during unloading. For a material point undergoing damage $\phi \leq \phi_c$, the resulting stress-strain curves will be nonlinear, with the amount of departure from linearity dependent on the specific form of the degradation function. Nonetheless having $\phi_c > 0$ allows the stress paths for subsequent unloading and reloading to coincide with the initial loading curve, which cannot be achieved otherwise. For the present work, we have chosen to set $\phi_c = 0.5$ in all of our simulations.

The coupled system described in \eqref{eq:govEq_StrongForm} is implemented in a classical finite element framework, with the primary unknowns being the displacement $\bm{u}$ and phase-field $\phi$. In a 2-dimensional setting, these are expressed in terms of the corresponding nodal degrees of freedom as
\begin{equation}
	\bm{u} = \sum\limits_{I=1}^m \bm{N}_I^{\bm{u}} \bm{u}_I \quad\text{and}\quad \phi = \sum\limits_{I=1}^m N_I \phi_I
\end{equation}
wherein
\begin{equation}
	\bm{N}_I =
	\left[ \begin{matrix}
		N_I & 0 \\[0.5em] 0 & N_I
	\end{matrix} \right]
\end{equation}
with $N_I = N_I \left( \bm{x} \right)$ denoting the shape function associated with node $I$, and $\bm{u}_I$ and $\phi_I$ the respective displacement and phase-field degrees of freedom at node $I$. The strain and phase-field gradient are given by
\begin{equation}
	\bm{\epsilon} = \sum\limits_{I=1}^m \bm{B}_I^{\bm{u}} \bm{u}_I \quad\text{and}\quad \nabla\phi = \sum\limits_{I=1}^m \bm{B}^\phi_I \phi_I
\end{equation}
in which
\begin{equation}
	\bm{B}_I^{\bm{u}} =
	\left[ \begin{matrix}
		N_{I,x} & 0 \\[0.5em] 0 & N_{I,y} \\[0.5em] N_{I,y} & N_{I,x}
	\end{matrix} \right]
	\quad\text{and}\quad
	\bm{B}_I^\phi =
	\left[ \begin{matrix}
		N_{I,x} \\[0.5em] N_{I,y}
	\end{matrix} \right].
\end{equation}
The former is the symmetrized gradient matrix associated with the Voigt form of the strain tensor. The test functions and their corresponding derivatives can be obtained from the above expressions by replacing $\bm{u}_I$ and $\phi_I$ with $\delta\bm{u}_I$ and $\delta\phi_I$ respectively. Due to the arbitrariness of the latter two quantities, numerical approximation of the weak form in \eqref{eq:weakForm} yields the following nonlinear system of equations at each node $I$:
\begin{subequations}
	\begin{align}
		\bm{r}^{\bm{u}}_I &= \int_\Omega g \left( \phi \right) \bm{B}_I^{\bm{u}T} \bm{\sigma} \dee\Omega - \int_\Omega \bm{N}_I^{\bm{u}T} \bm{b} \dee\Omega - \int_{\partial\Omega} \bm{N}_I^{\bm{u}T} \bm{t} \dee S = \bm{0} \label{eq:fem_linearMomentum} \\
		r^\phi_I &= \int_\Omega \left[ \mathcal{G}_c\ell\, \bm{B}_I^{\phi T} \nabla\phi + \frac{\mathcal{G}_c}{\ell} N_I \phi \right] \dee\Omega + \int_\Omega N_I g^\prime \left( \phi \right) \mathcal{H} \dee\Omega = 0 \label{eq:fem_phaseField}
	\end{align}
\end{subequations}
where $\mathcal{H}$ is the threshold-based history variable defined in \eqref{eq:histField_threshold}. Due to the non-convexity of \eqref{eq:totalEnergyPotential}, the above coupled system is solved using the alternate minimization algorithm outlined in \cite{Bourdin2000}. This involves cycling between \eqref{eq:fem_linearMomentum} and \eqref{eq:fem_phaseField}: the linear momentum equation is solved first using values of $\phi$ from the previous iteration. Next, the phase-field evolution equation is solved using the newly obtained values of $\bm{u}$. The iterations are carried out repeatedly until the prescribed criteria on the size of residuals and inter-iteration corrections on the unknowns are met.

For degradation functions in which the derivative $g^\prime \left( \phi \right)$ is nonlinear, the subsystem represented by \eqref{eq:fem_phaseField} also becomes nonlinear, resulting in the need to perform nested iterations. In our experience, a naive implementation of the alternate minimization algorithm wherein the linearized phase-field equation is solved only once before going back to linear momentum is generally unstable and leads to incorrect results. That is, in the subsystem
\begin{equation}
	\left\{ \phi \right\}_i^{m+1} = \left\{ \phi \right\}_i^m - \left[ \bm{K}^{\phi\phi} \left( \phi_i^m \right) \right]^{-1} \left\{ r^{\phi} \right\}_i^m
\end{equation}
corresponding to the $\left( m+1 \right)$th iteration within the $i$th time step, use of the exact Jacobian given by
\begin{equation}
	\bm{K}^{\phi\phi}_{IJ} \left( \phi_i^m \right) = \int_\Omega \left[ \mathcal{G}_c\ell\, \bm{B}_I^{\phi T} \bm{B}_J^\phi + \left( \frac{\mathcal{G}_c}{\ell} + \mathcal{H} g^{\prime\prime} \left( \phi_i^m \right) \right) N_I N_J \right] \dee\Omega
\end{equation}
often produces an incorrect evolution of the phase-field and eventual blow-up. However, we have found that such an approach can be made to work by replacing $g^{\prime\prime} \left( \phi_i^m \right)$ with an approximate expression derived from low order terms in a Taylor expansion. Assuming that the degradation function is smooth, we can obtain $g^\prime \left( \phi_1 \right)$ as an infinite sum of terms involving higher order derivatives of $g$ evaluated at $\phi_2 \in \left[ 0,1 \right]$:
\begin{equation}
	g^\prime \left( \phi_1 \right) = g^\prime \left( \phi_2 \right) + g^{\prime\prime} \left( \phi_2 \right) \left( \phi_1 - \phi_2 \right) + \frac{g^{\prime\prime\prime} \left( \phi_2 \right)}{2} \left( \phi_1 - \phi_2 \right)^2 + \ldots
\end{equation}
Now let $\phi_1 = 1$ so that $g^\prime \left( \phi_1 \right) = 0$. Dropping higher order terms as well as the subscript on $\phi_2$, we obtain
\begin{equation}
	0 \approx g^\prime \left( \phi \right) + g^{\prime\prime} \left( \phi \right) \left( 1 - \phi \right)
\end{equation}
which then gives us our approximation for the 2nd derivative of $g$:
\begin{equation}
	g^{\prime\prime}_\text{app} \left( \phi \right) = -\frac{g^\prime \left( \phi \right)}{1 - \phi}.
\end{equation}

\section{A curious case of crack nucleation}
Pre-existing cracks maybe accounted for in the phase-field model in two ways. The first is through initialization of the phase-field profile, the second by modeling the crack faces directly as internal boundaries in the discretized geometry. \cite{Sicsic2013} provide analytical results showing that the growth of fully developed fractures described via the phase-field obeys Griffith's law as the regularization parameter goes to zero. On the other hand it has been shown in numerical experiments that the same is not generally true for the extension of cracks built into the mesh. Recent studies \citep[e.g][]{Klinsmann2015,Nguyen2016} have observed that the simulated critical energy release rate (or analogously, the peak load) overshoots the correct value for sufficiently small $\ell$. This inconsistency becomes more understandable upon the realization that propagation of mesh-described cracks is actually a manifestation of nucleation rather than extension in the context of phase-field approaches, i.e. crack formation at a region where the phase-field is uniformly zero. Similar behavior can be observed in the case of crack nucleation at a notch or reentrant corner; the extension of mesh-modeled cracks is in fact a limiting case of the former where the notch angle is zero. Thus one can observe that the Francfort-Marigo-Bourdin phase-field model gives rise to three distinct types of simulated material response in connection with fracture: (a) propagation of phase-field-described fractures which is relatively well-understood, (b) crack nucleation in the \emph{absence} of stress singularities which will be discussed in \cref{sec:tensileStrength}, and (c) quasi-nucleation behavior associated with the extension of mesh-modeled cracks, which is our immediate concern in this section.

\subsection{Preliminary numerical example} \label{subsec:overshoot_example}
\begin{figure}
	\centering
	\begin{subfigure}[t]{0.23\textwidth}
		\centering
		\includegraphics[width=\textwidth]{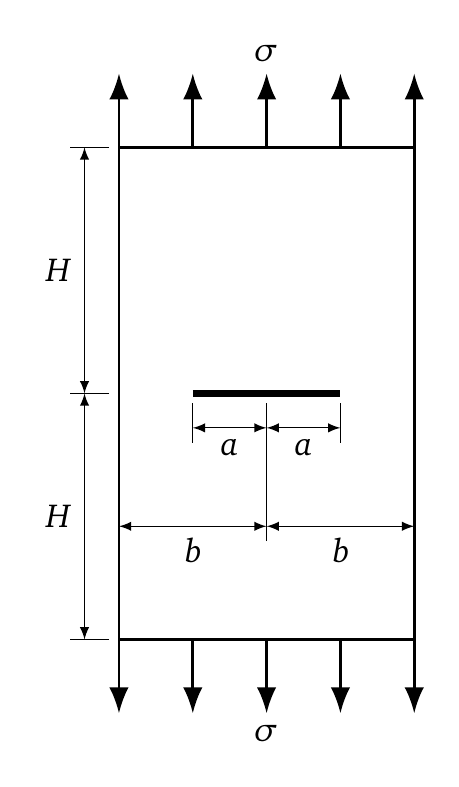}
		\caption{}
		\label{fig:CC_Specimen_Geometry}
	\end{subfigure} \hspace{1.0em}
	\begin{subfigure}[t]{0.17\textwidth}
		\centering
		\includegraphics[width=\textwidth]{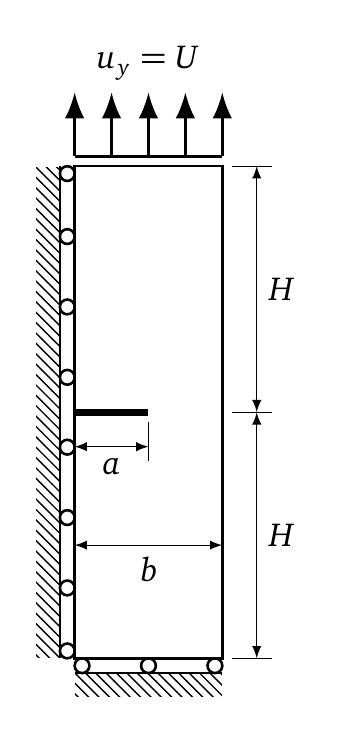}
		\caption{}
		\label{fig:CC_Specimen_CompDomain}
	\end{subfigure}
	\begin{subfigure}[t]{0.55\textwidth}
		\centering
		\includegraphics[width=\textwidth]{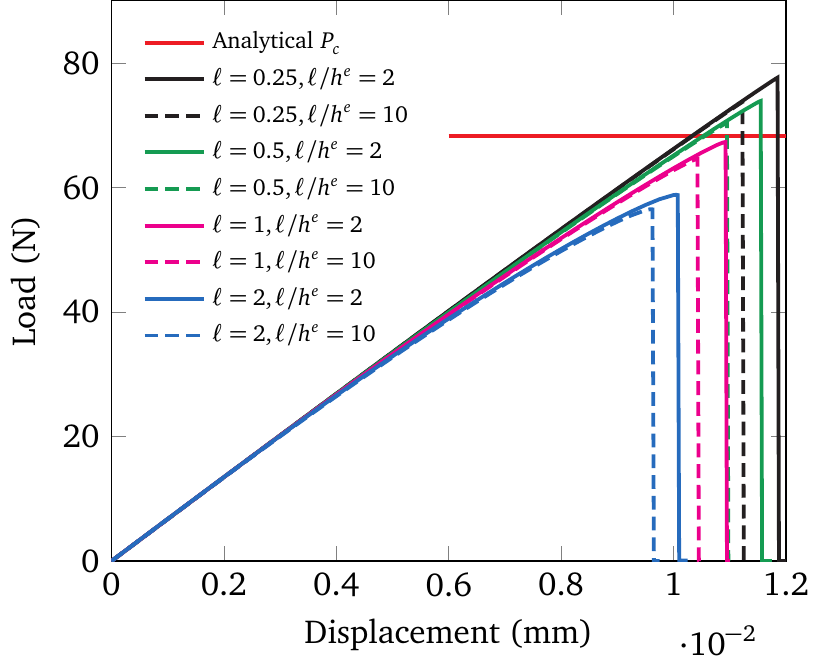}
		\caption{}
		\label{fig:CC_Specimen_Results}
	\end{subfigure}
	\caption{(a) center-cracked specimen subjected to uniform tension at the far-field, (b) computational domain and boundary conditions, and (c) Load-displacement curves for different values of $\ell$ and mesh refinement.}
	\label{fig:CC_Specimen}
\end{figure}

To illustrate the dependence of the material response on the phase-field length scale, we simulate fracture propagation in a homogeneous specimen containing a center crack and subjected to tensile loading as shown in \cref{fig:CC_Specimen_Geometry}.
Assuming that $H$ is taken large enough such that the tensile stresses at the boundary are acceptably uniform, the mode-I stress intensity factor can be computed for finite values of the ratio $a/b$ as
\begin{equation}
	\label{eq:SIF}
	K_I = \sigma \sqrt{\pi a} \ F \left( a/b \right)
\end{equation}
where $F \left( a/b \right)$ is a shape factor given by
\begin{equation}
	\label{eq:shape_factor}
	F \left( a/b \right) = \left[ 1 - 0.025 \left( \frac{a}{b} \right)^2 + 0.06 \left( \frac{a}{b} \right)^4 \right] \sqrt{\sec \frac{\pi a}{2b}}.
\end{equation}
The above formula has a reported accuracy of $0.1\%$ or better for any $a/b$ \citep{Tada2000}. We note that $a/b = 0$ and $H = \infty$ corresponds to the original fracture problem of \cite{Griffith1921}, for which $F \left( 0 \right) = 1$. For the plane strain case, the critical stress intensity factor and strain energy release rate are related by
\begin{equation}
	\mathcal{G}_c = K_{Ic}^2 \left( \frac{1 - \nu^2}{E} \right).
\end{equation}
Combining the above with \eqref{eq:SIF}, we obtain the following expression for the critical failure load:
\begin{equation}
	P_c = \frac{b}{F \left( a/b \right)} \left[ \dfrac{E\mathcal{G}_c}{\left( 1 - \nu^2 \right) \pi a} \right]^{1/2}.
\end{equation}
The actual computational domain is shown in \cref{fig:CC_Specimen_CompDomain} along with the relevant boundary conditions; due to symmetry, only half the geometry needs to be considered. Note that the initial crack $\Gamma$ of length $a$ in the computational domain is modeled as part of the geometry, and no initialization of the phase-field is performed to account for its presence. Actual dimensions used are $a = 10$mm, $b = 2a$ and $H = 10a$. Thus $a/b = 0.5$ and we obtain $F \left( a/b \right) = 1.1862$ from \eqref{eq:shape_factor}. The material constants are $E = 70,000$ MPa, $\nu = 0.22$ and $\mathcal{G}_c = 0.007$ N/mm. From the preceding equation, we obtain the critical failure load as $P_c = 68.26$ N. The simulation is carried out using monotonic displacement control with the specimen gradually stretched in increments of $\Delta U = 2.5\times 10^{-4}$ mm until failure occurs in the form of brutal cracking. In order to obtain a precise determination of the failure load, the increase in boundary displacements is carried out using smaller increments of $\Delta U = 2.5 \times 10^{-5}$ mm as failure is approached.  \Cref{fig:CC_Specimen_Results} shows the dependence of simulation results on the phase-field characteristic length as well as the relative mesh refinement, $\ell/h^e$. It can be seen that larger values of $\ell$ lead to an increase in deviation from linear behavior, whereas a smaller $\ell$ drives the peak load upwards. Furthermore there is an apparent lack of convergence with respect to the regularization parameter, since the load-displacement curve overshoots the true failure load when values of $\ell$ smaller than some threshold are used. As can be observed, the severity of this phenomenon is also influenced by the mesh refinement, and in particlar is greater for coarser meshes relative to $\ell$. We have found that different sets of material parameters give qualitatively the same behavior as what we have shown.

\subsection{Exploring the overshoot phenomenon} \label{subsec:overshoot}
The apparent lack of convergence in the material response with respect to $\ell$ in the above numerical example seems to contradict the $\Gamma$-convergence property of \eqref{eq:bourdinFunctional}, however this can be explained by the fact that Griffith's criterion does not actually involve the energy functional directly but rather its \emph{gradients} (Fr\'{e}chet derivatives). This nuance has no corresponding counterpart in image segmentation, and makes phase-field simulation of brittle fracture a fundamentally different problem from the former, despite the similarity of the Griffith energy to the Mumford-Shah functional. Thus while $\Gamma$-convergence of the regularized approximation to the sharp-boundary functional is by itself sufficient to produce physically meaningful results in an imaging context, this is no longer the case for brittle fracture.

At present, our understanding of the above phenomenon relies on numerical evidence obtained from analyzing problems such as the one presented in \cref{subsec:overshoot_example}. To elucidate further, recall that for a material which fractures according to Griffith's theory as summarized in \eqref{eq:griffith_principle}, the following inequality applies with regard to energy increments:
\begin{equation}
	-\delta\Psi_b^e \leq \mathcal{G}_c \delta\Gamma.
\end{equation}
for some arbitrary small crack extension $\delta\Gamma > 0$. Now $-\delta\Psi_b^e = G \delta\Gamma$ where $G$ is the energy release rate at the crack tip, so the strict inequality $-\delta\Psi_b^e < \mathcal{G}_c \delta\Gamma$ means that the crack must be stationary due to \eqref{eq:slacknessCondition}. If $-\delta\Psi_b^e = \mathcal{G}_c \delta\Gamma$, then a positive $\delta\Gamma$ is admissible and the crack can propagate stably. On the other hand, the reverse inequality $-\delta \Psi_b^e > \mathcal{G}_c \delta\Gamma$ is generally understood as corresponding to brutal cracking. In a quasi-static framework where dynamic effects are disregarded, fracture propagation simply continues until a state is reached wherein the condition $\delta\Psi_b^e < \mathcal{G}_c \delta\Gamma$ is once again satisfied, resulting in arrest of the crack. It can be shown that in many cases, such a condition cannot be satisfied for any length of crack advance which results in the fracture cutting through the entire width of the domain.

Similar behavior is manifested by the evolution of $\phi$ in the diffuse-crack model during brutal crack propagation, and can observed by scrutinizing successive iterations within the relevant time step. In contrast to the original theory however, the phase-field model contains only the equality part of Griffith's criterion in the phase-field evolution equation. That is, \eqref{eq:weakForm_phaseField} can be written as
\begin{equation}
	\label{eq:phaseFieldCriterion}
	-\delta\Psi_b^\text{app} = \mathcal{G}_c \delta\Gamma
\end{equation}
wherein
\begin{align}
	\delta\Psi_b^\text{app} &= \int_\Omega g^\prime \left( \phi \right) \psi \left( \bm{\epsilon} \right) \delta\phi \dee\Omega \\
	\delta\Gamma &= \int_\Omega \left( \frac{1}{\ell} \phi \, \delta\phi + \ell \nabla\phi \cdot \nabla \delta\phi \right) \dee\Omega
\end{align}
for some positive $\delta\Gamma$ that arises from an arbitrary incremental evolution of the phase-field, denoted by $\delta\phi$. The implications of this are immediately obvious when on looks at the strong form of the phase-field equation in \eqref{eq:evolution}: assuming that $g^\prime \left( \phi \right) < 0$ everywhere except at $\phi = 1$ (and this is in fact necessary for damage to evolve at all), then it is clear that $\phi$ must begin moving away from its initial value of 0 from the moment that nonzero stress is induced in the material. Furthermore, let $\eta_b$ denote the error arising from using $\Psi_b^\text{app}$ in place of $\Psi_b^e$, i.e.
\begin{equation}
	\eta_b = \Psi_b^\text{app} - \Psi_b^e.
\end{equation}
Plugging the above into \eqref{eq:phaseFieldCriterion} and writing $\delta\Psi_b^e$ in terms of $G$, we obtain the following relation:
\begin{equation}
	\label{eq:phaseField_prefracture}
	\delta\eta_b  = \left( G - \mathcal{G}_c \right) \delta\Gamma.
\end{equation}
From the previous numerical example, we can infer that at some critical loading $U_s$ brutal propagation of the crack will occur, presumably because now $-\Psi_b^\text{app} > \mathcal{G}_c \delta\Gamma$ for any $\delta\phi$.  The equivalent condition in terms of $\eta_b$ and $G$ is given by
\begin{equation}
	\label{eq:phaseField_brutalCracking}
	\delta\eta_b  < \left( G - \mathcal{G}_c \right) \delta\Gamma
\end{equation}
The different curves in \cref{fig:CC_Specimen_Results} demonstrate how the actual value of $U_s$ depends on $\ell$. The key idea here is that both $\delta\Gamma$ and $\delta\eta_b$ are influenced by $\ell$, but in varying degrees from one another. In particular, it is no longer just the quantity $G - \mathcal{G}_c$ that determines the onset of brutal cracking; as can be observed from \cref{fig:CC_Specimen_Results}, both undershoot and overshoot of the correct failure load are possible. The challenge is to have \eqref{eq:phaseField_brutalCracking} occur at the precise moment that $G$ exceeds  $\mathcal{G}_c$, so that brutal fracture occurs at the correct magnitude of loading. 

The current prevailing thought is that one can achieve the above scenario by some ``correct'' choice of the regularization parameter. However, the need to specify $\ell$ (and obviously $\ell > 0$) brings into question the benefit of having regularized approximations $\Gamma$-converge to the Griffith energy at all as $\ell$ goes to zero. One can argue that the removal of such a requirement is not a disadvantage since it lends more flexibility to the phase-field framework and likewise opens the door to other interesting and more exotic approximations of \eqref{eq:griffithFunctional}, such as the higher order formulations by \cite{Borden2014} and \cite{Li2015} that have so far not been proven to be $\Gamma$-convergent to Griffith's energy. More importantly, the main problem with relying on calibrating $\ell$ in order to obtain the correct instance of failure is that such a strategy is not guaranteed to succeed in all possible cases, in particular when the setup is very different from the one analyzed above. This is demonstrated in \cref{sec:numericalExamples}, where we study a problem for which the aforementioned technique does \emph{not} work at all, at least within practical limitations.

\subsection{Preserving linearity in the material response} \label{subsec:linearity}
An important consequence of \eqref{eq:phaseField_prefracture} is that the material response of the regularized model inevitably drifts from linear elastic behavior prior to fracture, since growth of $G$ as a result of increasing $U$ must be matched by a corresponding increase in the incremental error term $\delta\eta_b$. Since $\eta_b$ represents the discrepancy between approximate and the exact bulk energies, an ever-increasing increment in the error term means that the simulated material behavior deviates further and further from linear elasticity with increasing $U$ as evident in \cref{fig:CC_Specimen_Results}.  Some control on $\delta\eta_b$ can be exercised through the factor $\delta\Gamma$, i.e. we keep $\delta\eta_b$ small by keeping $\delta\Gamma$ small as well. However since $\delta\phi$ is arbitrary, we can accomplish this only by careful construction of either the degradation function (which affects the bulk energy), the crack length functional, or possibly both.

\section{A new family of degradation functions}
The ideas presented in \cref{subsec:overshoot,subsec:linearity} can be combined together to give us a set of properties for what we would consider an accurate phase-field model with regard to the extension of mesh-described cracks:
\begin{enumerate}[(a)]
	\item The simulated critical displacement should preferably be close to the correct value, and
	\item the accumulated error $\eta_b$ should be kept small prior to the occurrence of brutal fracture.
\end{enumerate}
Item (b) is quite straightforward, and is achieved by having brutal fracture occur at low values of the phase-field. Such behavior is readily observed with the alternatives to quadratic degradation that have appeared in the literature, for instance the quartic function
\begin{equation}
	\label{eq:quarticDegFcn}
	g_4 \left( \phi \right) = 4 \left( 1 - \phi \right)^3 - 3 \left( 1 - \phi \right)^4
\end{equation}
utilized by \cite{Karma2001} in conjunction with their own phase-field theory, and the cubic function
\begin{equation}
	\label{eq:cubicDegFcn}
	g_3 \left( \phi \right) = s \left[ \left( 1 - \phi \right)^3 - \left( 1 - \phi \right)^2 \right] + 3 \left( 1 - \phi \right)^2 - 2 \left( 1 - \phi \right)^3
\end{equation}
analyzed by \cite{BordenPhD}, in which the quantity $s$ controls the slope of the degradation function at the unbroken state. \cite{Kuhn2015} have shown that all three functions have similar post-failure behavior in stable crack growth, i.e. their differences lie primarily in the prediction of the level of strain or stress at which crack propagation occurs, and also in the amount of stiffness reduction observed prior to the onset of cracking.

Item (a) is more difficult to satisfy, in particular since quantities pertaining to the bulk energy are also dependent on material properties. The degradation function must then be parametric, in order to have the means of compensating for different values of these properties. We can see that none of the different functions mentioned above possess the latter property, so that one is instead forced to rely on tweaking $\ell$ as is done with the quadratic degradation function. From a conceptual standpoint this is not entirely satisfactory, since $\ell$ as a parameter belongs to the crack functional term and not the bulk energy. Furthermore a change in the regularization parameter leads to corresponding changes in \emph{both} the bulk and surface terms. It is our view that it is better to introduce parameters directly into the degradation function. In doing so one is able to alter the behavior of $\delta\eta_b$ independently of $\delta\Gamma$. Furthermore the resulting formulation does not force the interpretation of $\ell$ as a de facto material parameter, but is rather nearer to the original concept of \cite{Bourdin2000} where $\ell$ is purely a mathematical construct that arises in connection with the regularization of discontinuities.

\subsection{Exponential-type degradation} \label{sec:expTypeDeg}
\begin{figure}
	\centering
	\begin{minipage}{0.47\textwidth}
		\centering
		\includegraphics[width=\textwidth]{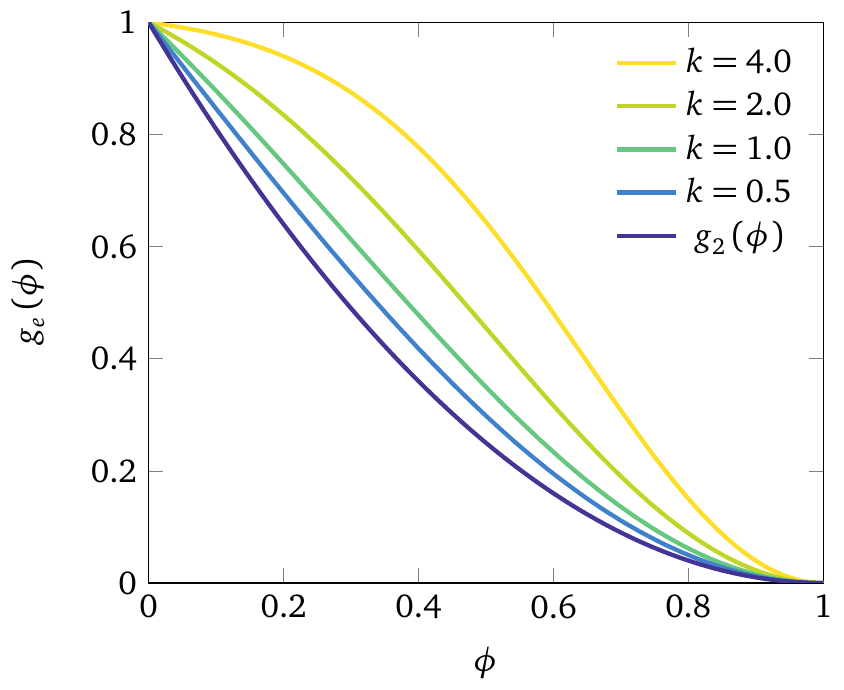}
		\captionof{figure}{$g_e \left( \phi \right)$ with $n=2$, showing the effect of parameter $k$.}
		\label{fig:ge_to_quadratic}
	\end{minipage} \hspace{1.0em}
	\begin{minipage}{0.47\textwidth}
		\centering
		\includegraphics[width=\textwidth]{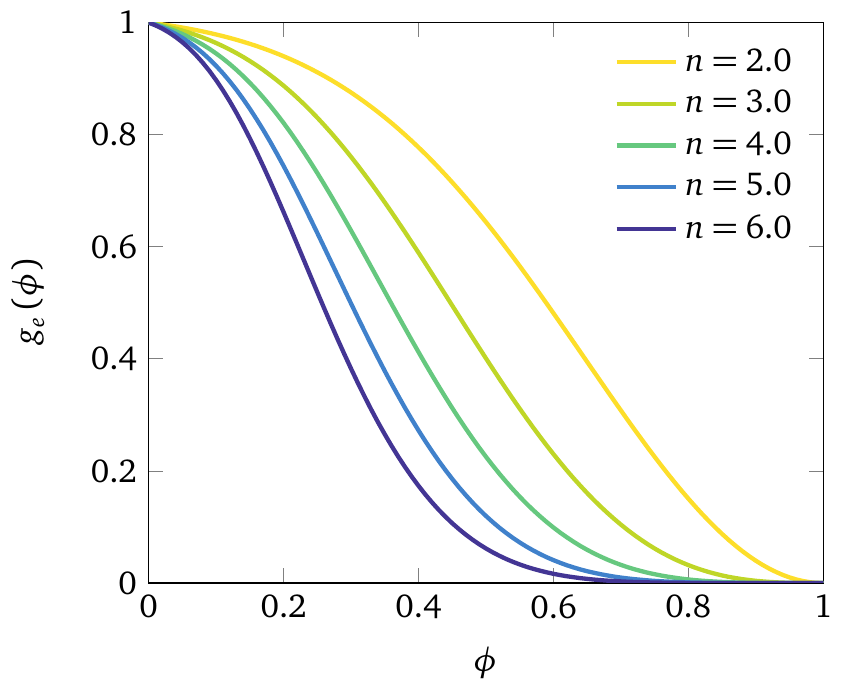}
		\captionof{figure}{$g_e \left( \phi \right)$ with $k=4$, showing the effect of parameter $n$.}
		\label{fig:ge_effect_of_n}
	\end{minipage}
\end{figure}

Consider now the family of degradation functions defined by the 3-parameter function
\begin{equation}
	\label{eq:2ParamDegFcn}
	g_e \left( \phi; k,n,w \right) = \left( 1-w \right) \frac{1 - e^{-k \left( 1 - \phi \right)^n}}{1 - e^{-k}} + wf_c \left( \phi \right)
\end{equation}
where $k$, $n$ and $w$ are real numbers such that $k > 0$, $n \geq 2$ and $w \in \left[ 0,1 \right]$. The function $f_c$ is a corrector term whose role shall be explored in the later discussions. For now let us assume that $w = 0$ so that \eqref{eq:2ParamDegFcn} has effectively only 2 free parameters. The resulting expression has the following properties:
\begin{enumerate}[a)]
	\item $g_e \left( \phi \right)$ is monotonically decreasing \label{enum:prop1},
	\item $g_e \left( 0 \right) = 1$, $g_e \left( 1 \right) = 0$, \label{enum:prop2}
	\item $g_e^\prime \left( 0 \right) < 0$, $g_e^\prime \left( 1 \right) = 0$. \label{enum:prop3}
\end{enumerate}
In choosing the form of \eqref{eq:2ParamDegFcn} we have aimed for a minimal but sufficient number of parameters that allows us to have some control in the overall shape of the function in order to restore proper balance between bulk and surface energy increments, as well as suppress unphysical stiffness reduction prior to fracture. Note that one obtains the function $\left( 1 - \phi \right)^n$ in the limit as $k$ approaches 0, as shown in \cref{fig:ge_to_quadratic}.
On the other hand, increasing $n$ has the effect of flattening $g_e \left( \phi \right)$ as $\phi$ goes to 1, shown in \cref{fig:ge_effect_of_n}. The parameters $k$ and $n$ must be chosen such that crack propagation occurs at the right energy release rate for some given $E$, $\mathcal{G}_c$ and $\ell$. Of prime importance here is the shape of $g_e \left( \phi \right)$ at the vicinity of $\phi = 1$ which controls the amount of elastic bulk energy that is dissipated at the diffuse crack tip. On the other hand, spurious stiffness reduction prior to fracture is connected to the behavior of $g^\prime \left( \phi \right)$ at $\phi = 0$; we want to keep $g^\prime \left( 0 \right)$ small which is equivalent to setting $k$ to be large. However, choosing an excessively large value for $k$ also results in undesirable stress-strain behavior. In the following analysis, we show that it is possible to eliminate one parameter in \eqref{eq:2ParamDegFcn} by selecting the largest values of $k$ (given some $n$) for which the resulting stress-strain relationships is considered acceptable.

\subsection{Analytic model behavior in 1D}
\begin{figure}
	\centering
	\includegraphics[width=0.8\textwidth]{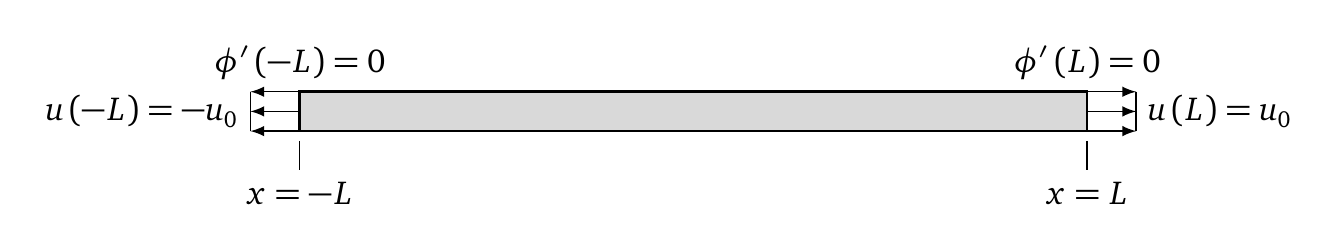}
	\caption{Domain and boundary conditions for 1-dimensional homogeneous bar subjected to tension.}
	\label{fig:bar_geometry}
\end{figure}

\begin{figure}
	\centering
	\includegraphics[width=0.6\textwidth]{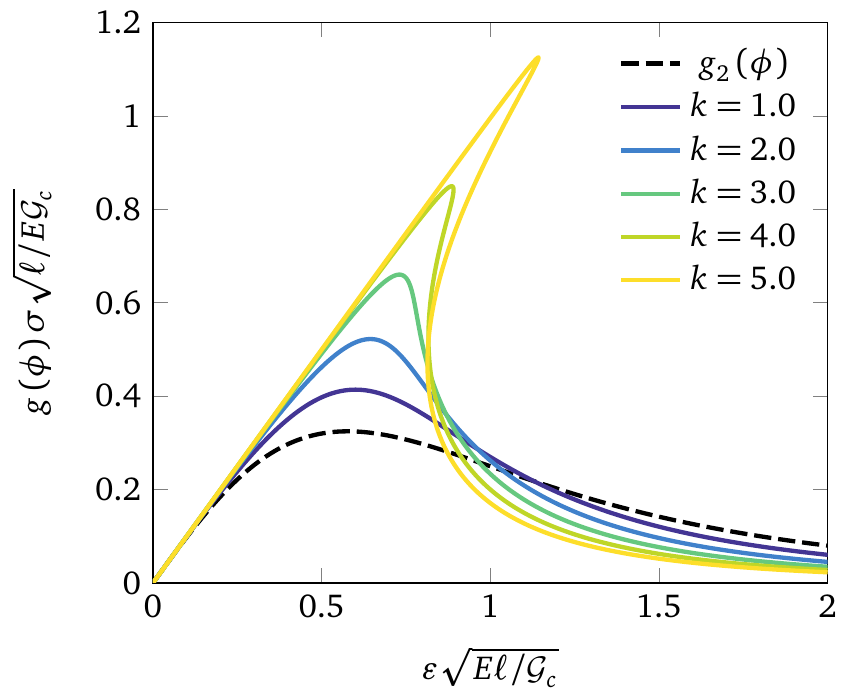}
	\caption{Stress-strain curves for the 1-dimensional bar using $g_e \left( \phi \right)$ with $n = 2$. Larger values of $k$ suppress the deviation from linear elastic behavior but also lead to the development of snap-back behavior.}
	\label{fig:snapback}
\end{figure}

\begin{figure}
	\centering
	\begin{subfigure}{0.49\textwidth}
		\centering
		\includegraphics[width=\textwidth]{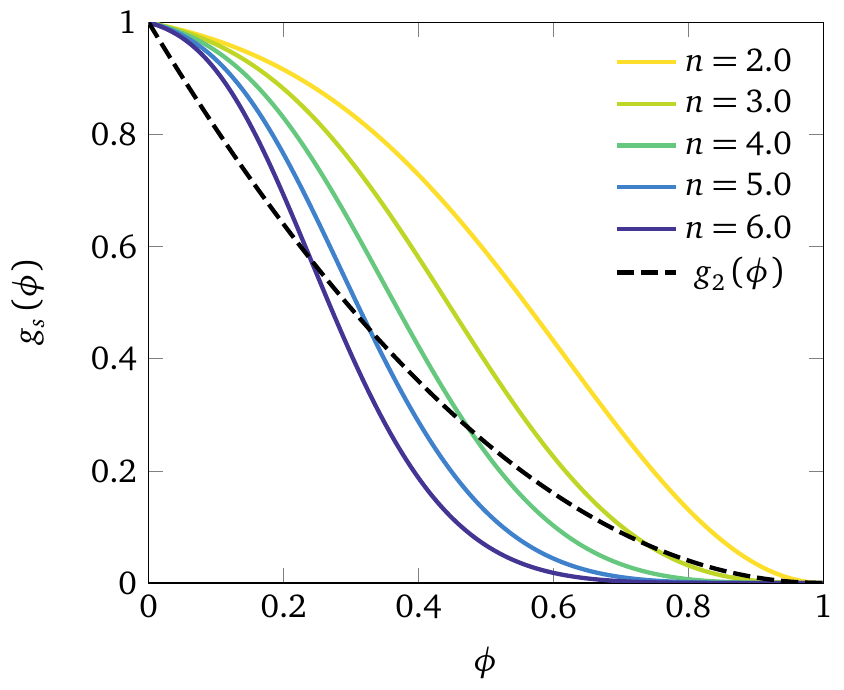}
		\caption{}
	\end{subfigure} \hspace{0.4em}
	\begin{subfigure}{0.49\textwidth}
		\centering
		\includegraphics[width=\textwidth]{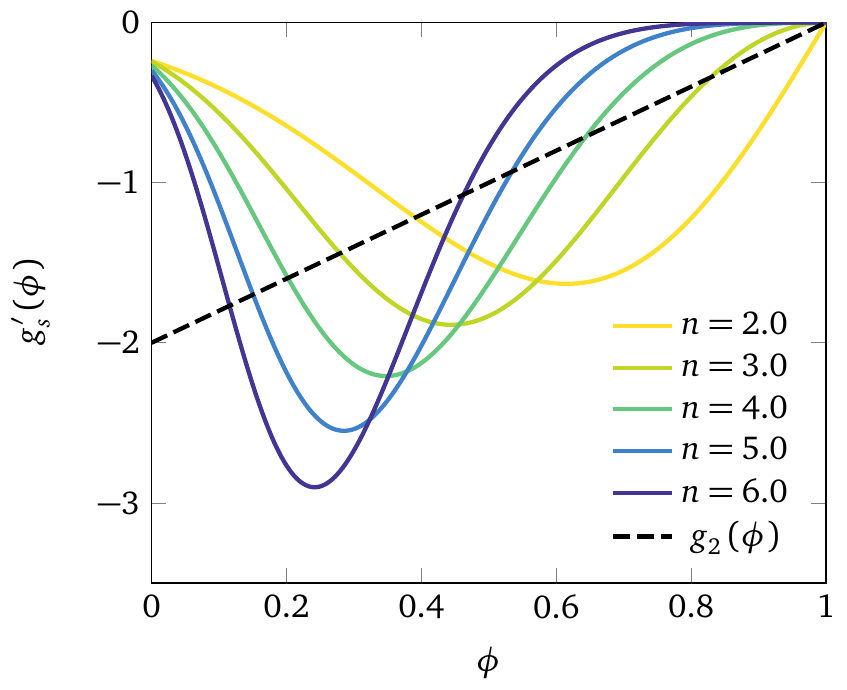}
		\caption{}
	\end{subfigure}
	\caption{Effect of parameter $n$ on (a) the single-parameter degradation function $g_s \left( \phi \right)$, and (b) its derivative.}
	\label{fig:1param_degfcn}
\end{figure}

\begin{figure}
	\centering
	\begin{subfigure}{0.49\textwidth}
		\centering
		\includegraphics[width=\textwidth]{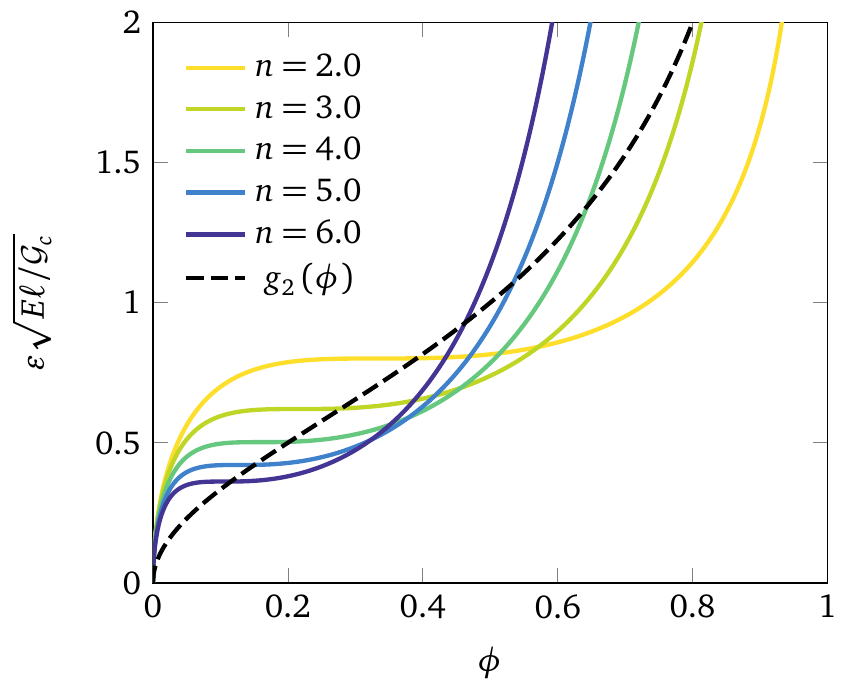}
		\caption{}
	\end{subfigure} \hspace{0.4em}
	\begin{subfigure}{0.49\textwidth}
		\centering
		\includegraphics[width=\textwidth]{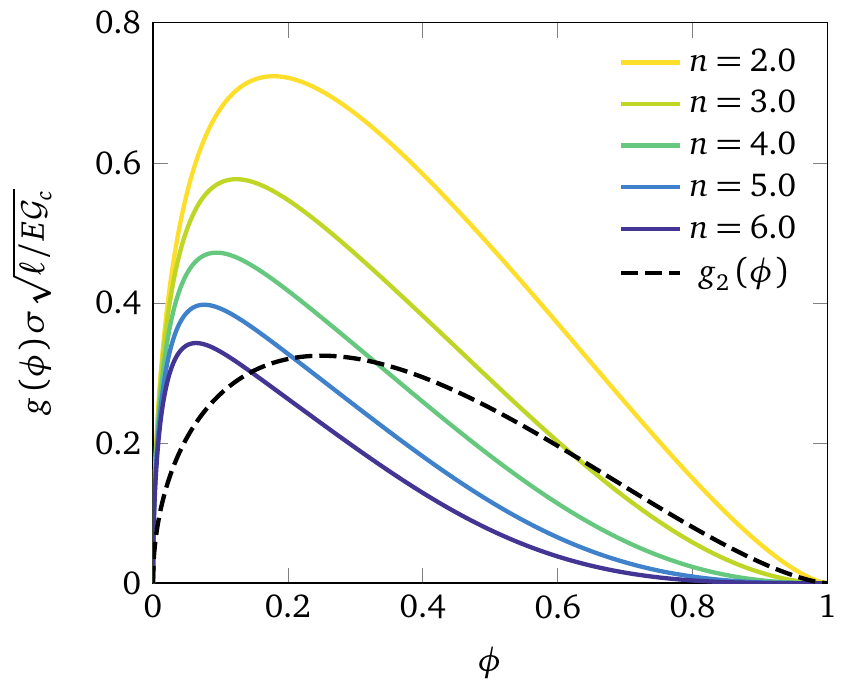}
		\caption{}
		\label{fig:stress_pf_relationship}
	\end{subfigure}
	\caption{Dependence of (a) strain and (b) stress on the phase-field arising from the adoption of $g_s \left( \phi \right)$ in modeling fracture of a 1-dimensional homogeneous bar subjected to tension.}
	\label{fig:pf_relationships}
\end{figure}

\begin{figure}
	\centering
	\includegraphics[width=0.5\textwidth]{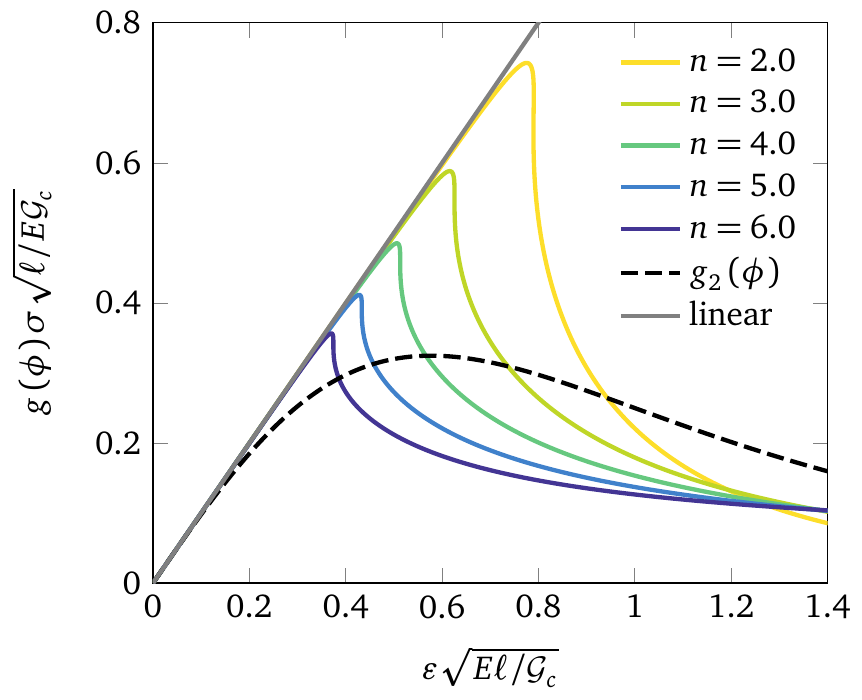}
	\caption{Stress-strain curves resulting from phase-field modeling of tensile fracture in a 1-dimensional homogeneous bar, utilizing $g_s \left( \phi \right)$ as degradation function.}
	\label{fig:stress_strain_curves}
\end{figure}

In order to study the effect the parameters $k$ and $n$ in our proposed family of functions, we take a look at the 1-dimensional case of a materially homogeneous bar with uniform cross section and length equal to $2L$. The bar is subjected to the boundary conditions $u \left( \pm L \right) = \pm u_0$ and $\phi^\prime \left( \pm L \right) = 0$ as shown in \cref{fig:bar_geometry}.
Assuming zero body forces, the governing equations in \eqref{eq:govEq_StrongForm} reduce to
\begin{subequations}
	\label{eq:GovEq_StrongForm_1D}
	\begin{align}
		\frac{d}{dx} \left[ g \left( \phi \right) \sigma \left( \epsilon \right) \right] &= 0 \label{eq:equil_strongForm_1D} \\
		\mathcal{G}_c \ell_0 \frac{d^2 \phi}{dx^2} - \frac{\mathcal{G}_c}{\ell_0} \phi &= g^\prime \left( \phi \right) \psi \left( \epsilon \right)
		\label{eq:phaseField_strongForm_1D}
	\end{align}
\end{subequations}
in which $\epsilon = du/dx$, $\sigma = E \epsilon$ and $\psi = \frac{1}{2}\sigma\varepsilon$. We focus on spatially homogeneous solutions for the phase-field, $\phi \left( x \right) \equiv \phi_0$ which implies that the stress is also spatially uniform, i.e. $\sigma \equiv \sigma_0 = E \epsilon_0$. As $\phi$ is no longer a function of $x$, \eqref{eq:phaseField_strongForm_1D} simplifies to
\begin{equation}
	\label{eq:SimplifiedEvolution}
	-\frac{\mathcal{G}_c}{\ell_0}\phi = \frac{1}{2} g^\prime \left( \phi \right) E \varepsilon^2.
\end{equation}
While it is physically more correct to express $\phi$ as a function of $\epsilon$ (since crack formation is driven by the mechanical response), for complicated forms of $g \left( \phi \right)$ it becomes more convenient to adopt the opposite order of dependence. Hence we obtain
\begin{equation}
	\label{eq:strain1D}
	\epsilon \left( \phi \right) = \left[ \frac{-2\mathcal{G}_c \phi}{\ell_0 Eg^\prime \left( \phi \right)} \right]^\frac{1}{2}
\end{equation}
with the corresponding derivative given by
\begin{equation}
	\label{eq:strainDerivative1D}
	\frac{\dee \epsilon}{\dee\phi} = -\frac{\mathcal{G}_c}{\ell_0 E} \left[ \frac{-2\mathcal{G}_c \phi}{\ell_0 Eg^\prime \left( \phi \right)} \right]^{-\frac{1}{2}} \left\{ \frac{g^\prime \left( \phi \right) - \phi g^{\prime\prime} \left( \phi \right)}{\left[ g^\prime \left( \phi \right) \right]^2} \right\}.
\end{equation}
Consequently the derivative of the damaged-reduced stress can be obtained with respect to the phase-field as
\begin{equation}
	\frac{\dee}{\dee\phi} \left[ g \left( \phi \right) \sigma \right] = g^\prime \left( \phi \right) E \varepsilon \left( \phi \right) + g \left( \phi \right) E \frac{\dee\epsilon}{\dee\phi}.
\end{equation}
The effective stress-strain curve accounting for damage due to the phase-field can then be defined as
\begin{align}
	\frac{\dee}{\dee\epsilon} \left[ g \left( \phi \right) \sigma \right] &= \frac{\dee}{\dee\phi} \left[ g \left( \phi \right) \sigma \right] \frac{\dee\phi}{\dee\epsilon} = g^\prime \left( \phi \right) E \epsilon \left( \phi \right) \frac{\dee\phi}{\dee\epsilon} + g \left( \phi \right) E \nonumber \\
	&= \left[ g^\prime \left( \phi \right)\epsilon \left( \phi \right) \frac{\dee\phi}{\dee\epsilon} + g \left( \phi \right) \right] E.
\end{align}
Combining the last equation above with \eqref{eq:strain1D} and \eqref{eq:strainDerivative1D}, we obtain after further manipulation the expression
\begin{equation}
	\frac{\dee}{\dee\epsilon} \left[ g \left( \phi \right) \sigma \right] = \frac{2\phi \left[ g^\prime \left( \phi \right) \right]^2 + g \left( \phi \right) \left[ g^\prime \left( \phi \right) - \phi g^{\prime\prime} \left( \phi \right) \right]}{g^\prime \left( \phi \right) - \phi g^{\prime\prime} \left( \phi \right)} E.
	\label{eq:stress_strain_behavior}
\end{equation}
Now $g^\prime \left( 0 \right) < 0$ by construction for $\phi < 1$, and if $g \left( \phi \right)$ has monotonically increasing slope (i.e. $g^{\prime\prime} \left( \phi \right) \geq 0$) then the above expression is well defined for $\phi \in \left[ 0,1 \right]$. However for degradation functions of the form given by \eqref{eq:2ParamDegFcn}, the existence of an inflection point means that the denominator in \eqref{eq:stress_strain_behavior} may become zero at some point, implying the existence of a vertical tangent in the $\sigma$--$\epsilon$ curve and possibly also snap-back behavior. This phenomenon is more pronounced for larger values of $k$, as illustrated in \cref{fig:snapback}.
However it can be seen that a high value of $k$ also acts to suppress the undesired deviation from linear elastic behavior. Hence we want to choose this parameter as large as possible in order to minimize the said effect, but still small enough so as not to generate snap-back. This implies that $k = k \left( n \right)$, and the specific relationship is found by considering the limiting case where the denominator in \eqref{eq:stress_strain_behavior} goes to zero. This yields the expression
\begin{equation}
	k \left( n \right) = \frac{\left( n-2 \right)\phi^\star + 1}{n\phi^\star \left( 1-\phi^\star \right)^n}
	\label{eq:limit_k}
\end{equation}
where
\begin{equation}
	\phi^\star = \left\{ \begin{array}{cl}
	\dfrac{1}{3}, & n = 2 \\[1.5em]
	\dfrac{-\left( n+1 \right) + \sqrt{5n^2 - 6n + 1}}{2\left( n^2 - 2n \right)}, & \text{otherwise} \\
	\end{array} \right.
\end{equation}
with the relevant calculations given in \ref{sec:k_derivation}. Plugging the above results into \eqref{eq:2ParamDegFcn} gives the reduced-parameter degradation function
\begin{equation}
	g_s \left( \phi; n, w \right) = \left( 1-w \right) \frac{1 - e^{-k \left( n \right) \left( 1 - \phi \right)^n}}{1 - e^{-k \left( n \right)}} + w f_c \left( \phi \right)
	\label{eq:1ParamDegFcn}
\end{equation}
where again for the meantime we take $w = 0$. The profile of $g_s \left( \phi \right)$ and its derivative are shown in Fig. \ref{fig:1param_degfcn} for several $n$.
Due to the fact that $g^\prime_s \left( 0 \right) < 0$, growth of the phase-field takes place naturally in the presence of local stress gradients. This is in contrast to degradation functions where $g^\prime \left( 0 \right) = 0$, for which special solution procedures are required to trigger the evolution of $\phi$ away from an undamaged state. Furthermore it has been shown \citep[e.g.][]{Kuhn2015} that for certain configurations of polynomial degradation functions, Eq.\ \eqref{eq:strain1D} may predict inadmissible values of the phase-field (e.g.\ $\phi \notin \left[ 0,1 \right]$) at low strains implying a bifurcation-type behavior with $\phi$ remaining at the undamaged state until the point of bifurcation. However said point does not generally coincide with the onset of fracture, so that some stiffness reduction still occurs prior to the realization of peak loads. On the other hand, the family of degradation functions represented by Eq.\ \eqref{eq:1ParamDegFcn} give rise to smooth $\varepsilon$-$\phi$ and $\sigma$-$\phi$ relationships as shown in \cref{fig:pf_relationships}.
It follows that for these type of functions, the material stress-strain behavior will exhibit elastic stiffness reduction, albeit in much more reduced magnitudes compared to the quadratic degradation function (see \cref{fig:stress_strain_curves}).
Likewise an important result is that for some given (fully determined) degradation function, the resulting normalized $\sigma$-$\varepsilon$ curve is unique so that the \emph{actual} failure stress and strain are dependent on $\ell$ as well as the material parameters. This implies that there is no single degradation function that works for the entire range of values of $E$, $\mathcal{G}$ and $\ell$. Otherwise, the regularization parameter $\ell$ cannot be freely chosen but rather must be determined from the other material parameters in order to give the correct failure stress. The latter condition imposes a severe limitation on the phase-field method, particularly when viewed in the context of multi-physics simulations where one might desire to have control over the amount of crack regularization in order to satisfy requirements stemming from physics external to the mechanics and fracture propagation.

\subsection{Role of $w$ and $f_c \left( \phi \right)$}
\begin{figure}
	\centering
	\begin{subfigure}{0.49\textwidth}
		\centering
		\includegraphics[width=\textwidth]{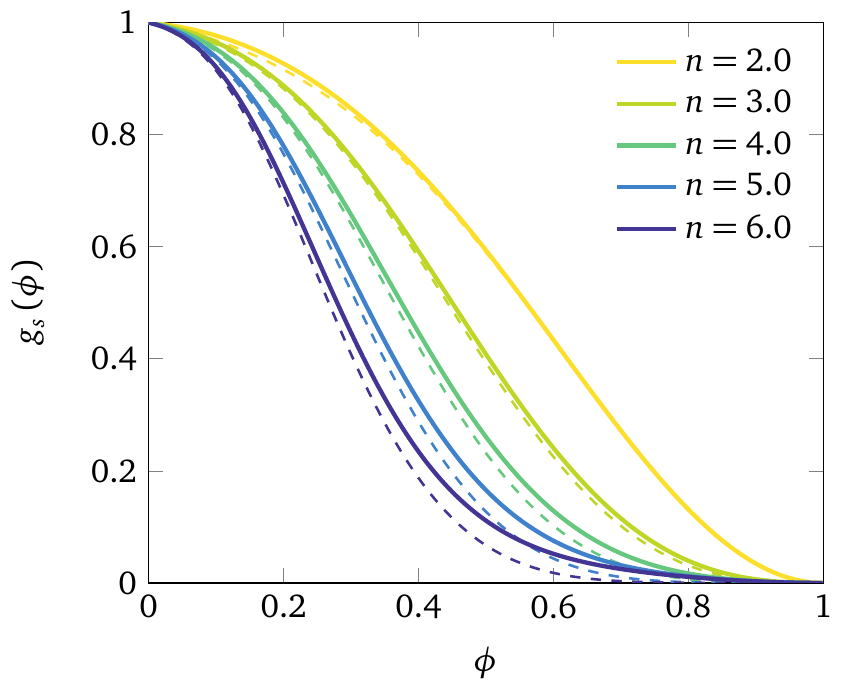}
		\caption{}
	\end{subfigure} \hspace{0.4em}
	\begin{subfigure}{0.49\textwidth}
		\centering
		\includegraphics[width=\textwidth]{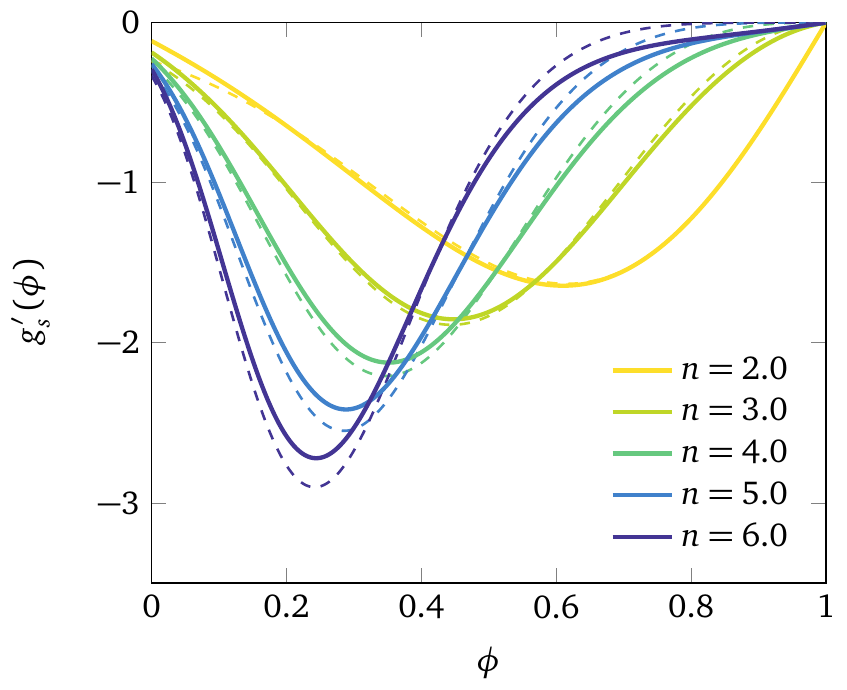}
		\caption{}
	\end{subfigure}
	\caption{Plots of the single-parameter degradation function $g_s \left( \phi \right)$ and its derivative for different values of $n$, showing the influence of the correction term $f_c \left( \phi \right)$. The solid plots are obtained by setting $w = 0.1$, while the dashed plots have $w = 0$ yielding the original uncorrected functions.}
	\label{fig:1param_degfcn_corrected}
\end{figure}

Unfortunately, the simplified form of \eqref{eq:1ParamDegFcn} with $w = 0$ is not entirely adequate due to the fact that for higher values of $n$, the flattened shape of $g_s \left( \phi \right)$ means that near-total annihilation of the material stiffness already occurs at values of $\phi$ significantly less than 1. As a consequence, the phase-field stagnates below unity even though the material is fully damaged. As a result, calculation of crack lengths via evaluation of $\Gamma \left( \phi \right)$ will yield incorrect results. The above shortcoming can be remedied through $f_c \left( \phi \right)$, which acts as a correction term influencing how $g \left( \phi \right)$ goes to zero as $\phi \rightarrow 1$. Its purpose is to impart a residual gradient to $g_s \left( \phi \right)$ that is independent of $n$, so that $g_s^\prime \left( \phi \right)$ is always sufficiently below zero for $\phi < 1$. A suitable expression satisfying the properties enumerated in \cref{sec:expTypeDeg} is
\begin{equation}
	\label{eq:correctionTerm}
	f_c \left( \phi \right) = a_2 \left( 1-\phi \right)^2 + a_3 \left( 1-\phi \right)^3.
\end{equation}
In order to fully determine the constants $a_2$ and $a_3$, we impose two conditions. The first is that $f_c^\prime \left( \phi^\star \right) - \phi^\star f_c^{\prime\prime} \left( \phi^\star \right) = 0$ in order to retain validity of expressions obtained based on $\phi^\star$ in \ref{sec:k_derivation}. The second is that $f_c \left( 0 \right) = 1$. This yields the following expressions for the constants:
\begin{equation}
	a_2 = \frac{3 \left( \phi^\star \right)^2 - 3}{3 \left( \phi^\star \right)^2 - 1}, \qquad a_3 = \frac{2}{3 \left( \phi^\star \right)^2 - 1}.
\end{equation}
We note that $f_c \left( \phi \right)$ itself is in general \emph{not} a degradation function since for sufficiently large $\phi^\star$ it may be that $f_c \left( \phi \right) > 1$ at certain values of $\phi$. Thus $w$ should be kept small, otherwise the correction term dominates. We have found that setting $w = 0.1$ imparts a sufficient residual in the gradient of $g_s \left( \phi \right)$ while still satisfying that requirements given in \cref{sec:expTypeDeg}. The resulting plots for $g \left( \phi \right)$ and $g^\prime \left( \phi \right)$ are shown in \cref{fig:1param_degfcn_corrected}.
An additional benefit of having the correction term in the form of \eqref{eq:correctionTerm} is that for small values of $w$ the material response prior to fracture is closer to linear.

\subsection{Fracture initiation based on tensile strength}
\label{sec:tensileStrength}
\begin{figure}
	\centering
	\includegraphics[width=0.6\textwidth]{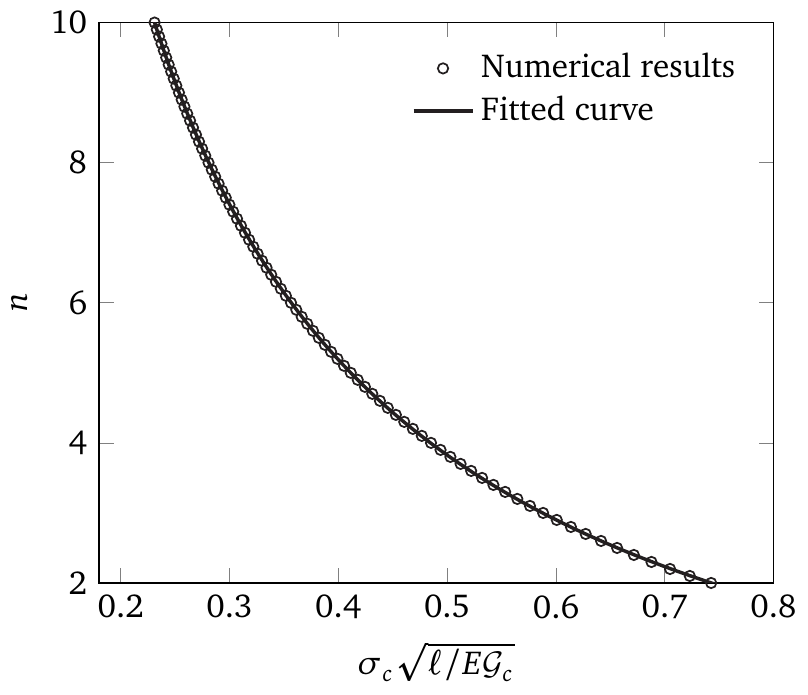}
	\caption{Relationship between normalized peak stress and parameter $n$ for the 1-dimensional tension test assuming uniform stress and damage.}
	\label{fig:peakStressRelation}
\end{figure}

The ability to initiate cracks in the absence of stress singularities requires the notion of strength in the form of a critical tensile stress $\sigma_c$ that is absent in the original theory of Griffith. Following the approach of \cite{Pham2011} and \cite{Bourdin2014}, let us assume that this coincides with the peak stress in the stress-strain curve associated with the 1-d homogeneous-stress model described above. For the case of the quadratic degradation function, the peak stress is reached at a phase-field value of 0.25, leading to the relation \citep{Nguyen2016}
\begin{equation}
	\ell = \frac{27 E \mathcal{G}_c}{256 \sigma_c^2}.
	\label{eq:ell_quadratic}
\end{equation}
For the family of degradation functions defined by \eqref{eq:1ParamDegFcn}, the above expression is further dependent on $n$ as evident from \cref{fig:stress_pf_relationship}. An explicit expression relating $n$ to the material parameters including $\ell$ is not easily obtained owing to the complicated form of \eqref{eq:1ParamDegFcn}. Instead we can utilize an approximate expression made by fitting a function to numerical evaluations of the peak stress for different values of $n$ as shown in \cref{fig:peakStressRelation}.
This function is expressed in terms of the dimensionless quantity $\sigma_{nd} = \sigma_c \sqrt{\ell / E \mathcal{G}_c}$ and is of the form
\begin{equation}
	\label{eq:fittedExpression}
	n \left( \sigma_{nd} \right) = c_0 + c_1 \sigma_{nd}^{-1} + c_2 \sigma_{nd}^{-2} + c_3 \sigma_{nd}^{-3}.
\end{equation}
With the weighting factor $w$ set to 0.1, the resulting values of the coefficients $c_0$ to $c_3$ are as follows:
\begin{equation}
	\begin{split}
		c_0 &= -1.96837\;16827 \\
		c_1 &= +3.07254\;12764 \\
		c_2 &= -0.10199\;57566 \\
		c_3 &= +0.00719\;48119
	\end{split}
\end{equation}
It should be emphasized however that \eqref{eq:fittedExpression} much like \eqref{eq:ell_quadratic} is valid only for the case where there are no stress gradients, and therefore has very limited applicability to cases where fracture nucleates from a stress concentration. Furthermore, these two equations do not account for the dependence that $n$ or $\ell$ must have on the mesh refinement when stresses are no longer uniform. On the other hand when stress concentrations are \emph{finite}, it is straightforward to check via inspection of numerical results whether critical stresses have been exceeded, and thus model calibration in such a case is much easier compared to one where the fracture emanates from a stress singularity.

\section{Numerical Examples} \label{sec:numericalExamples}
In this section we examine the performance of the proposed single-parameter degradation function relative to the conventional quadratic model through several examples. Our particular interest is in examining its ability to accurately capture the onset of fracture in the case of (a) a phase-field crack initiating at a location of stress singularity, and (b) one where a crack nucleates due to a nonsingular stress concentration reaching the prescribed material strength. The first numerical example is a recalculation of the problem presented in \cref{subsec:overshoot_example} using the new degradation function. It demonstrates how to determine the proper value of the parameter $n$ and also explores the effect of mesh refinement. The second example provides numerical evidence that the parameter tuning for $n$ becomes increasingly robust as the phase field parameter $\ell$ is reduced. The third example deals with a problem featuring strength-based crack initiation and also subsequent branching in a bi-material specimen. It highlights the need to carefully scrutinize numerical results and also the danger in blindly utilizing ready-made formulas for determining $\ell$ or $n$ which do not account for the specific local stress distributions in the problem at hand. In the final example, we investigate the new degradation function's potential to accurately model the stable propagation of an initial crack that is explicitly modeled in the geometry.

Numerical computations were carried out within a finite element framework implemented in our in-house C++ code, which utilizes OpenMP to achieve shared-memory parallelization on a desktop machine having a multi-core processor. For all problems, the relevant domains are discretized using 3-node triangles having linear shape functions and assume plane strain conditions. Our code allows the combination of elements having a different number of primary unknowns, and this feature is utilized in some of the examples below. In such cases, additional boundary conditions have to be implemented at element interfaces in order to have proper closure of the governing equations. In using \eqref{eq:1ParamDegFcn}, we have set $w = 0.1$ leaving $n$ as the sole free parameter subject to calibration/tuning. The coupled system of equations is solved using the alternate minimization algorithm, where we apply the linear approximation described at the end of \cref{sec:equationsAndNumerics} for the portion of the Jacobian matrix pertaining to the phase-field equations. With the aforementioned technique, very little difference is observed in computation times (e.g. number of iterations per step) between the simulations which utilize the quadratic degradation function and those which make use of our proposed alternative that is significantly more nonlinear.

\subsection{Brutal crack propagation in center-cracked specimen} \label{subsec:numex_ccspecimen}
\begin{figure}
	\begin{minipage}{0.47\textwidth}
		\centering
		\includegraphics[width=0.8\textwidth]{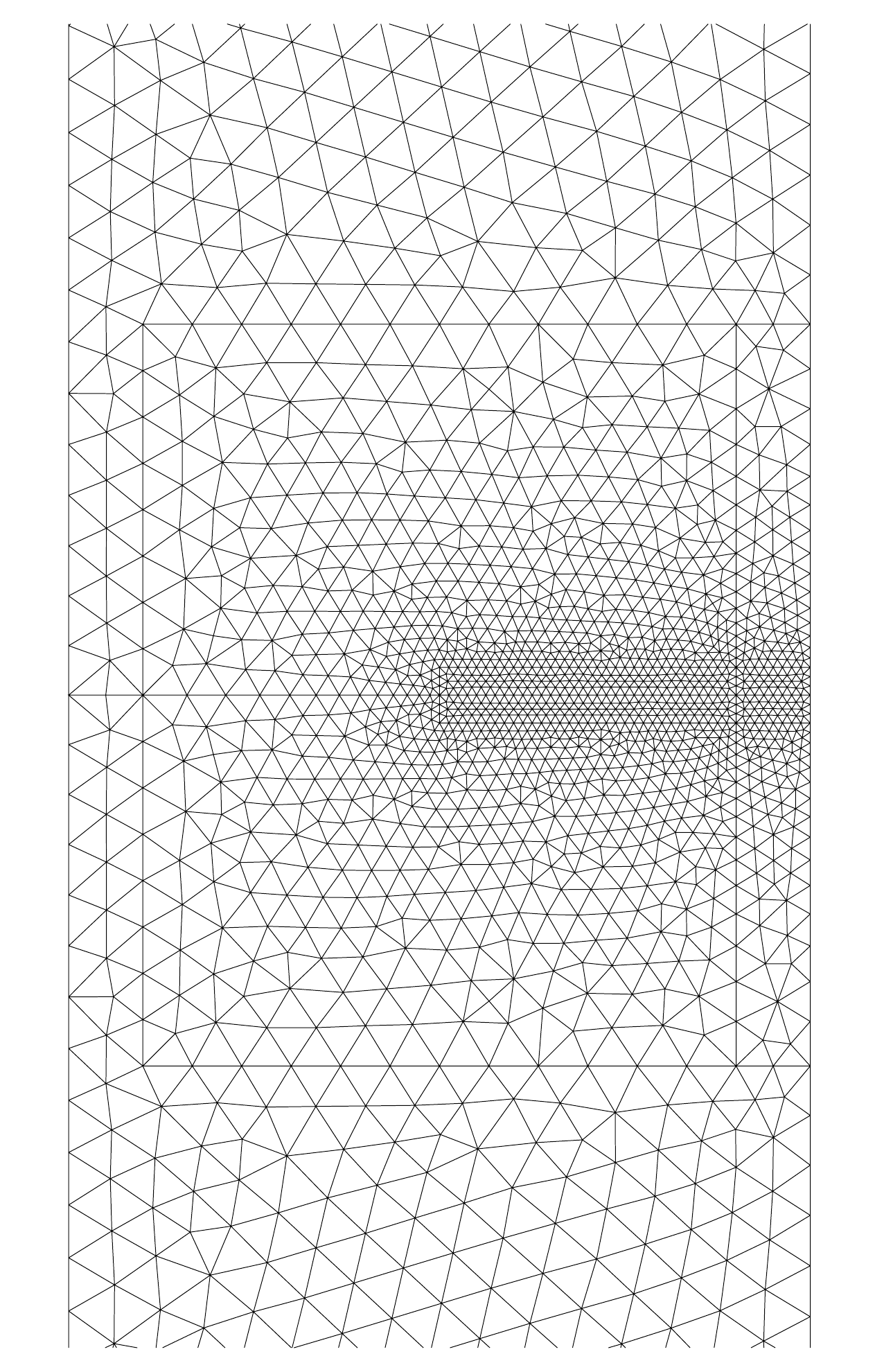}
		\captionof{figure}{Mesh refinement along projected fracture propagation path for center-cracked specimen.}
		\label{fig:numex1_meshdetail}
	\end{minipage} \hspace{1.0em}
	\begin{minipage}{0.47\textwidth}
		\centering
		\includegraphics[width=\textwidth]{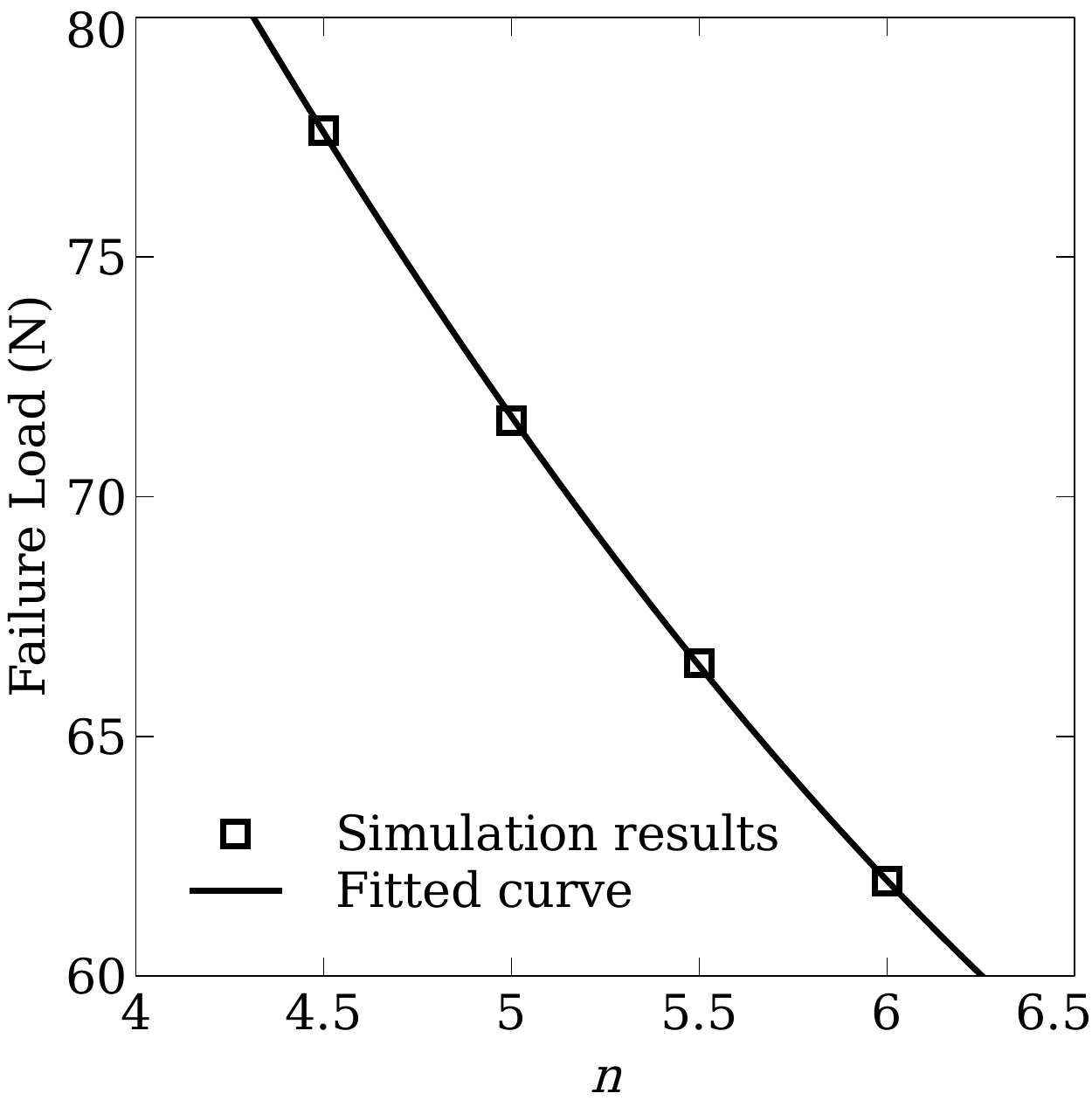}
		\captionof{figure}{Influence of degradation parameter $n$ on the failure load for the center-cracked specimen.}
		\label{fig:n_load_relationship}
	\end{minipage}
\end{figure}

\begin{figure}
	\begin{minipage}{0.47\textwidth}
		\centering
		\includegraphics[width=\textwidth]{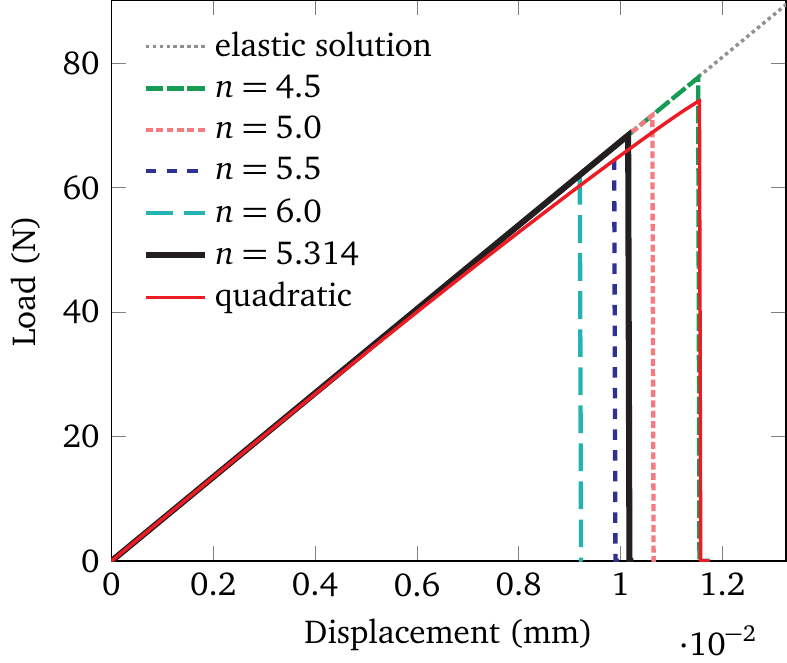}
		\captionof{figure}{Load displacement curves corresponding to different values of $n$.}
		\label{fig:n_dependence}
	\end{minipage} \hspace{1.0em}
	\begin{minipage}{0.47\textwidth}
		\centering
		\includegraphics[width=\textwidth]{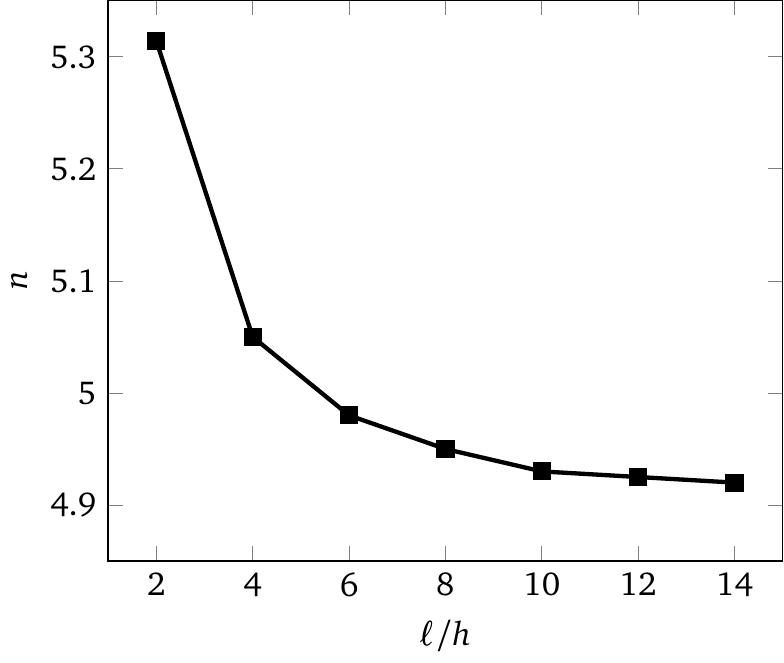}
		\captionof{figure}{Effect of mesh refinement on the degradation parameter.}
		\label{fig:ratio_vs_n}
	\end{minipage}	
\end{figure}

We revisit the brutal cracking problem of \cref{subsec:overshoot_example} involving a center-cracked specimen loaded in tension. As the analytical failure load is known for such a setup, it is useful not only for comparing the effect of our proposed single-parameter degradation function on the model behavior versus the original quadratic, but also as a means of calibrating the former by determining the proper value of $n$. Using the same specimen dimensions and material properties as before, along with a characteristic length of $\ell = 0.5$ mm and critical mesh refinement ratio of $\ell / h^e = 2$ (see \cref{fig:numex1_meshdetail} for detail of meshing in the crack vicinity), we resolve the problem utilizing our newly proposed degradation function given by \eqref{eq:1ParamDegFcn} with $w = 0.1$ as earlier recommended.
Initially, the prescribed upward displacement at the top boundary (see \cref{fig:CC_Specimen_CompDomain}) is increased using constant increments of $\Delta u_\text{coarse} = 2.5 \times 10^{-4}$ mm to determine the approximate displacement $u_\text{crit}$ at which failure occurs, after which the simulation is rerun with displacement increments refined to $2.5 \times 10^{-5}$ mm between $u_\text{crit} + \Delta u_\text{coarse}$ in order to achieve higher precision in the simulated failure load. \cref{fig:n_load_relationship} shows the results obtained from using different values of $n$ between 4.5 and 6. The proper value of the degradation function parameter corresponding to the desired critical load of $P_c = 68.26$ N is obtained via polynomial curve fitting, which yields a value of $n = 5.314$. Incorporating this into the simulation produces a failure load of 68.38 N, representing an error of 0.18\% with respect to the benchmark solution. While accuracy of the calculated load may be further improved by employing smaller $\Delta u_\text{fine}$ in addition to adjusting the value of $n$, the curve fitting procedure employed above nonetheless serves as a simple and straightforward means of achieving a reasonably accurate calibration of our proposed degradation function. An important property of \eqref{eq:1ParamDegFcn} evident from \cref{fig:n_load_relationship} is that a higher value of $n$ always leads to lower simulated failure load. Plots of the load-displacement curves for different $n$ are shown in \cref{fig:n_dependence}. We observe that the results are reasonably robust in terms of the exponent $n$ in that all choices of $n$ lead to an accurate representation of the linear regime prior to onset of fracture, in contrast to the classical quadratic degradation function. Furthermore, for the specific value of $\ell$ employed in the simulations, even an inaccurate calibration of $n$ having around 10\% deviation from the optimal value still leads to a more accurate failure load than predicted by the quadratic model. 

We also investigate the influence of the mesh refinement on the numerical results, as it is well known that the discretization of the domain close to the cracks must satisfy certain requirements on element sizes with respect to $\ell$ in order to properly resolve the exponential character of the phase-field. Specifically, $h < \beta \ell$ where $h$ is the length of element edges at the fracture vicinity and $\beta$ is a factor typically set to 1/2 in the literature based on results from \cite{Miehe2010_ijnme}. However this estimate was based on a setup where the crack is aligned with element edges, allowing for the natural reproduction of the gradient discontinuity that occurs at $\phi = 1$. In practice, the peak of the phase-field profile must occur at element Gauss points in order to effect a full degradation of the material stiffness. This implies that for constant gradient elements, this peak actually exists as a plateau of width $h$, which is an additional source of error when calculating the functional $\Gamma \left( \phi \right)$. Hence it may be necessary to choose a smaller value of $\beta$. Keeping the value of $\ell = 0.5$ mm constant for the above problem, we determine $n$ for different values of the effective element size at the critical zone. The resulting plot is shown in \cref{fig:ratio_vs_n}. It can be observed that the change in $n$ becomes significantly smaller for $h \leq \ell/10$, indicating that we see numerical convergence with respect to the ratio $\ell/h$. Unfortunately, a full convergence study of $n$ with respect to mesh refinement is limited by the accuracy of the approximate analytical solution in equation \eqref{eq:shape_factor}.

\subsection{Four-point bending test} \label{subsec:fourPointBend}
\begin{figure}
	\centering
	\includegraphics[width=\textwidth]{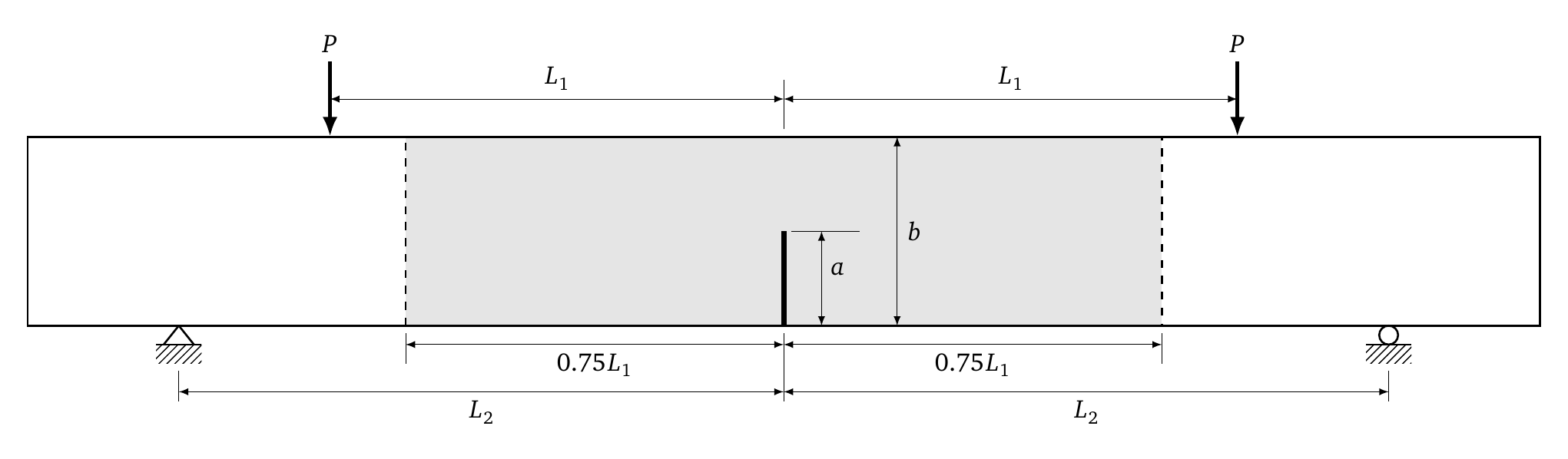}
	\caption{Beam with initial crack under four-point bending.}
	\label{fig:fourPointBeam}
\end{figure}

For the second example, we simulate fracture propagation in a beam having an initial crack of length $a$ and subjected to four-point bending as shown in \cref{fig:fourPointBeam}.
Our aim is to investigate the robustness of the degradation function parameter $n$ obtained in the previous section, by solving auxiliary problems that involve loading configurations fundamentally different to those in the main problem. To this end, we use the same values for the material parameters as given in \cref{subsec:overshoot_example}. Likewise, we treat Example \ref{subsec:numex_ccspecimen} as a prior calibration step.

The particular loading configuration investigated in this section produces a uniform internal moment between the inner applied loads, and by setting $a = 10$ mm, $b = 2a$ and $L_1 = 10a$ for the specimen dimensions we end with what is essentially the same computational domain as the previous example, albeit subjected to pure bending in the central beam portion of length $2L_1$. The internal bending moment at this region has a magnitude of $\left( L_2 - L_1 \right) P$, and for the current example we have set the moment arm $L_2 - L_1$ equal to 50 mm. However since the loading consists of concentrated forces and support reactions, we have found it necessary to model as non-fracturing the beam portions where the forces are applied in order to avoid spurious damage evolution at these locations. The regions colored white in \cref{fig:fourPointBeam} indicate portions of the domain that are modeled as linear elastic with only the displacement field $\bm{u}$ as the primary unknown, whereas the gray region has both $\bm{u}$ and $\phi$. Thus a boundary condition for the phase-field must be specified at the interface between fracturing and non-fracturing regions. For the current example, this is the Neumann condition $\nabla\phi \cdot \bm{n} = 0$, with $\bm{n}$ denoting the unit normal vector to the interface. 

A semi-analytical solution for the critical moment corresponding to an energy release rate of $\mathcal{G}_c$ at the crack tip can be computed as
\begin{equation}
\label{eq:critMoment}
M_c = \frac{b^2}{6 F \left( a/b \right)}  \left[ \frac{E \mathcal{G}_c}{\left( 1-\nu^2 \right)\pi a} \right]^{1/2},
\end{equation}
where for pure bending the shape factor $F \left( a/b \right)$ has the form
\begin{equation}
F \left( a/b \right) = 1.122 - 1.40 \left( a/b \right) + 7.33 \left( a/b \right)^2 - 13.08 \left( a/b \right)^3 + 14.0 \left( a/b \right)^4
\end{equation}
with a reported accuracy of 0.2\% in the stress intensity factor $K_I$ for $a/b \leq 0.6$ \citep{Tada2000}. For the current specimen, $a/b = 0.5$ and the above formula gives $F \left( a/b \right)  = 1.4945$. Plugging this into \eqref{eq:critMoment} yields a critical bending moment of 180.60 N-mm, which we designate as the benchmark solution for the problem.

Four simulation runs were carried out for comparison with the benchmark solution given above.  For the first, we employ the standard quadratic degradation function with $\ell$ calibrated to a value of 0.94 by matching the simulated critical load to the benchmark solution for the center-cracked specimen (see \cref{subsec:overshoot_example}). The second simulation run makes use of our proposed degradation function, where we have set $\ell = 0.94$ in order to compare results of different degradation functions given the same regularization length scale. The corresponding value of $n$ for this case is found to be 5.26 based on calibration runs using  the CC-specimen setup. In the third run, we set $\ell = 0.5$ which allows us to directly use the result of \cref{subsec:numex_ccspecimen}. The fourth simulation uses $\ell = 0.3$, with the obligatory calibration step yielding a value of 5.325 for the parameter $n$. In all four cases, the loading was applied in the form of prescribed downward displacements, first at increments of $\Delta U = -0.0025$ mm per step and then later refined to $-0.0001$ mm per step near the onset of crack propagation.

\begin{figure}
	\centering
	\includegraphics[width=0.5\textwidth]{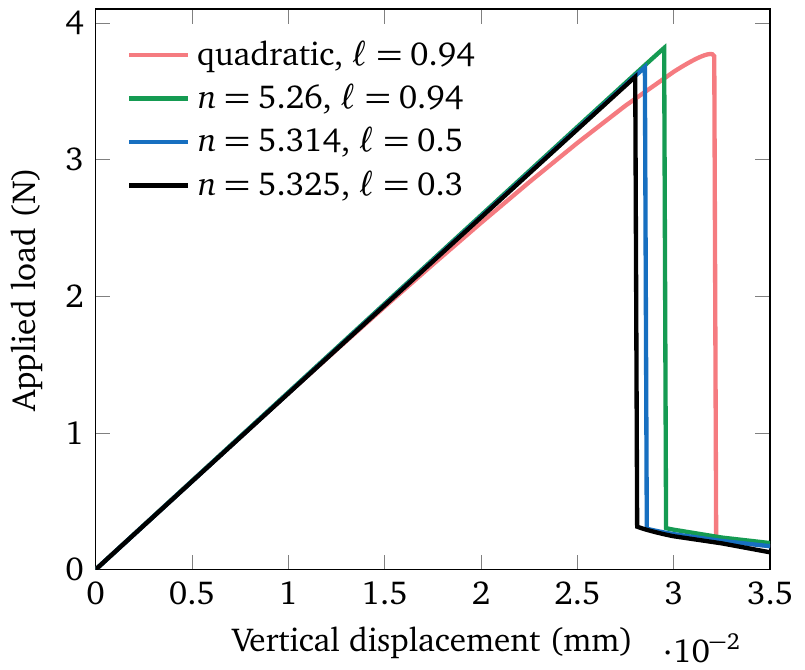}
	\caption{Load-displacement curves for the four-point bending specimen. The vertical axis gives the magnitude of the downward force $P$ at each of the two loading points shown in \cref{fig:fourPointBeam}; the horizontal axis gives the magnitude of vertical displacement at these locations.}
	\label{fig:fourPointLoadDisp}
\end{figure}

A summary of results for the four simulations is given in \cref{tab:fourPointBendingResults}, where the relative error of a quantity $Q$ with respect to its benchmark value is computed as
\begin{equation}
	RE = \left| \frac{Q^\text{simulated} - Q^\text{benchmark}}{Q^\text{benchmark}} \right| \times 100 \%.
\end{equation}
The corresponding load-displacement curves are shown in \cref{fig:fourPointLoadDisp}.
\begin{table}
	\centering
	\caption{Simulated critical internal bending moment for the four-point bending specimen. $RE$ denotes relative error.}
	\label{tab:fourPointBendingResults}
	\begin{tabular}{cccccccc}
		Simulation & Description & $\ell$ & $\ell / h$& $RE_{P_c^\text{calib}}$ & $U_c$ & $M_c$ & $RE_{M_c}$ \\
		run & & (mm) & & (\%) & (mm) & (N-mm)  & (\%) \vspace{0.5em} \\ \hline \\[-0.5em]
		1 & quadratic & 0.94 & 2.0 & $0.09$ & $-0.0321$ & 187.80 & $3.99$ \\
		2 & $n = 5.26$ & 0.94 & 2.0 & $0.21$ & $-0.0298$ & 192.75 & $6.72$ \\
		3 & $n = 5.314$ & 0.5 & 2.0 & $0.18$ & $-0.0285$ & 183.75 & $1.74$ \\ 
		4 & $n=5.325$ & 0.3 & 2.0 & $0.16$ & $-0.0280$ & 180.25 & $0.194$ \vspace{0.2em} \\ \hline \\[-0.5em]
	\end{tabular}
\end{table}
We can see that for the two runs with a coarse length scale of $\ell=0.94$, the relative errors for $M_c$ are significantly different from those for $P_c$ in the auxiliary problem used for calibration. These discrepancies show that the stress distributions around crack tips have a non-negligible influence on the model behavior, regardless of the form used for the degradation function. This is an unavoidable consequence of the diffuse approaches since the energy release rate at a crack tip is obtained via a nonlocal calculation. One should note that in this case the relative errors themselves are not definitive of a particular model's accuracy since they are influenced by the size of load increments (i.e., time steps), and also because the model parameter can often simply be re-tuned to give better results although this latter step was not done in the current example. It is however evident from comparing the relative errors obtained during calibration and those for the main setup that parameter values are not automatically transferable from one problem to another, and that for the degradation function proposed in the present work such transferability is affected by the value of $\ell$ used (presumably in relation to the material parameters $E$ and $\mathcal{G}_c$). 

In contrast, the results of the last three simulations seem to indicate that transferability of values for $n$ improves as $\ell$ is decreased. This is indeed an interesting outcome, and is consistent with the understanding that it is the diffuse representation of the crack tip which influences the energy release rates. Without drawing too broad conclusions from a single example, it appears that the new degradation function proposed here is less sensitive to calibration than the traditional degradation function. Nonetheless it is clear from this exercise that one must be careful in designing calibration procedures for any kind of degradation function, particularly when $\ell$ is large.

\subsection{Crack initiation and branching}
\begin{figure}
	\centering
	\begin{subfigure}{0.49\textwidth}
		\includegraphics[width=\textwidth]{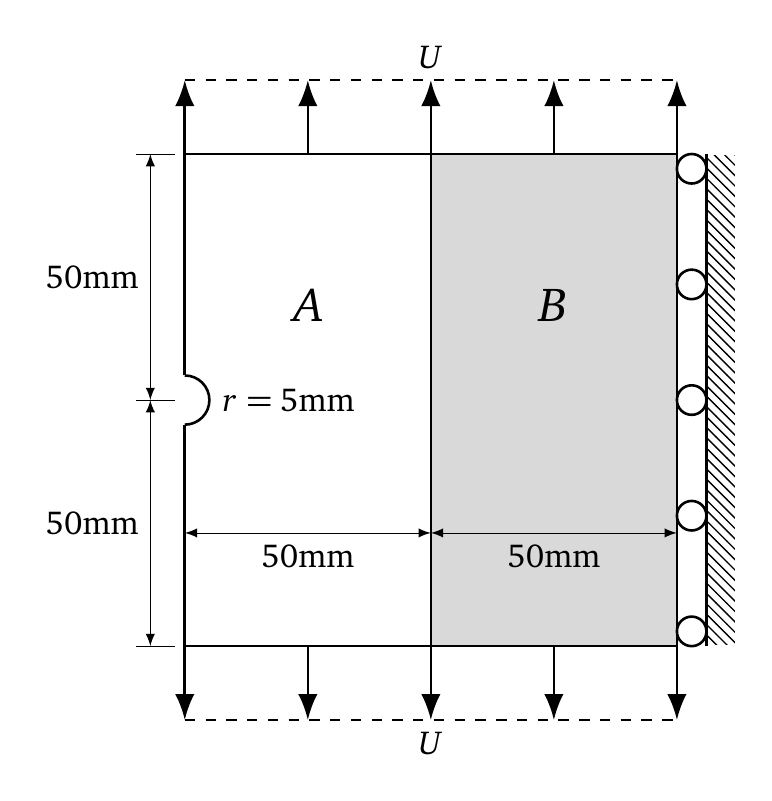}
		\caption{}
		\label{fig:bimaterial_geometry}
	\end{subfigure}
	\begin{subfigure}{0.49\textwidth}
		\includegraphics[width=\textwidth]{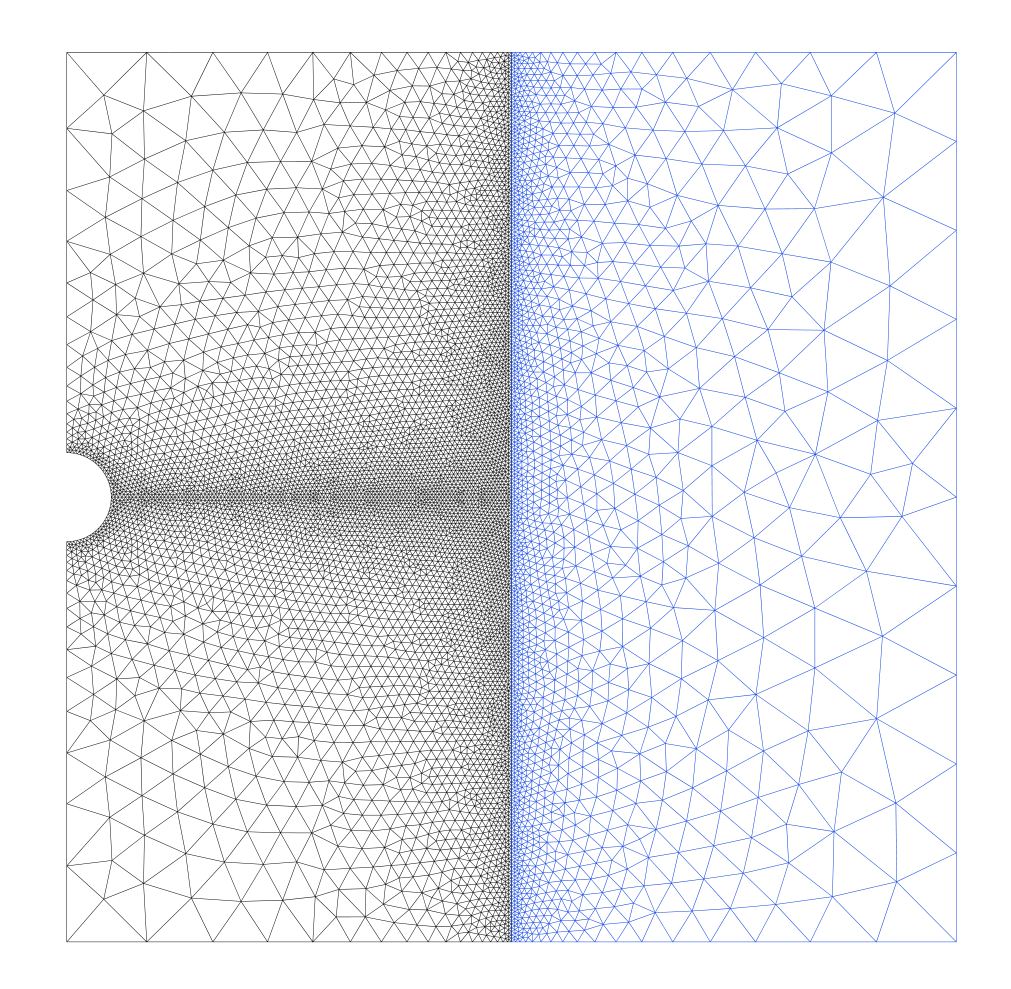}
		\caption{}
		\label{fig:bimaterial_mesh}
	\end{subfigure}
	\caption{(a) Geometry and applied loading for the bi-material specimen, and (b) finite element discretization.}
\end{figure}

\begin{figure}
	\centering
	\includegraphics[width=0.7\textwidth]{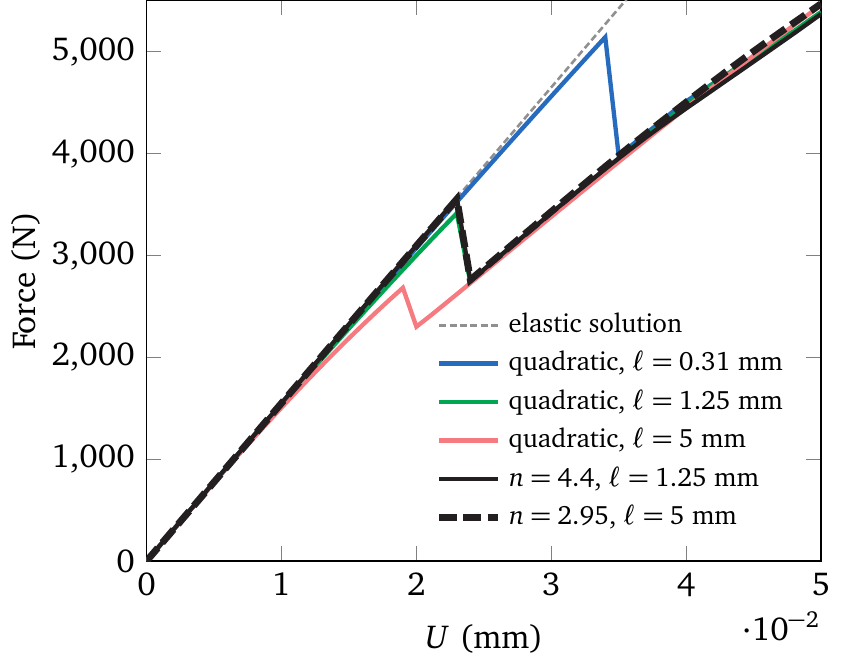}
	\caption{Total force on top boundary versus magnitude of applied displacement for the bi-material problem.}
	\label{fig:bimaterial_combinedPlots}
\end{figure}

\begin{figure}
	\begin{subfigure}{0.24\textwidth}
		\centering
		\includegraphics[width=\textwidth]{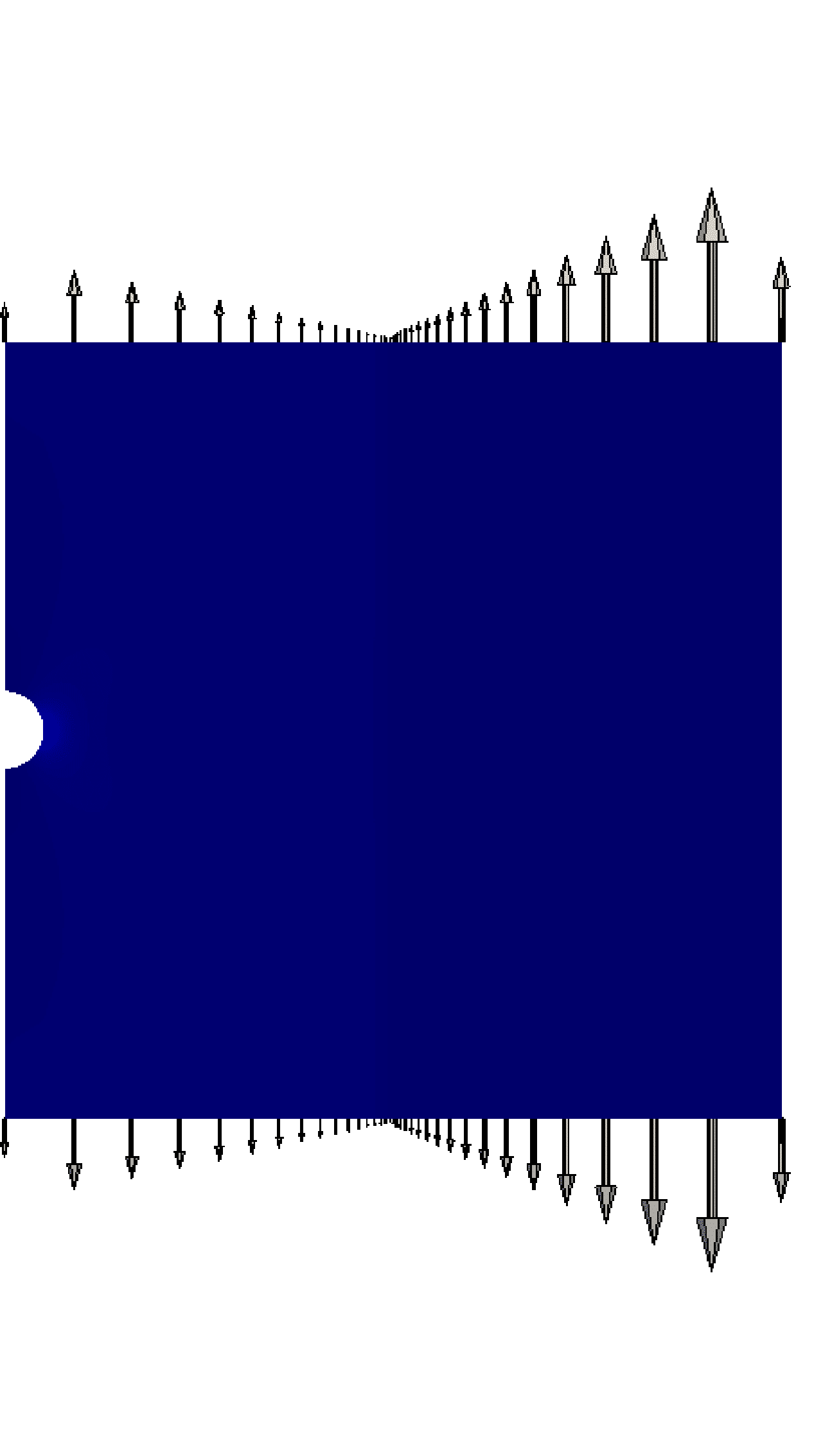}
		\caption{$U = 0.023$ mm}
	\end{subfigure}
	\begin{subfigure}{0.24\textwidth}
		\centering
		\includegraphics[width=\textwidth]{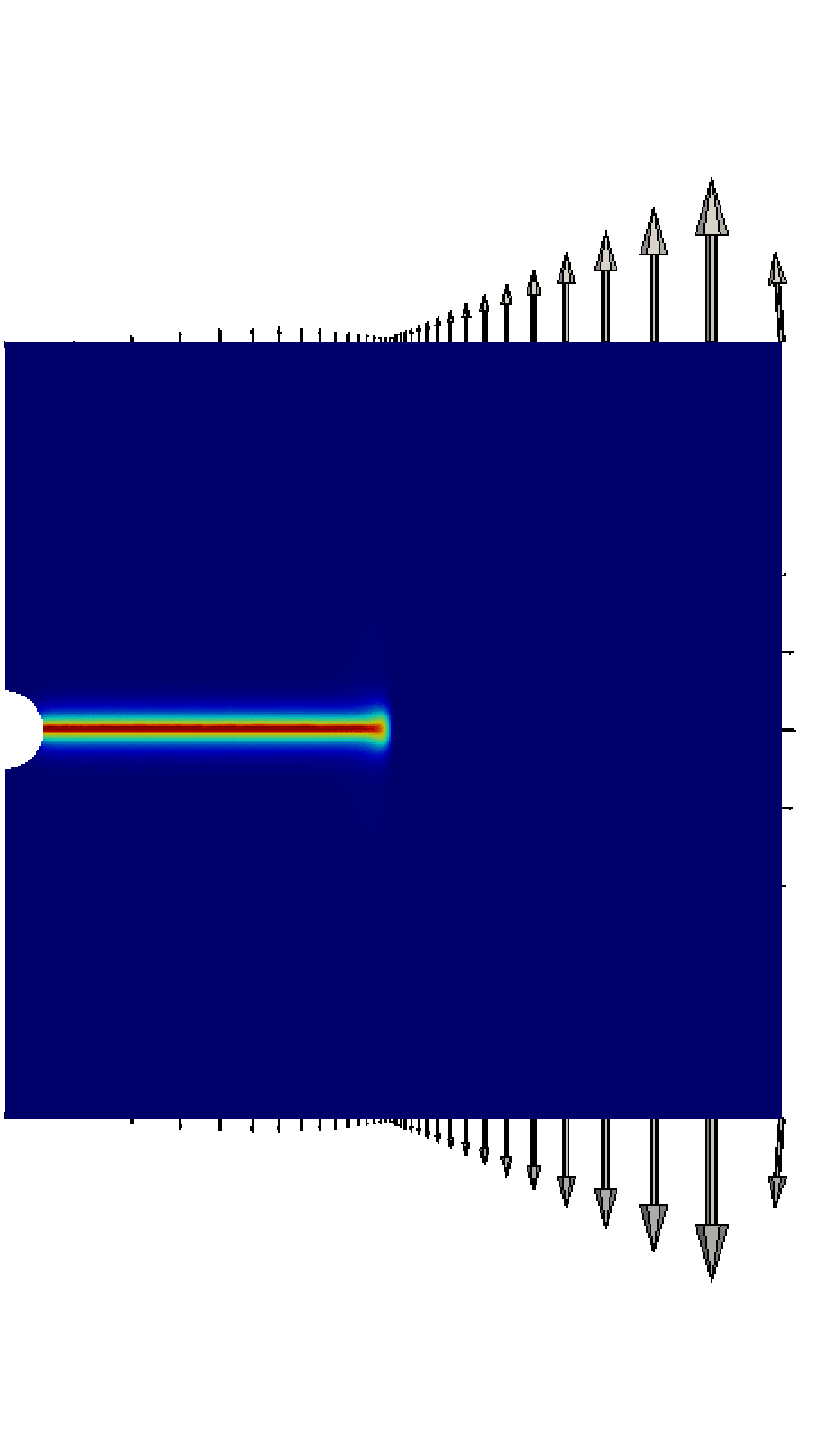}
		\caption{$U = 0.024$ mm}
	\end{subfigure}
	\begin{subfigure}{0.24\textwidth}
		\centering
		\includegraphics[width=\textwidth]{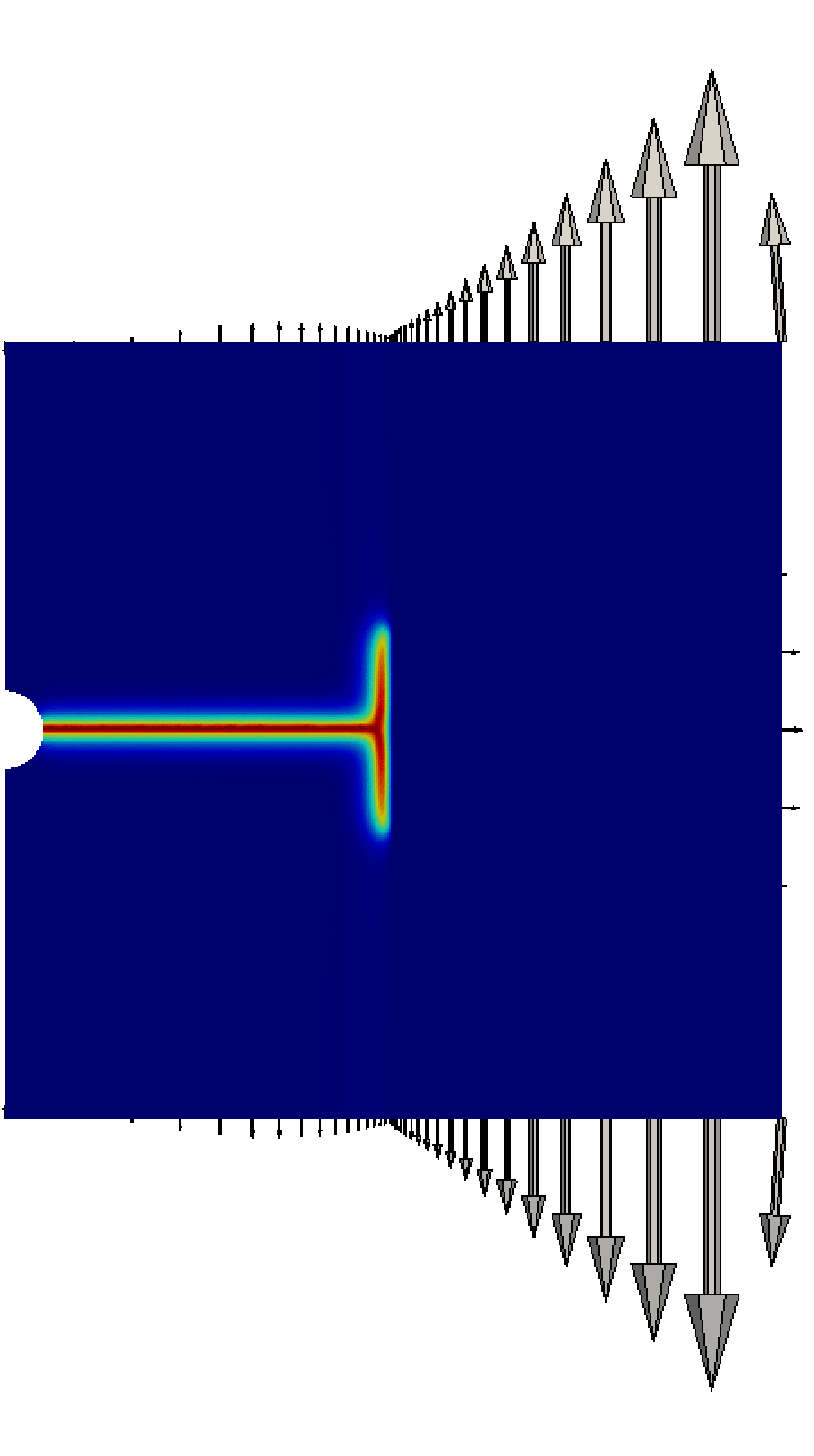}
		\caption{$U = 0.04$ mm}
	\end{subfigure}
	\begin{subfigure}{0.24\textwidth}
		\centering
		\includegraphics[width=\textwidth]{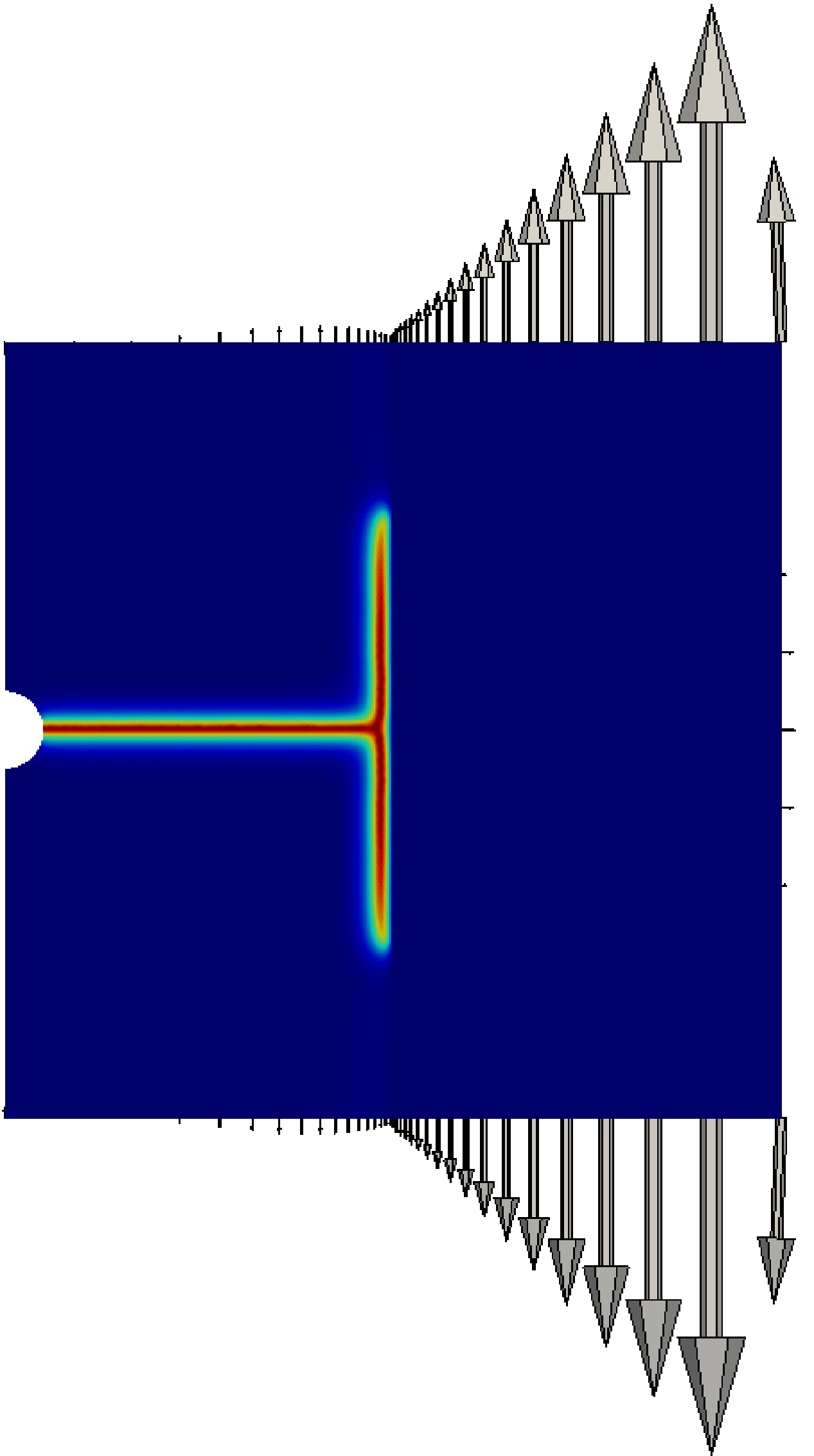}
		\caption{$U = 0.05$ mm}
	\end{subfigure}
	\caption{Element-wise values of the phase-field and boundary nodal forces pertaining to simulation 4 ($n = 4.4$, $\ell = 1.25$ mm).}
	\label{fig:force_evolution}
\end{figure}

\begin{figure}
	\centering
	\begin{subfigure}{0.35\textwidth}
		\includegraphics[width=\textwidth]{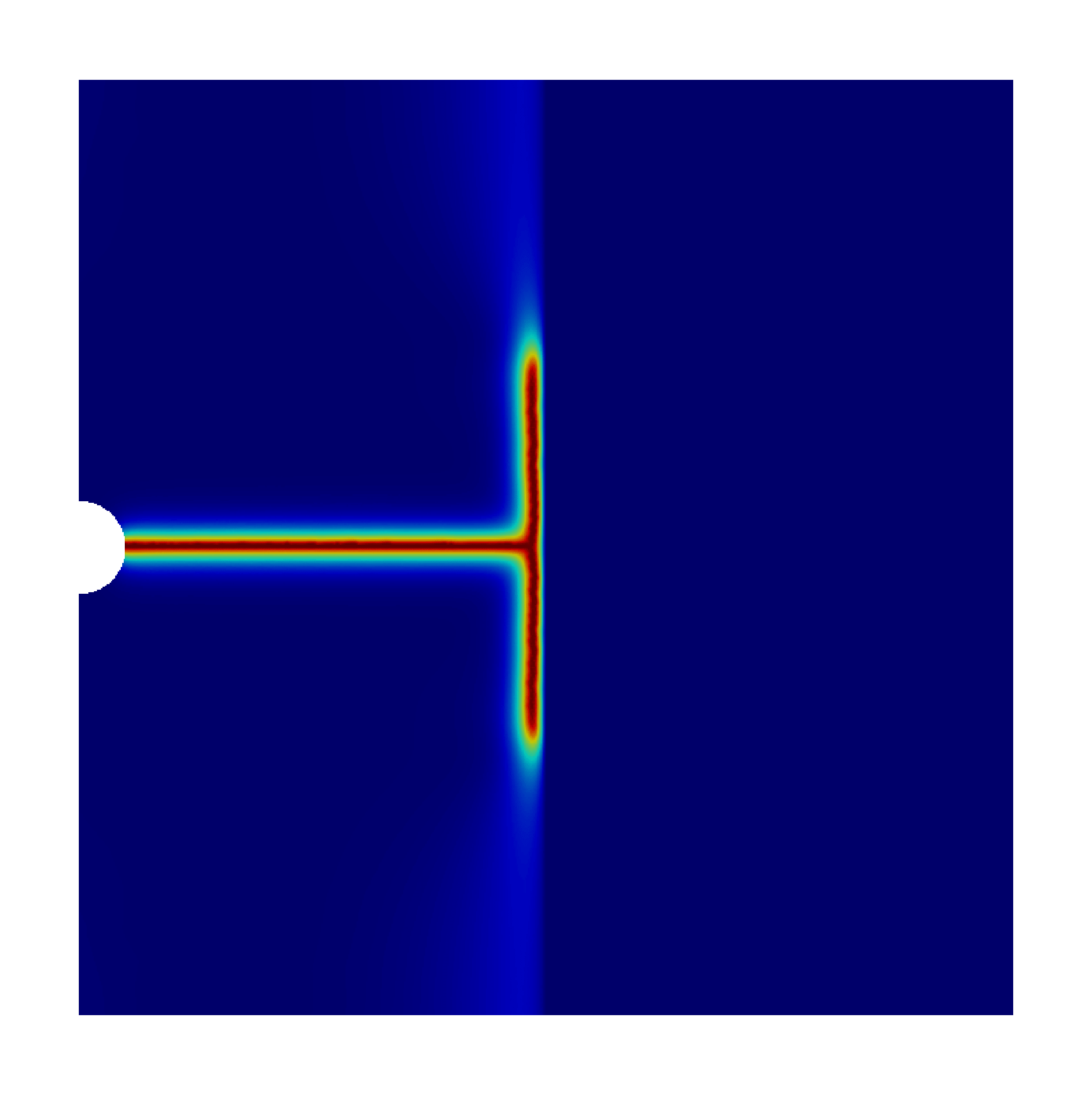}
		\caption{Quadratic degradation, $\ell = 1.25$ mm}
	\end{subfigure}
	\begin{subfigure}{0.35\textwidth}
		\includegraphics[width=\textwidth]{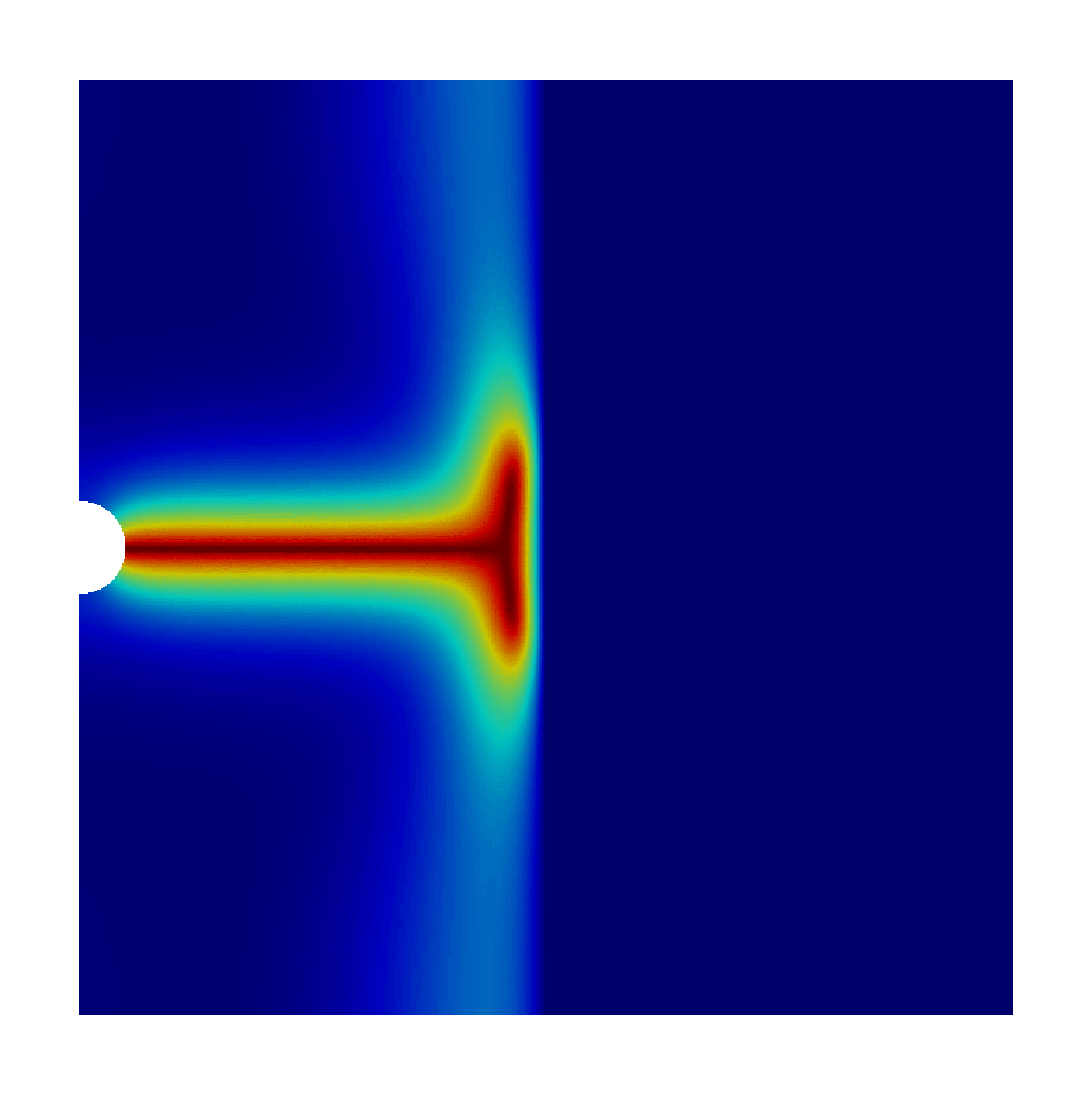}
		\caption{Quadratic degradation, $\ell = 5.0$ mm}
	\end{subfigure} \\
	\begin{subfigure}{0.35\textwidth}
		\includegraphics[width=\textwidth]{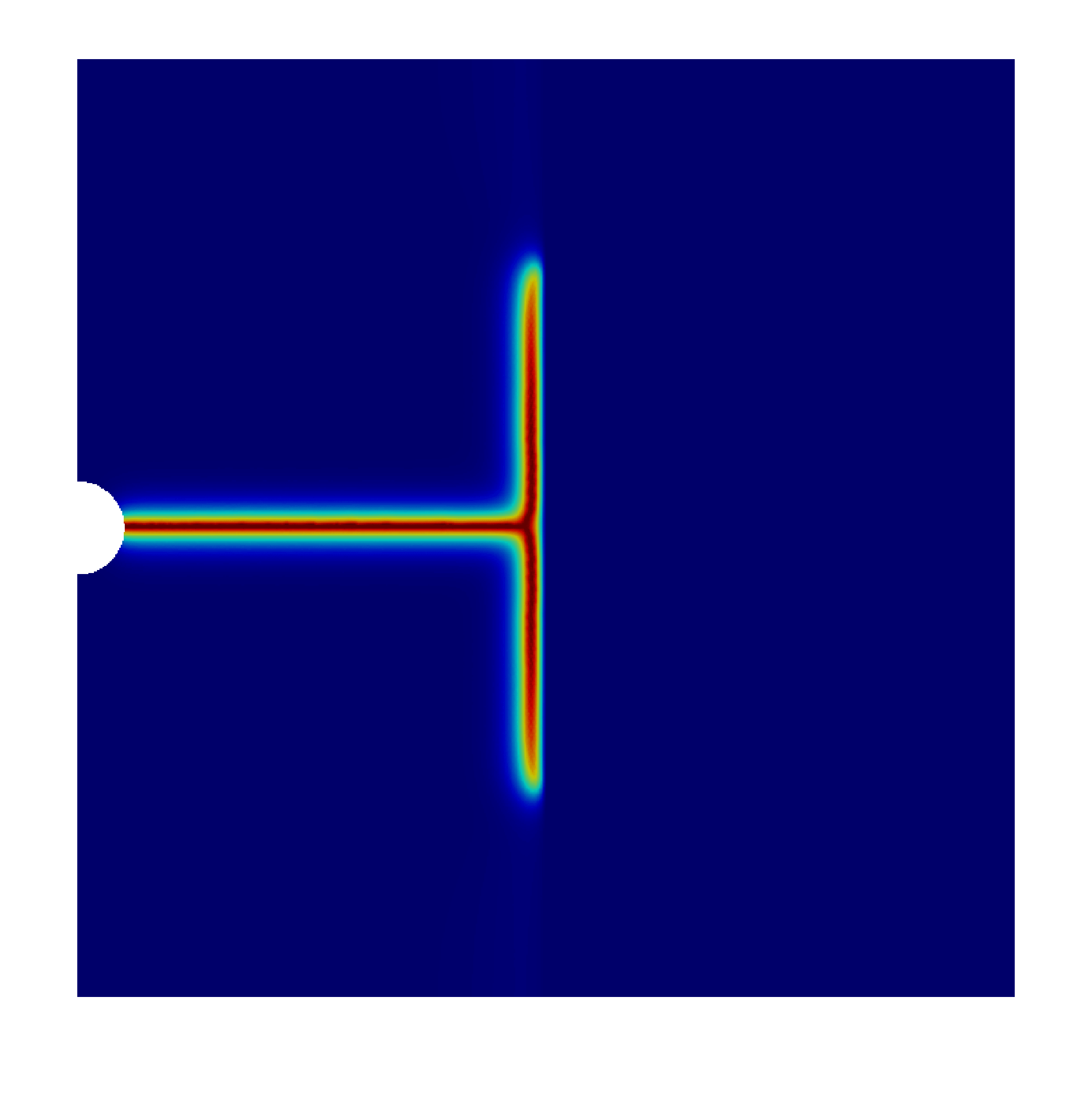}
		\caption{$n = 4.4$, $\ell = 1.25$ mm}
	\end{subfigure}
	\begin{subfigure}{0.35\textwidth}
		\includegraphics[width=\textwidth]{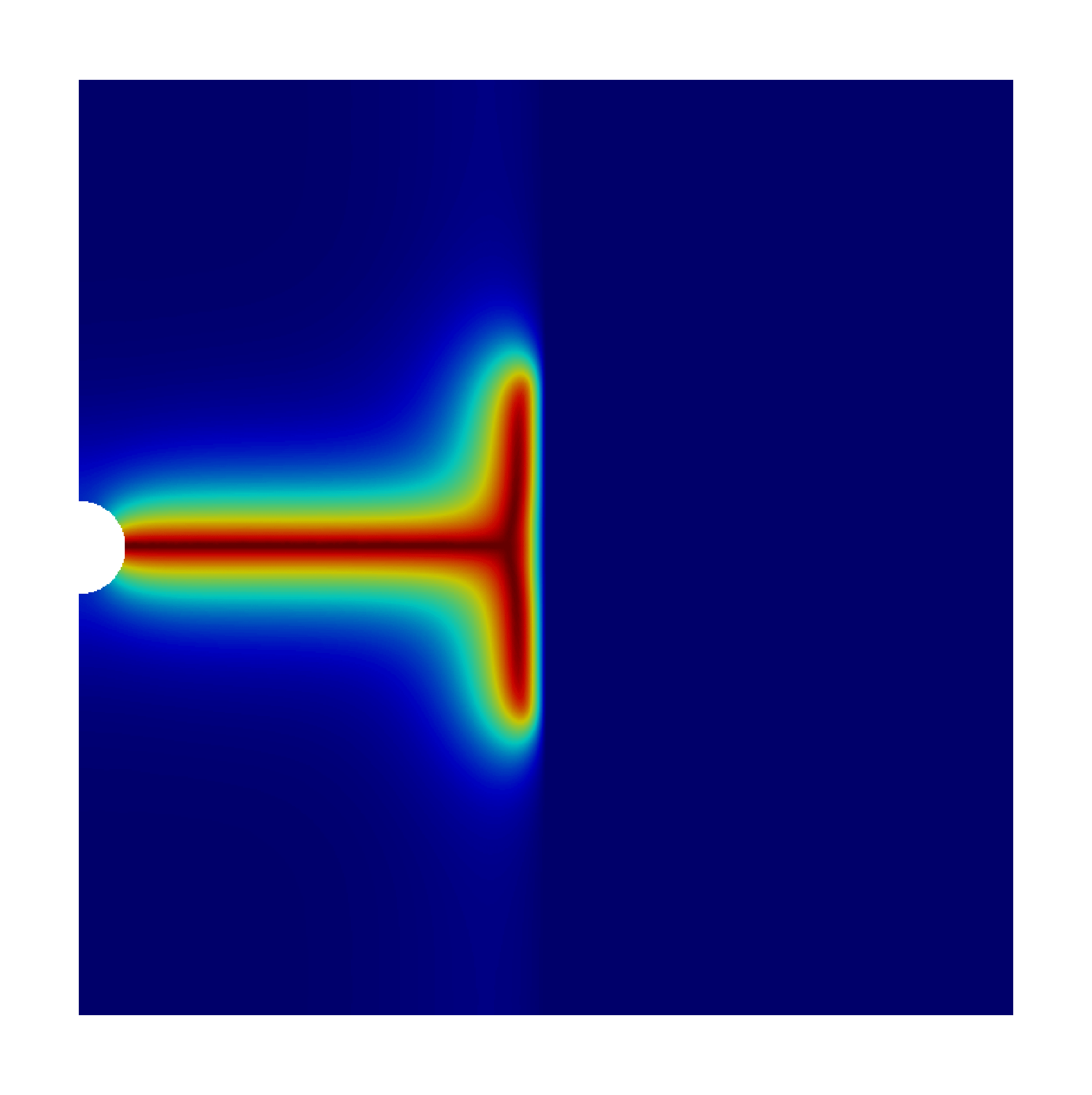}
		\caption{$n = 2.9$, $\ell = 5.0$ mm}
	\end{subfigure}
	\caption{Phase-field profile corresponding to an applied displacement magnitude of $U = 0.05$ mm.}
	\label{fig:branching_finalcrack}
\end{figure}

Our third example involves a bi-material specimen that is loaded in tension as shown in \cref{fig:bimaterial_geometry}.
The material properties corresponding to the regions designated as $A$ and $B$ in the figure are given in \cref{tab:matProp}, where it can be seen that material $B$ is stiffer than the other and is also non-fracturing.
\begin{table}
	\centering
	\caption{Material properties for the bi-material specimen.}
	\label{tab:matProp}
	\begin{tabular}{ccc}
		Region & $A$ & $B$ \vspace{0.2em} \\ \hline \\[-0.5em]
		$E$ & 100 GPa & 200 GPa \\
		$\nu$ & 0.2 & 0.2 \\
		$\mathcal{G}_c$ & 0.1 N/mm & $-$ \\
		$\sigma_c$ & 70 MPa & $-$ \\[0.5em] \hline
	\end{tabular} \vspace{1em}
\end{table}
We thus adopt the approach employed in the previous example: region $B$ is modeled as a linear elastic material with only displacement degrees of freedom, while in region $A$ we incorporate additional unknowns pertaining to the phase-field. In contrast to \cref{subsec:fourPointBend} however, here we impose the homogeneous Dirichlet condition $\phi = 0$ on the interface separating between the two regions. This is done to ensure that the resulting phase-field profile is meaningful with respect to crack length calculations. Prescribed uniform vertical displacements of magnitude $U = 0.05$ mm are applied at the top and bottom boundaries in increments of $\Delta U = 0.001$ mm. We compare simulation results obtained from using our proposed new degradation function to that of the classical model employing quadratic energy degradation for two values of the phase-field regularization parameter, namely $\ell = 1.25$ mm and $\ell = 5$ mm. All four cases utilize the same discretization of the problem domain shown in \cref{fig:bimaterial_mesh}, where the effective size of element edges along the anticipated path of crack propagation have been set to $h = 0.4$mm. In addition a fifth simulation run was carried out with $\ell$ set to 0.31 mm on a finer discretization having $h = 0.15$ mm; this corresponds to the case where failure occurs at the specified value fo $\sigma_c$ in connection with a quadratic degradation model. Force-displacement curves for the five cases are displayed in \cref{fig:bimaterial_combinedPlots}, while values of specific quantities of interest at crack initiation are listed in \cref{tab:bimaterial_quantities}.
\begin{table}
	\centering
	\caption{Details of simulation results pertaining to the bi-material problem: magnitude of applied displacement at crack initiation ($U_c$), total vertical force at top boundary  ($F_c$), maximum tensile stress ($\sigma_\text{max}$), and phase-field value in the critical element ($\phi_c$).}
	\label{tab:bimaterial_quantities}
	\begin{tabular}{ccccccc}
		Simulation & Description & $\ell$ (mm) & $U_c$ (mm) & $F_c$ (N) & $\sigma_\text{max}$ (MPa) & $\phi_c$ \vspace{0.2em} \\ \hline \\[-0.5em]
		1 & quadratic & $0.31$ & 0.034 & 5142 & 71.02 & 0.4430 \\
		2 & quadratic & $1.25$ & 0.023 & 3415 & 48.06 & 0.4611 \\
		3 & quadratic & $5.0$ & 0.019 & 2682 & 39.71 & 0.4135 \\
		4 & $n = 4.4$ & $1.25$ & 0.023 & 3562 & 69.71& 0.1103 \\
		5 & $n = 2.95$ & $5.0$ & 0.023 & 3553 & 70.17 & 0.1235 \\[0.5em] \hline
	\end{tabular}
\end{table}
Due to the fact that boundary displacements are applied in constant increments without refinement near the instance of fracture initiation as done in the previous example, it is not possible to reproduce exactly the specified critical stress of 70 MPa during crack nucleation. Simulations 1 to 3 were carried out using the classical quadratic degradation function, while 4 and 5 utilize the new single-parameter degradation function given in \eqref{eq:1ParamDegFcn}. For the former, an estimate for the required magnitude of $\ell$ corresponding to $\sigma_c = 70$ MPa may be obtained from \eqref{eq:ell_quadratic}. This yields $\ell = 0.215$, however as observed from \cref{fig:bimaterial_combinedPlots}, the correct value of the regularization length for the quadratic case is nearer to 0.31. We also note that simulations 4 and 5 produce virtually identical results with respect to the peak load, demonstrating the ability of the proposed new degradation function to properly compensate for different magnitudes of crack regularization. It can be observed that past the initial crack formation which is represented by the sudden drop in the force-displacement curve, all simulations display essentially the same behavior. This is not surprising, since the for all cases, the initial crack traverses the entire width of region $A$, and so the subsequent residual force comes mainly from the resultant stresses in the non-fracturing part of the specimen as shown in \cref{fig:force_evolution}.
Meanwhile, the final crack trajectories corresponding to $U = 0.05$ mm obtained from simulations 2 to 4 are shown in \cref{fig:branching_finalcrack}.
We point the reader to a particular nuance of the current numerical example, namely that it is not immediately obvious simply from looking at the combined force-displacement plots in \cref{fig:bimaterial_combinedPlots} which curve represents the correct specimen behavior under the given loading conditions.

\begin{figure}
	\centering
	\begin{subfigure}{0.48\textwidth}
		\centering
		\includegraphics[width=0.9\textwidth]{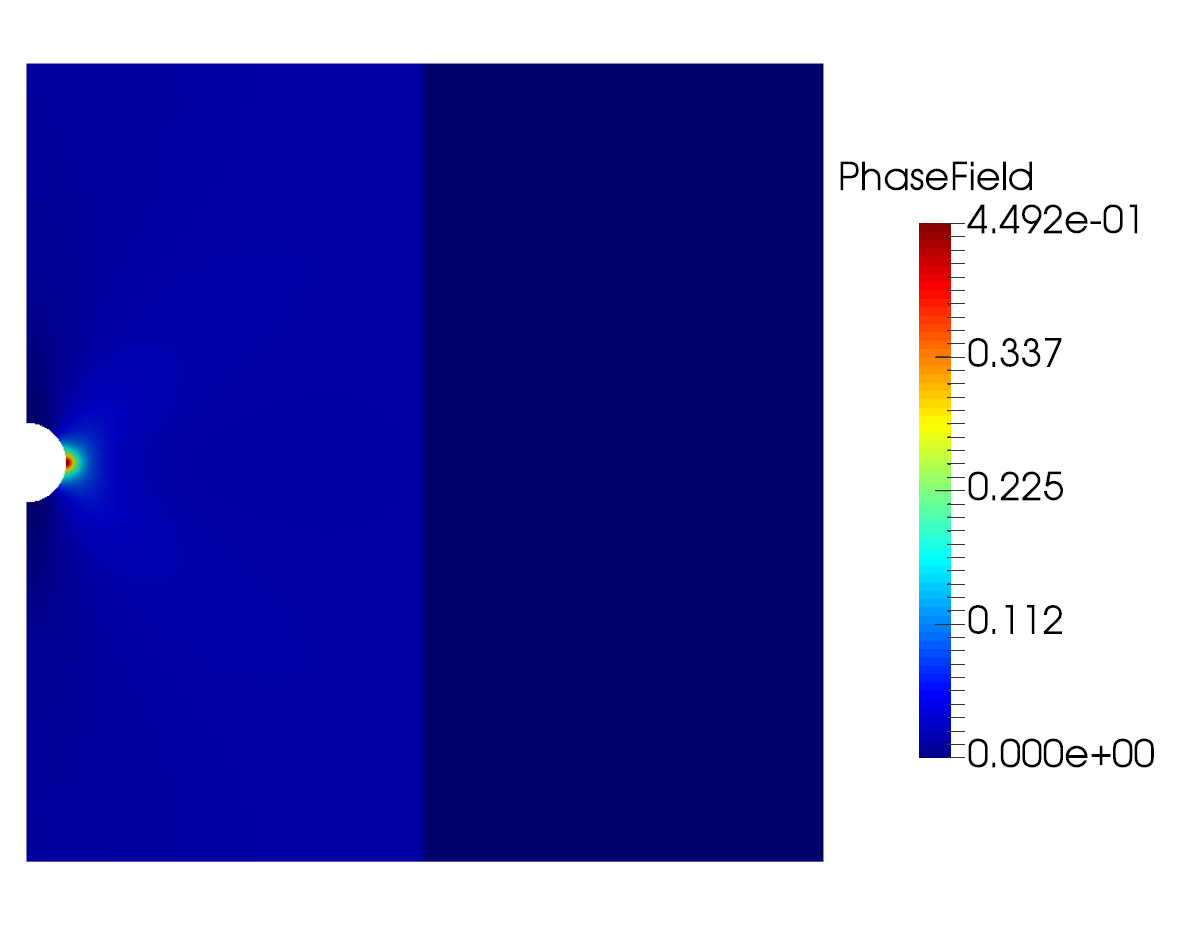}
		\caption{$U_c = 0.034$ from simulation 1}
	\end{subfigure}
	\begin{subfigure}{0.48\textwidth}
		\centering
		\includegraphics[width=0.9\textwidth]{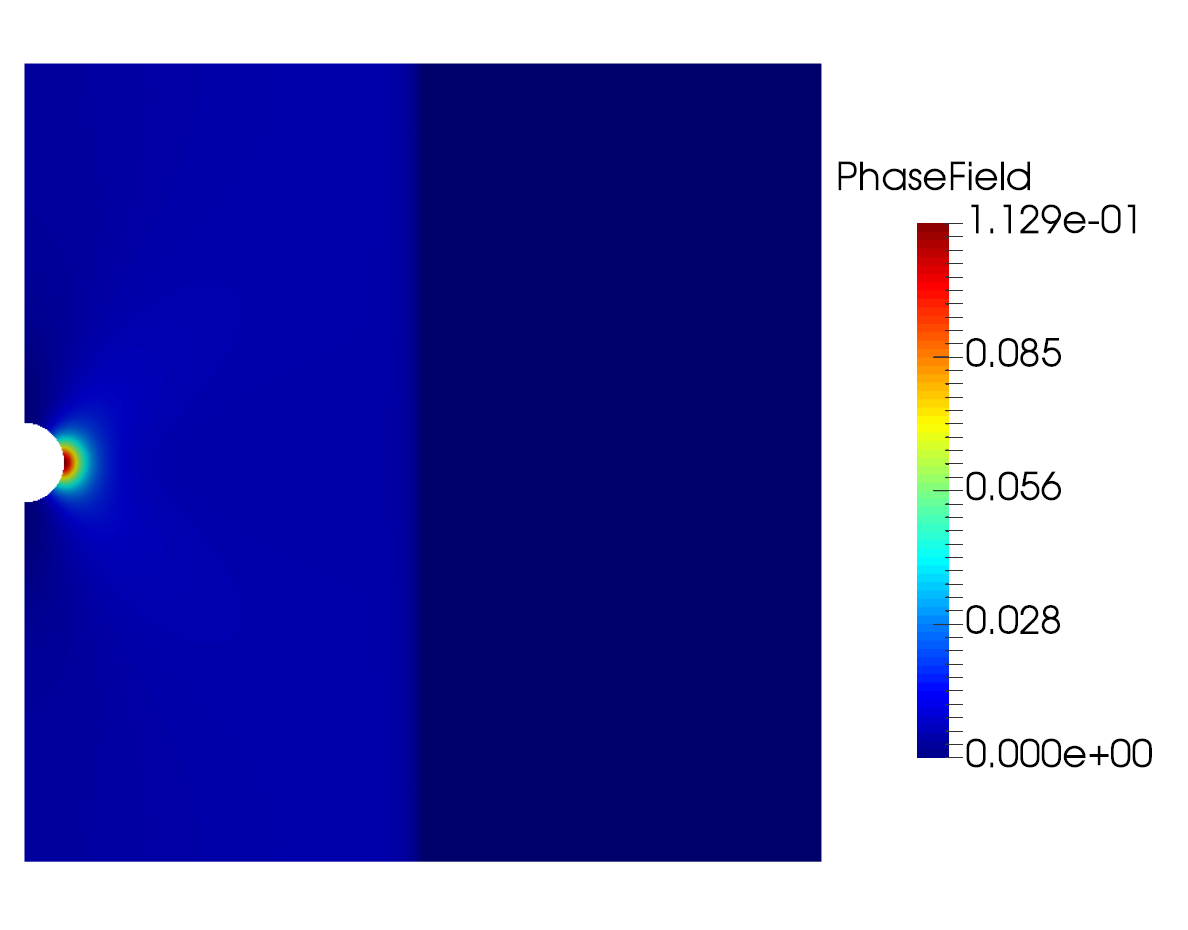}
		\caption{$U_c = 0.023$ from simulation 4}
	\end{subfigure}
	\caption{Plots of the phase-field profile at critical displacements based on nodal values (see \cref{tab:bimaterial_quantities} for references to simulation numbers). Note that color maps are scaled based on the respective maximum values of the phase-field occurring in each case.}
	\label{fig:critical_damage}
\end{figure}

An important insight can be found by examining the value of $\phi$ in the critical element at which the stress is maximum. From \cref{tab:bimaterial_quantities} we see that for simulation 1 this equal to 0.443 which corresponds to a degradation factor of $g_2 \left( 0.443 \right) = 0.310$. This means that just prior to fracture, the critical element has a stiffness of only slightly more than a third of its original value. This leads to a severe under-calculation of the critical stress, with the simulation reporting a value of $\sigma= 71.02$ MPa at the critical element whereas a separate simulation assuming linear elastic behavior of the whole domain produces a stress of  106.1 MPa at the same location. This amounts to an overshoot of more than 50\% of the true tensile strength. However since the damaged region comprises only a small fraction of the specimen's total area (see \cref{fig:critical_damage}), the linear elastic behavior exhibited by the undamaged region dominates the specimen response leading to deceptively small deviations in the force-displacement curves.
In contrast, simulation 4 has $\phi = 0.1103$ at the critical element prior to failure. Coupled with the form of \eqref{eq:1ParamDegFcn} that minimizes stress degradation for small values of the phase-field, we obtain $g_s \left( 0.1103 \right) = 0.938$ for $n = 4.4$ and $w = 0.1$ which gives rise to much less distortion of the stress compared to the classical quadratic degradation function. Indeed at a displacement magnitude of $U_c = 0.023$ mm, the phase-field model predicts a tensile stress of 69.71 MPa which is much closer to the value of 71.75 MPa obtained from assuming purely elastic material behavior. Additionally, we note that the values of $\ell$ and $n$ which lead to what may be considered as the ``correct'' model response in the case of using respectively the quadratic and exponential degradation functions are significantly different from the estimates obtained by using the formulas given in \cref{sec:tensileStrength}. This is due to the incompatibility between actual stress states at the crack initiation region for the current example (which are already localized prior to crack initiation) and the assumption of homogeneous stress pertaining to the 1-d case that was used in deriving the expressions in the aforementioned section.

\subsection{Stable crack growth in a homogeneous medium}
\begin{figure}
	\centering
	\includegraphics[width=0.5\textwidth]{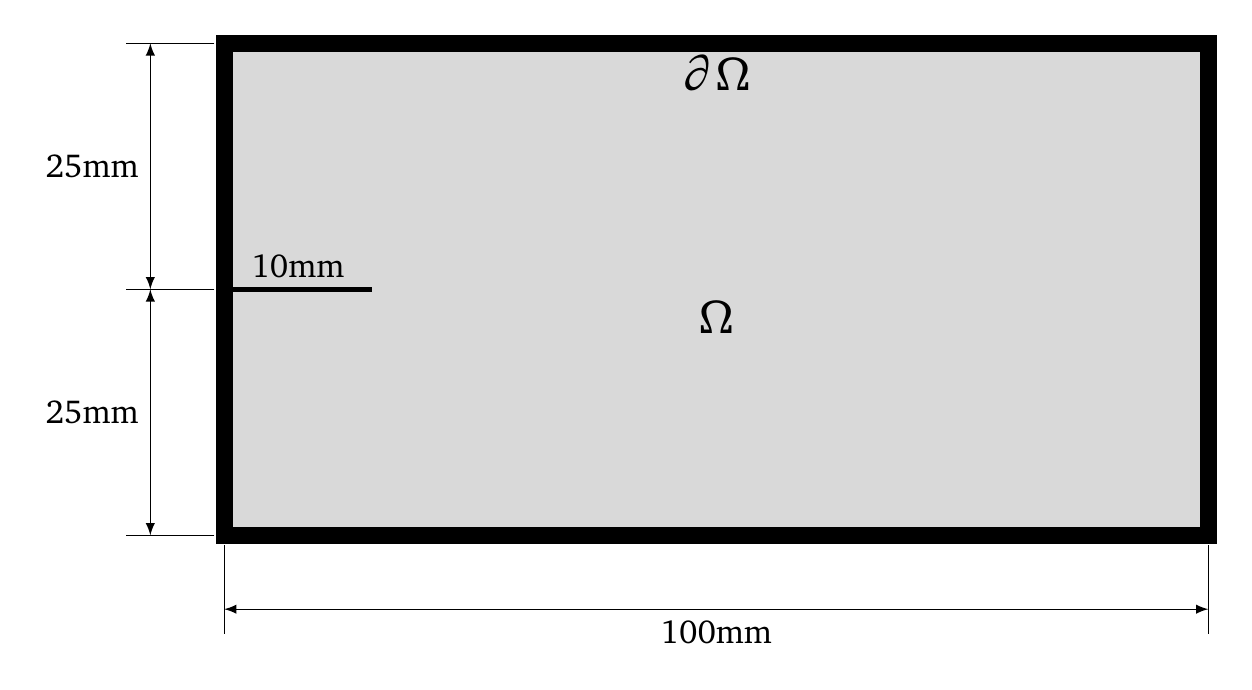}
	\caption{Specimen geometry and dimensions for the surfing problem. Displacements are prescribed on $\partial\Omega$ indicated by the bold lines, but not on the faces of the initial crack.}
	\label{fig:surfing_geometry}
\end{figure}

The problem of a rectangular specimen subjected to so-called \emph{surfing boundary conditions} was initially used by \cite{Hossain2014} for studying the effective toughness of heterogeneous media and later adopted by \cite{Kuhn2016} in the context of configurational forces. Details of the specimen geometry together with the initial crack are given in \cref{fig:surfing_geometry}.
In both of the works mentioned, the phase-field profile is initialized such that $\phi = 1$ at all points in the crack locus, decaying with the proper gradients towards zero away from the crack. In contrast for the current example, no such initialization is carried out in order to simulate the transition from crack initiation at a location of stress singularity towards propagation of a fracture that is fully described by the phase-field. The Dirichlet boundary conditions are derived from a $K_I$-controlled displacement field corresponding to a crack under mode-I loading, given by
\begin{equation}
	\left\{ \begin{array}{c} U_x \\ U_y \end{array} \right\} =
	K_I \frac{1 + \nu}{E} \sqrt{\frac{r}{2\pi}} \left( \kappa - \cos\theta \right)
	\left\{ \begin{array}{c} \cos \left( \theta / 2 \right) \\ \sin \left( \theta / 2 \right) \end{array} \right\}
\end{equation}
in terms of polar coordinates $r$ and $\theta$, with the crack extending infinitely along the line $\theta = \pi$ from a tip located at $r = 0$. The quantity $\kappa$ is Kolosov's constant which is equal to $3 - 4\nu$ for the case of plane strain. Crack propagation is achieved by translation of the above coordinate system with respect to the original configuration of the specimen resulting in the horizontal motion of the crack tip. Letting $x_{K_I} \left( t \right) = vt$ and $y_{K_I} \left( t \right) = 0$ be the Cartesian coordinates of the crack tip for some fictitious time $t$ and positive constant $v$, we obtain
\begin{equation}
	\begin{split}
	r \left( t \right) &= \sqrt{ \left( x - vt \right)^2 + y^2} \\
	\theta \left( t \right) &= \arctan \left( \frac{y}{x - vt} \right), \quad \theta \in \left[ -\pi, \pi \right]
	\end{split}
\end{equation}
For simplicity, we have chosen to let $v = 1$. The material properties used for the specimen are $E = 210$ GPa, $\nu = 0.3$ and $\mathcal{G}_c = 2.7$ N/mm. Finally, we set $K_I$ to a constant value of $\sqrt{E \mathcal{G}_c}$ and run the simulation from $t = 5$ up to $t = 30$ in increments of $\Delta t = 0.5$. The analytical response of the specimen can be understood as follows: for $t \in \left[ 5, 10 \right)$, the energy release rate at the crack tip is smaller than $\mathcal{G}_c$, so that the crack does not propagate. At $t = 10$, this quantity is exactly equal to $\mathcal{G}_c$, allowing the crack to growth. Henceforth for $t > 10$, the crack tip moves to the $K$-field center denoted by $x_{K_I} \left( t \right)$. 

\begin{figure}
	\begin{subfigure}{0.48\textwidth}
		\centering
		\includegraphics[width=\textwidth]{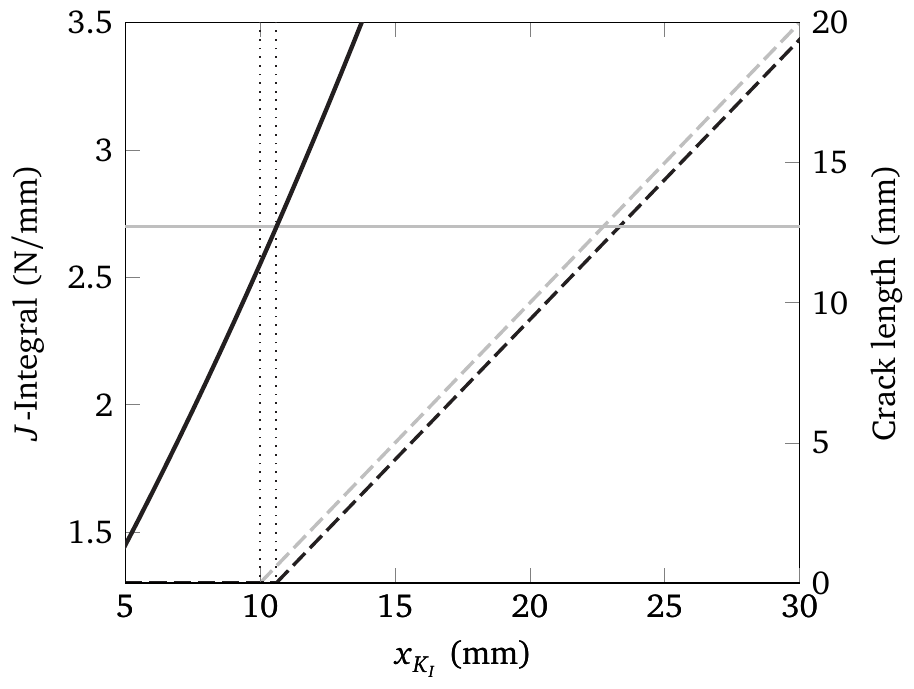}
		\caption{Constructed solution based on linear elasticity.}
		\label{fig:surfing_elastic}
	\end{subfigure}
	\begin{subfigure}{0.48\textwidth}
		\centering
		\includegraphics[width=\textwidth]{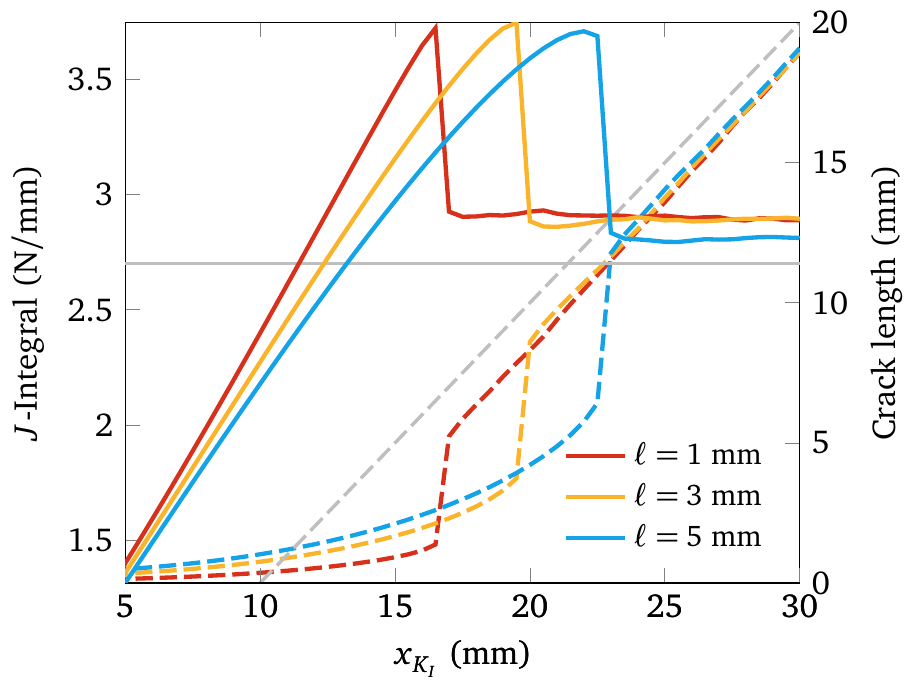}
		\caption{Results for different $\ell$ obtained using quadratic degradation.}
		\label{fig:surfing_quad}
	\end{subfigure}
	\caption{Evolution of $J$-integral and crack length. The solid gray line represents the specified fracture toughness of the material ($\mathcal{G}_c$), while the dashed gray line represents the analytical length of crack extension.}
\end{figure}

To gain insight on the numerical behavior of the fracturing specimen, we conduct a preliminary simulation assuming plain linear elastic response without fracturing, the results for which are shown in \cref{fig:surfing_elastic}.
The energy release rate at the crack tip is obtained by calculating the $J$-integral over the contour defined by the specimen boundary, $\partial \Omega$. One can see that this is underestimated in the numerical solution, i.e. the $J$-integral is less than $\mathcal{G}_c$ at $t = 10$. Consequently, location of the crack tip predicted by the numerical simulation lags behind the analytical location as illustrated by the black and gray dashed lines in \cref{fig:surfing_elastic}. 

\begin{figure}
	\centering
	\begin{subfigure}{0.32\textwidth}
		\includegraphics[width=\textwidth]{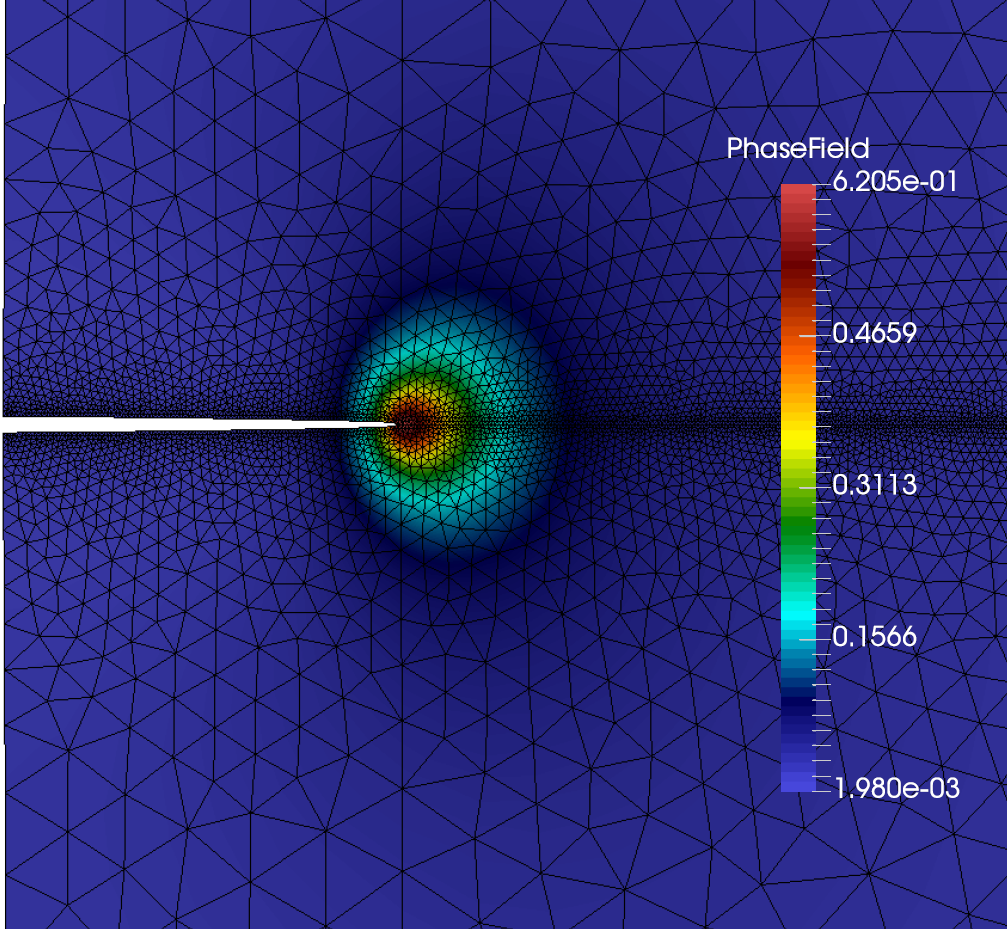}
		\caption{$\ell = 1$ mm}
	\end{subfigure}
	\begin{subfigure}{0.32\textwidth}
		\includegraphics[width=\textwidth]{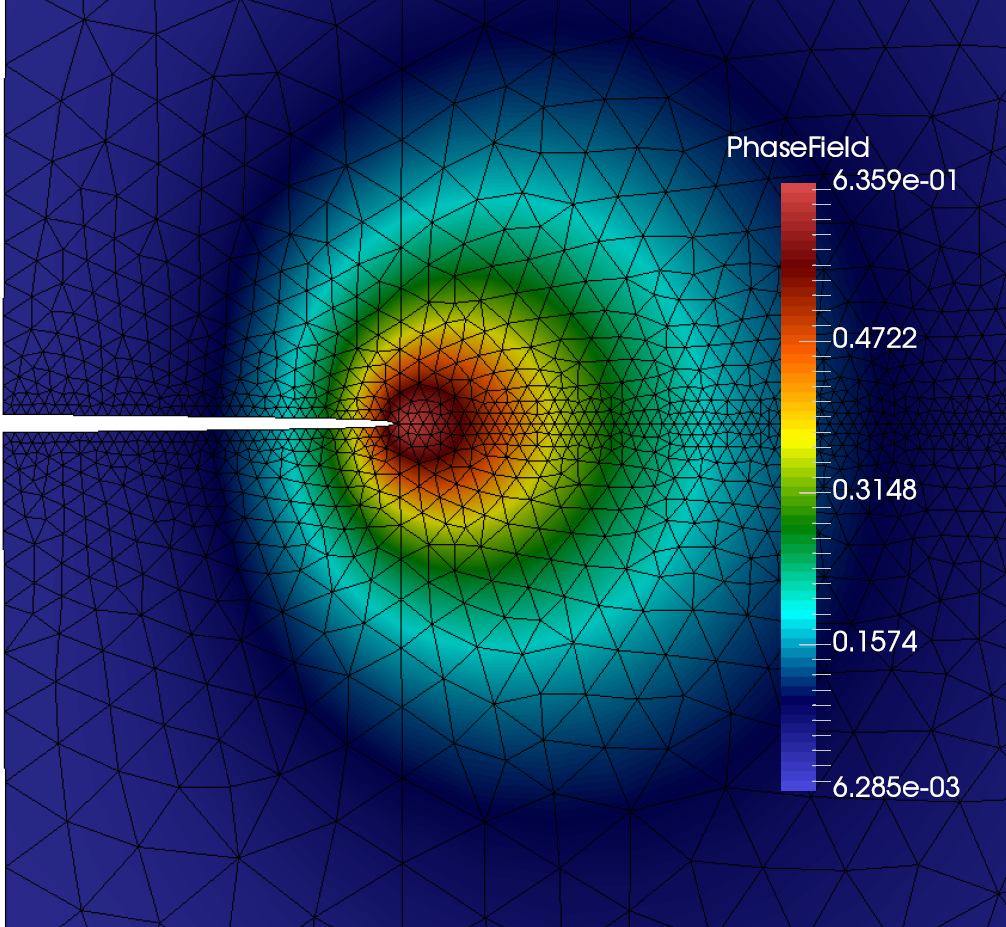}
		\caption{$\ell = 3$ mm}
	\end{subfigure}
	\begin{subfigure}{0.32\textwidth}
		\includegraphics[width=\textwidth]{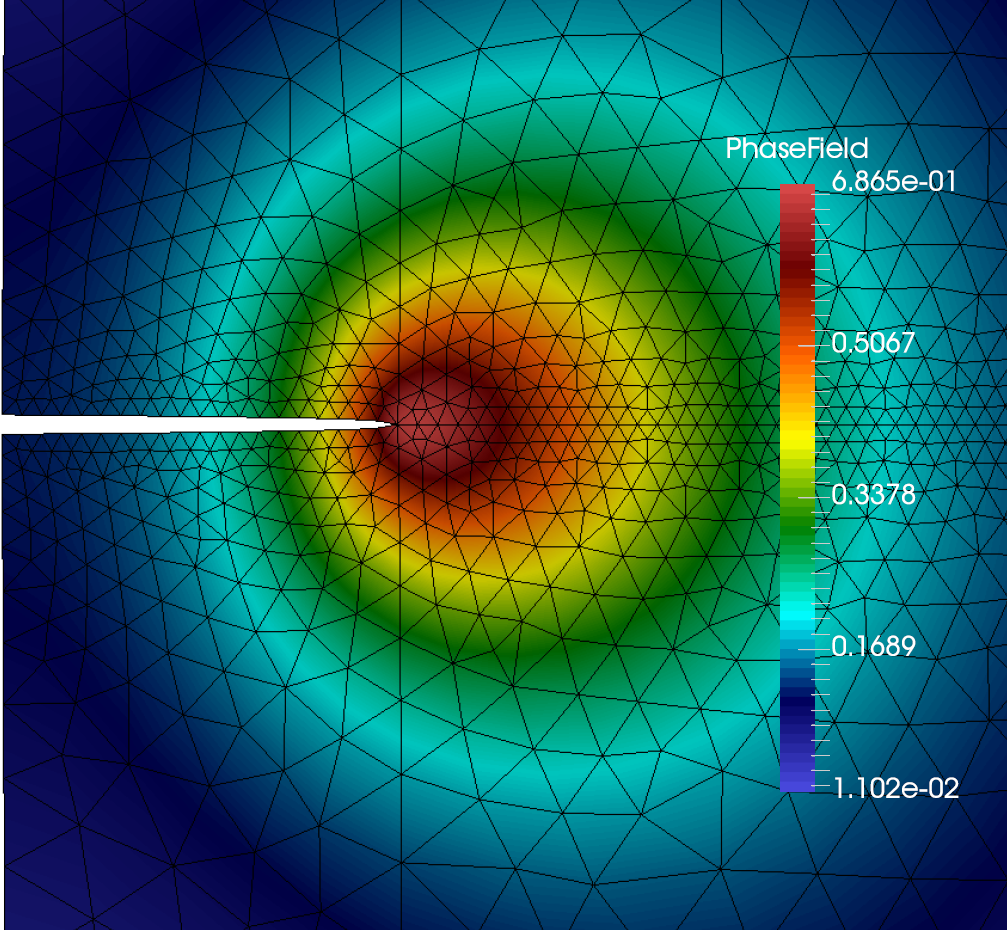}
		\caption{$\ell = 5$ mm}
	\end{subfigure}
	\caption{Phase-field profile at the crack tip prior to the occurrence of brutal cracking in the surfing problem utilizing quadratic degradation.}
	\label{fig:diffuse_tip}
\end{figure}

\begin{figure}
	\centering
	\begin{subfigure}{0.45\textwidth}
		\includegraphics[width=\textwidth]{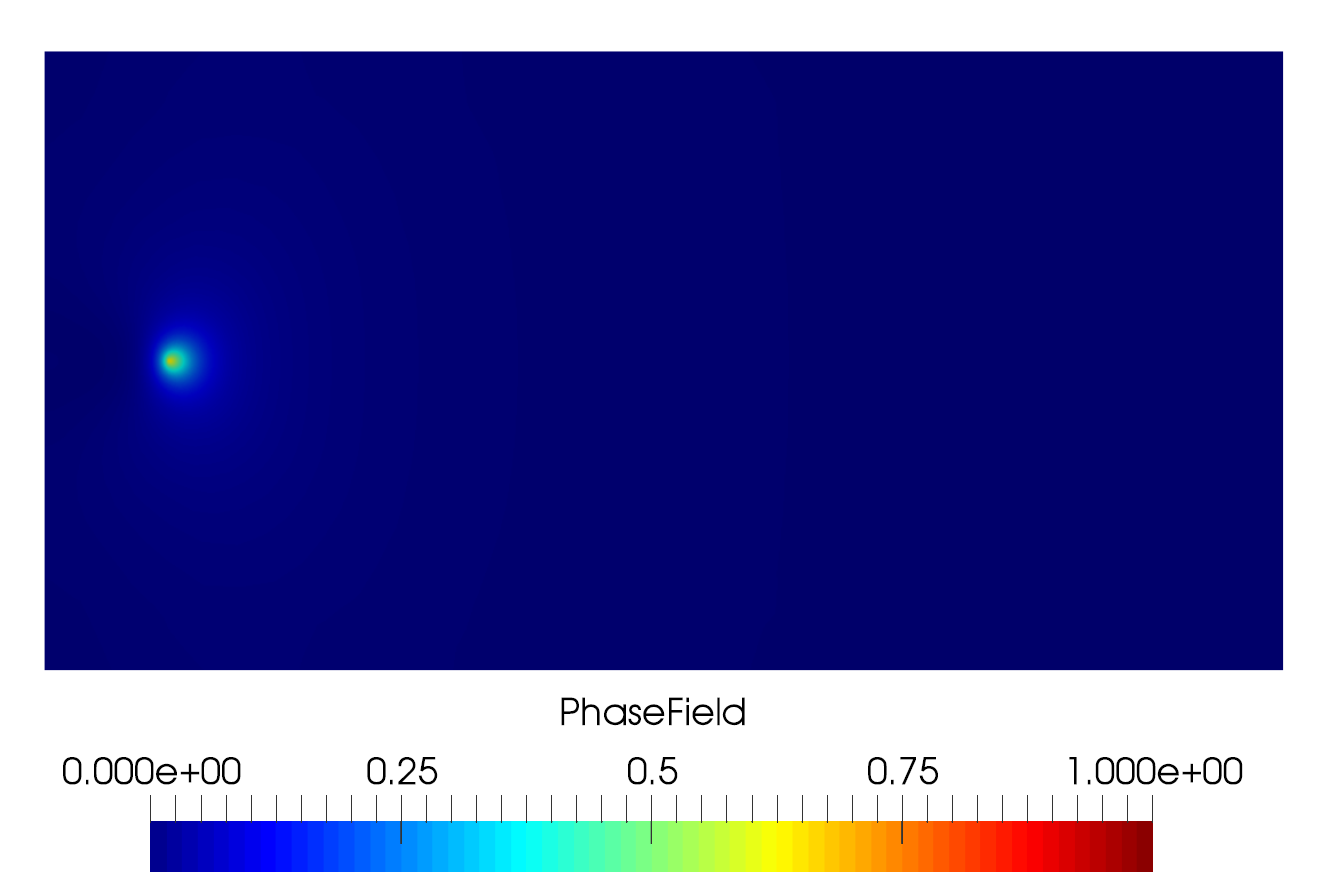}
		\caption{$t = 16.5$}
	\end{subfigure}
	\begin{subfigure}{0.45\textwidth}
		\includegraphics[width=\textwidth]{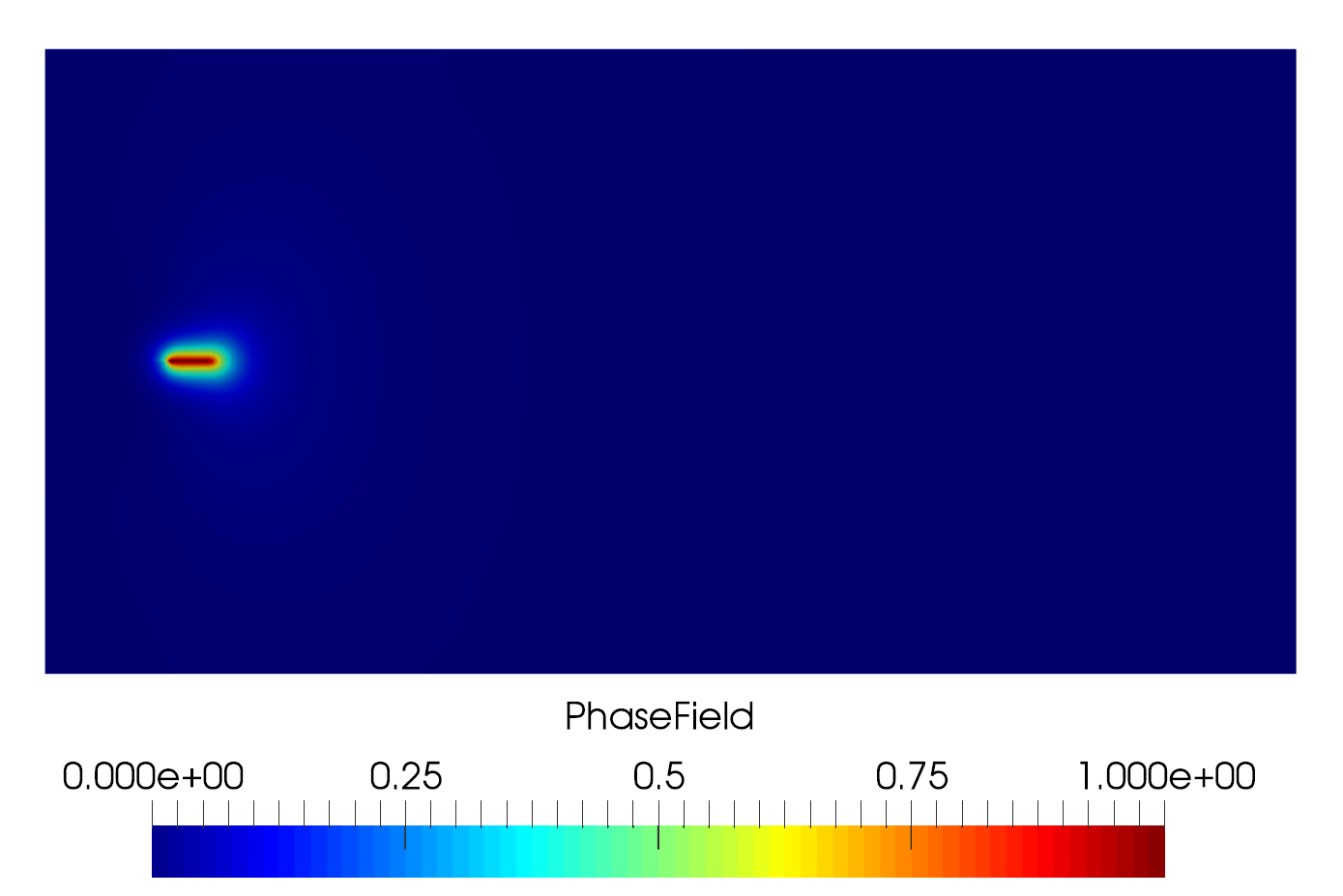}
		\caption{$t = 17$}
	\end{subfigure}
	\caption{Phase-field profiles before and after the crack length jump in the surfing problem utilizing quadratic degradation ($\ell = 1$ mm).}
	\label{fig:brutal_cracking}
\end{figure}

We now examine phase-field model behavior in connection with the quadratic degradation function for three different values of the regularization parameter, namely $\ell = 1$ mm, 3 mm and 5 mm. Again this is done by looking at two quantities of interest: the energy release rate at the crack tip represented by the $J$-integral, and the length of crack extension described by the phase-field that is obtained by evaluating the functional $\Gamma \left( \phi \right) = \int_\Omega \left( \frac{1}{2\ell} \phi^2 + \frac{\ell}{2} \nabla\phi \cdot \nabla\phi \right) \dee\Omega$. The results for different values of $\ell$ are summarized in \cref{fig:surfing_quad} and exhibit similar behavior. We first observe a zone of premature crack growth where the crack length is seen to increase at rates less than $v$. This preliminary growth is not related to any physical extension of the crack but is in fact due to the evolution of the phase-field profile representing the diffuse crack tip, illustrated in \cref{fig:diffuse_tip}.
This is followed by brutal cracking represented by a sudden increase in the crack length (see \cref{fig:brutal_cracking}), after which the fracture grows stably at a rate more or less equal to $v$.
As expected, the numerical location of the crack tip trails the $K$-field center associated with the applied boundary condition. In addition, the critical energy release rate is overestimated in the region of stable crack growth, i.e. $\mathcal{G}_c^\text{num} > \mathcal{G}_c$. This phenomenon has been previously reported in the literature, and we refer the reader to \cite{Hossain2014} and \cite{Kuhn2016} for more detailed discussions on the matter. Our main focus at the moment is the overshoot that occurs in the $J$-integral prior to the onset of fracture, which results in even further delay of the actual crack extension. Such behavior is obviously unphysical, and more so does not occur when the initial crack is described by the phase-field as demonstrated in various numerical examples from the aforementioned literature. It is our belief that this artifact is heavily dependent on the specific form of the degradation function. More importantly in this case, the overshoot does not decrease with smaller $\ell$. The results shown in \cref{fig:surfing_quad} provide evidence that one cannot in general rely on the strategy of calibrating $\ell$ in order to obtain correct model behavior, and furthermore casts doubt on the notion that $\ell$ should be viewed as a material parameter, particularly in connection with the reproduction of $\mathcal{G}_c$. As previously mentioned, the overshoot of the critical energy release rate results in a delay of actual crack extension such that by the time it occurs, there is an excess in bulk energy that must be dissipated. Upon the onset of fracture, instantaneous catch-up growth occurs resulting in a \emph{finite} increase of the crack length as shown in \cref{fig:brutal_cracking}.

\begin{figure}
	\begin{subfigure}{0.48\textwidth}
		\centering
		\includegraphics[width=\textwidth]{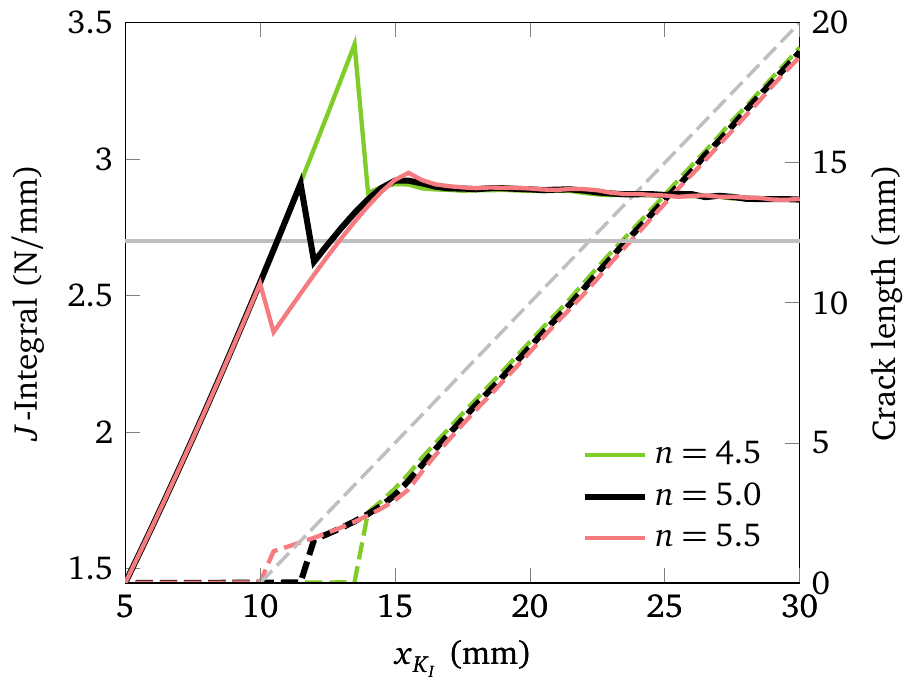}
		\caption{$\ell = 1$ mm}
		\label{fig:surfing_exp_1}
	\end{subfigure}
	\begin{subfigure}{0.48\textwidth}
		\centering
		\includegraphics[width=\textwidth]{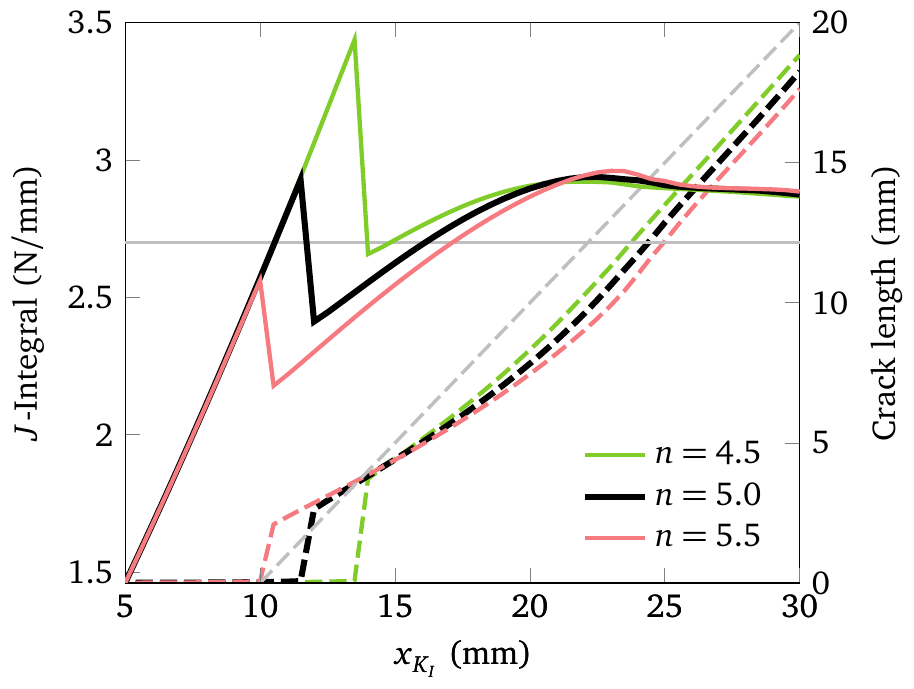}
		\caption{$\ell = 3$ mm}
		\label{fig:surfing_exp_3}
	\end{subfigure}
	\caption{Evolution of $J$-integral and crack length. The solid gray line represents the specified fracture toughness of the material ($\mathcal{G}_c$), while the dashed gray line represents the analytical length of crack extension.}
	\label{fig:surfing_exp}
\end{figure}

\begin{figure}
	\centering
	\begin{subfigure}{0.45\textwidth}
		\centering
		\includegraphics[width=\textwidth]{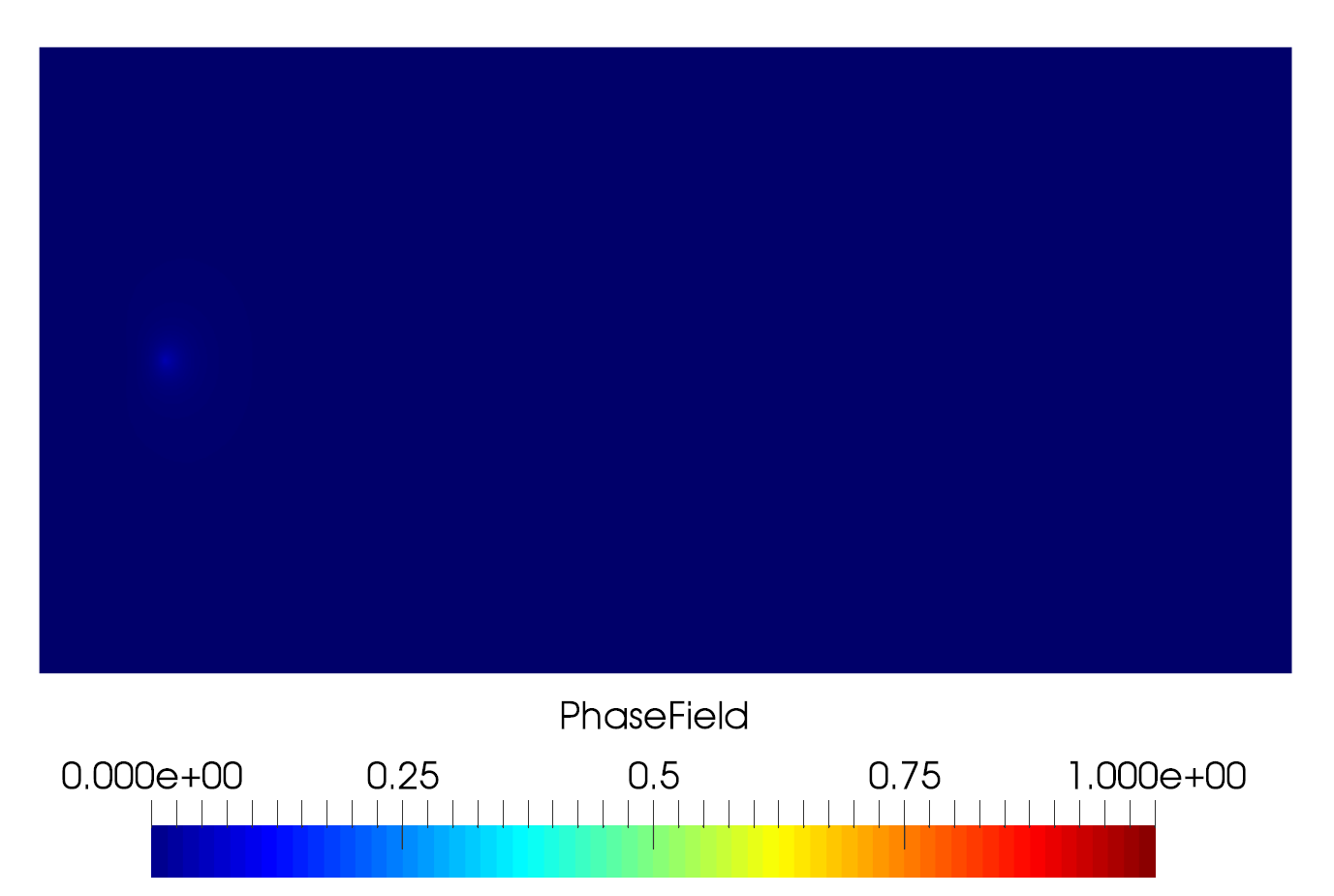}
		\caption{$t = 11.5$}
	\end{subfigure}
	\begin{subfigure}{0.45\textwidth}
		\centering
		\includegraphics[width=\textwidth]{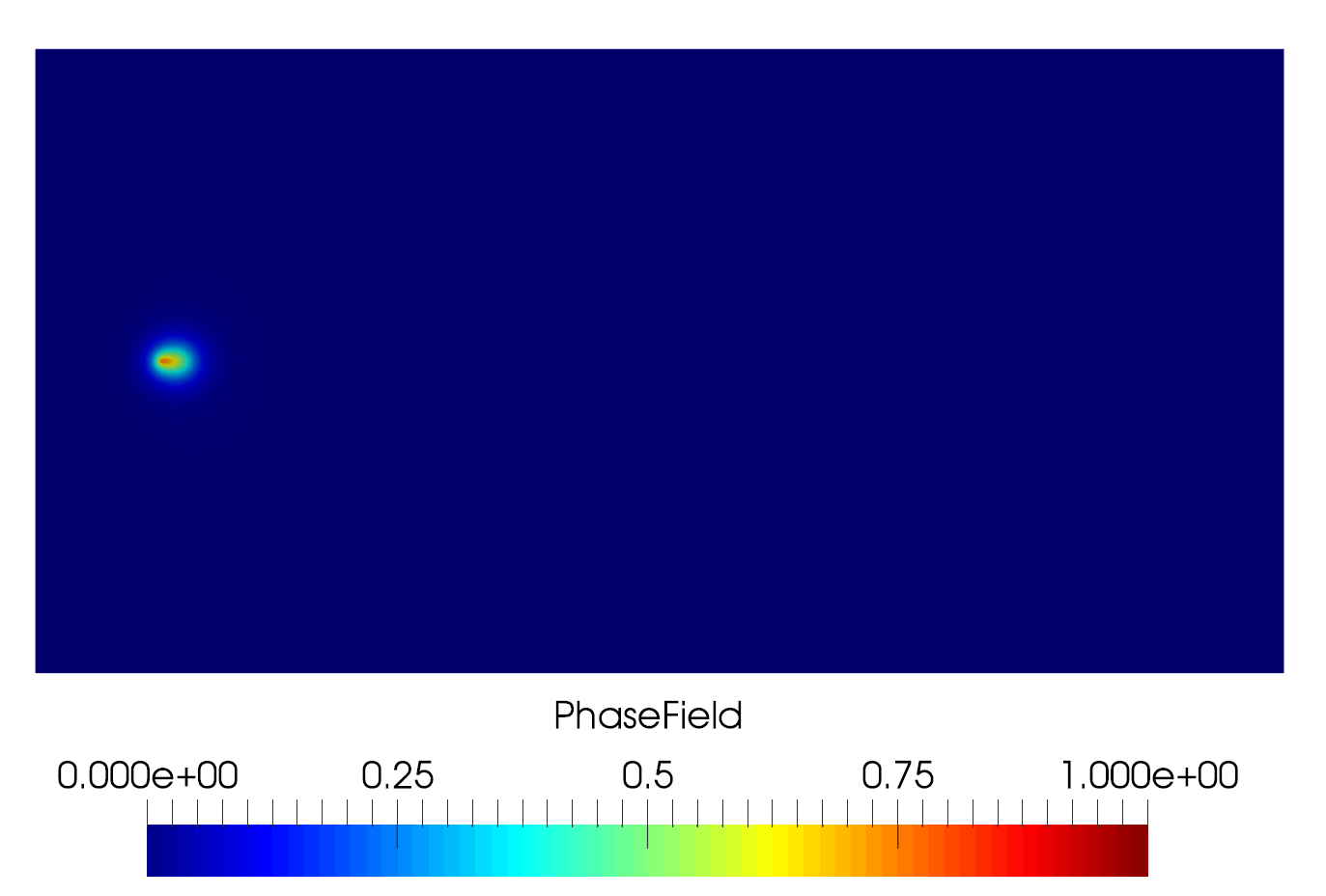}
		\caption{$t = 12$}
	\end{subfigure}
	\caption{Evolution of phase-field at the crack tip for the surfing problem, using the proposed single-parameter degradation function ($n = 5.0$, $\ell = 1$ mm).}
	\label{fig:tip_formation}
\end{figure}

\begin{figure}
	\centering
	\begin{subfigure}{0.45\textwidth}
		\centering
		\includegraphics[width=\textwidth]{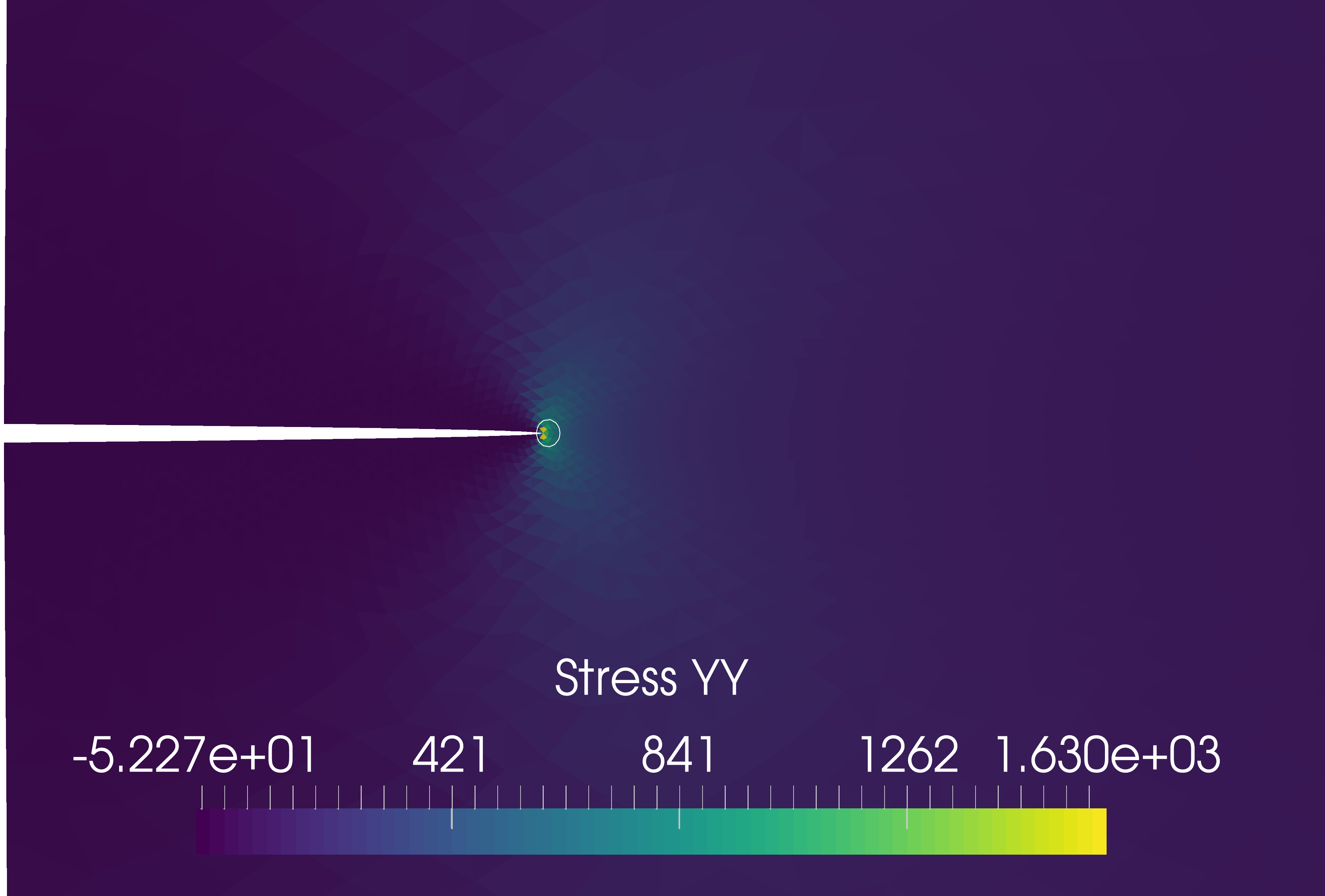}
		\caption{$t = 11.5$}
	\end{subfigure}
	\begin{subfigure}{0.45\textwidth}
		\centering
		\includegraphics[width=\textwidth]{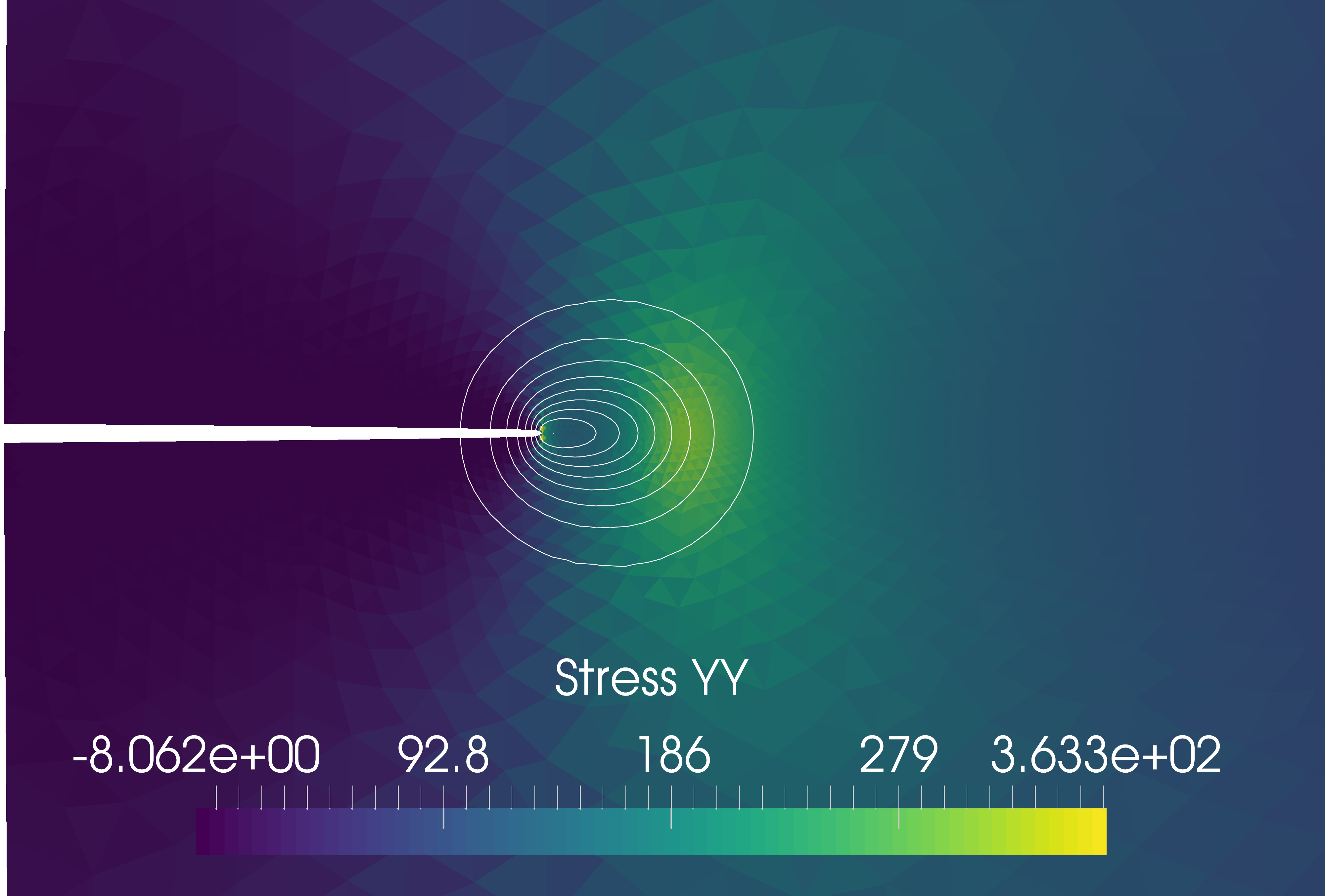}
		\caption{$t = 12$}
	\end{subfigure}
	\caption{Plots of $\sigma_{yy}$ at the crack tip vicinity immediately before and after the start of crack growth in the surfing problem, using the proposed single-parameter degradation function with $n = 5.0$ and $\ell = 1$ mm. The phase-field profile is indicated by the superimposed contours.}
	\label{fig:stress_unloading}
\end{figure}

On the other hand, simulations utilizing the proposed single-parameter degradation show behavior that more closely reflects the physics as shown in \cref{fig:surfing_exp}.
As with the previous results pertaining to quadratic degradation, we can observe that for the same value of the degradation parameter $n$, varying the magnitude of $\ell$ has little effect on the amount of spurious overshoot in the energy release rate prior to crack extension. Rather, the parameter $n$ itself is effective in controlling this feature, and thus can be calibrated such that crack extension occurs when the $J$-integral reaches a value of $\mathcal{G}_c^\text{num}$. Additionally the phase-field remains at very low values before the onset of fracture, resulting in negligible increase of the crack length prior to the actual onset of crack growth as seen in \cref{fig:tip_formation}.
Nonetheless, we can observe from \cref{fig:surfing_exp} that setting $n$ too low results in an overshoot behavior similar to the case of quadratic degradation, while for sufficiently high values of $n$ a dip occurs in the $J$-integral following onset of crack extension. The latter is also a numerical artifact, with an underlying mechanism that is converse to what occurs for overshooting. That is, $\phi$ experiences a jump in value at the crack tip around which an exponentially decaying profile is enforced by the evolution equation for the phase-field. This leads to the crack extension being too big, so that now the numerical crack tip may be understood to have jumped \emph{ahead} of $x_{K_I}$. The result is a virtual unloading at the crack tip vicinity evidenced by the decrease in maximum tensile stress shown in \cref{fig:stress_unloading}.
However since the crack tip diffusion is controlled by the phase-field length scale, the aforementioned dip may be reduced by simply decreasing the magnitude of $\ell$.

\section{Concluding remarks}
In this paper, we have introduced a novel family of energy degradation functions aimed at overcoming major drawbacks of the standard phase-field model in simulating fracture nucleation in brittle materials. A key feature of these functions is their dependence on a set of parameters, permitting us to effect minute changes to their shape. This allows for a more detailed study on how the form of the degradation function influences the phase-field model response independent of the regularization parameter $\ell$. Of particular interest is the discovery that use of the standard quadratic degradation function leads to a delay in the onset of crack propagation, leading to an overshoot in predicted critical loads which \emph{cannot} be ameliorated by adjusting the value of $\ell$. This finding is remarkable, since such a strategy was previously thought to be adequate for recovering correct failure loads based on prior numerical examples found in the literature. Furthermore, it casts doubt on the idea that $\ell$ should be viewed as a material parameter, at least for the case involving brittle fracture of linear elastic materials. On the other hand with the proposed family of degradation functions, it is possible to obtain significantly more accurate simulations provided that proper calibration of the function parameters is carried out. The computational overhead resulting from the consequent nonlinearity of the phase-field evolution equation can be virtually eliminated by employing suitable linear approximations  for the tangent matrices which then allows for straightforward application of the alternate minimization algorithm. 

An important consideration for the proposed family of degradation functions is the actual number of independent parameters that must be specified, since this directly affects the difficulty or ease of calibration. In this paper we have chosen to work with a function that has only one parameter to be calibrated out of an initial four, in the belief that more would render the model unappealing for use in an industry setting. Consequently, we do not take full advantage of the flexibility of our model. Furthermore the elimination of extra, unwanted parameters was done based on a rationale that prioritized the preservation of linear elastic response prior to fracture. Looking at results of the numerical examples we can see that this objective has been sufficiently accomplished, however the price to pay is a spurious dip in the bulk energy after the initial crack nucleation which occurs even with proper calibration as seen in the surfing problem. In the current model, this can only be alleviated by reducing the phase-field regularization which in turn increases computational expense due to meshing requirements along the crack trajectories. As an alternative, one can allow damage to occur gradually in the vicinity of the crack nucleation point prior to failure, however this requires a degradation of the bulk energy to preserve energy balance and runs counter to the rationale mentioned above. It is thus outside the scope of the present paper, and will be explored in a future work.

\section*{Acknowledgements}
This work was funded by the Research Council of Norway through grant no. 228832/E20 and Statoil ASA through the Akademia agreement. The authors likewise express their gratitude to Prof.\ Blaise Bourdin for suggesting the use of the surfing boundary problem as a test case for model comparison.

\appendix
\section{Derivation of $k \left( n \right)$} \label{sec:k_derivation}
\begin{figure}
	\centering
	\includegraphics[width=0.5\textwidth]{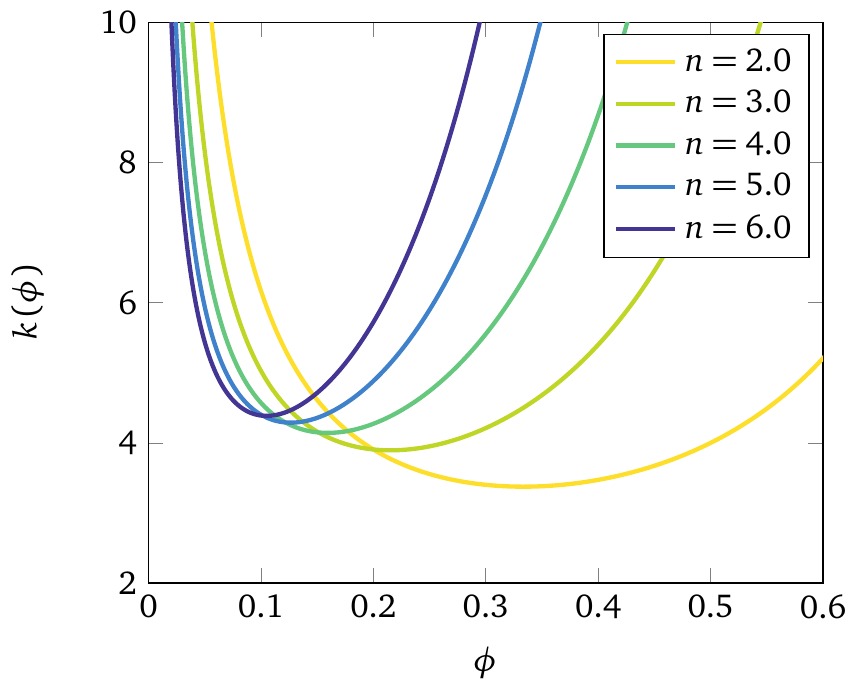}
	\caption{Behavior of $k \left( \phi, n \right)$.}
	\label{fig:k_dependence}
\end{figure}

Let $g \left( \phi \right)$ be defined according to Eq.\ \eqref{eq:2ParamDegFcn}. The corresponding first and second derivatives are given by
\begin{align}
	\label{eq:gprime}
	g^\prime \left( \phi \right) &= \frac{-kn}{1-e^{-k}} \left( 1-\phi \right)^{n-1} e^{-k \left( 1-\phi \right)^n} - 2q \left( 1-\phi \right) \\
	\label{eq:gdoubleprime}
	g^{\prime\prime} \left( \phi \right) &= \frac{-kn}{1-e^{-k}} \left( 1 - \phi \right)^{n-2} \left[ kn \left( 1 - \phi \right)^n - n + 1 \right] e^{-k \left( 1 - \phi \right)^n} + 2q.
\end{align}
We are interested in the limiting scenario where Eq.\ \eqref{eq:stress_strain_behavior} becomes infinite, i.e.
\begin{equation}
	g^\prime \left( \phi \right) - \phi g^{\prime\prime} \left( \phi \right) = 0.
\end{equation}
Additionally, we will consider only the case where $\phi < 1$, since setting $\phi = 1$ results in a zero numerator in Eq.\ \eqref{eq:stress_strain_behavior}. Plugging Eqs.\ \eqref{eq:gprime} and \eqref{eq:gdoubleprime} into the above expression gives
\begin{align*}
	0 &= \frac{-kn}{1-e^{-k}} \left( 1 - \phi \right)^{n-1} e^{-k \left( 1 - \phi \right)^n} - \frac{-kn}{1-e^{-k}} \phi \left( 1 - \phi \right)^{n-2} \left[ kn \left( 1 - \phi \right)^n - n + 1 \right] e^{-k \left( 1 - \phi \right)^n} - 2q \\
	&= \frac{-kn e^{-k \left( 1 - \phi \right)^n}}{1-e^{-k}} \left( 1 - \phi \right)^{n-2} \left\{ 1 - \phi - \phi \left[ kn \left( 1 - \phi \right)^n - n + 1 \right] \right\} - 2q
\end{align*}
which simplifies to
\begin{equation}
	1 - \phi - \phi \left[ kn \left( 1 - \phi \right)^n - n + 1 \right] = 0.
\end{equation}
Solving for $k$ in the above equation, we obtain
\begin{equation}
	k \left( \phi, n \right) = \frac{\left( n - 2 \right)\phi + 1}{n\phi \left( 1 - \phi \right)^n}.
	\label{eq:k_dependence}
\end{equation}
The behavior of $k \left( \phi, n \right)$ is shown in Figure \ref{fig:k_dependence} for several values of $n$.
We are interested in the minimum possible value of $k$ for each given $n$, hence the relevant condition to consider is
\begin{equation}
	\frac{\partial k}{\partial \phi} = \frac{\left( n^2 - 2n \right)\phi^2 + \left( n+1 \right)\phi - 1}{n\phi^2 \left( 1 - \phi \right)^{n+1}} = 0.
\end{equation}
Assuming that the denominator does not equal zero, the above equation reduces to
\begin{equation}
	\left( n^2 - 2n \right)\phi^2 + \left( n+1 \right)\phi - 1 = 0
\end{equation}
whereupon we obtain the positive root
\begin{equation}
	\phi^\star = \frac{-n-1 + \sqrt{5n^2 - 6n + 1}}{2\left( n^2 -2n \right)}
\end{equation}
via the quadratic formula. The above result is applicable for $n \neq 2$. For the case where $n = 2$, Eq.\ \eqref{eq:k_dependence} becomes
\begin{equation}
	k \left( \phi,2 \right) = \frac{1}{2\phi \left( 1-\phi \right)^2}.
\end{equation}
Proceeding similarly to the previous case, we have
\begin{equation}
	\frac{\partial k }{\partial\phi} = 0 = -\frac{2 \left( 1-\phi \right)^2 - 4\phi \left( 1 - \phi \right)}{4\phi^2 \left( 1-\phi \right)^4} = -\frac{1 - 3\phi}{2\phi^2 \left( 1-\phi \right)^3}
\end{equation}
so that the solution is
\begin{equation}
	\phi^\star = \frac{1}{3}.
\end{equation}

\section*{References}
\bibliographystyle{elsarticle-harv}
\bibliography{Reference}

\end{document}